\documentclass[nofootinbib,preprint,onecolumn,11pt,prd]{revtex4-1}

\usepackage{color}
\usepackage{amsmath,bm,amsfonts}
\usepackage{graphicx}
\usepackage{pgfplots,pgfplotstable}
\usepgfplotslibrary{fillbetween}
\usepgfplotslibrary{groupplots}
\usetikzlibrary{intersections}
\pgfplotsset{compat=newest}

\pgfplotsset{every axis/.style={
    width=12cm,
    height=10cm,
    grid=both,
    scaled ticks=false,
    yticklabel style={/pgf/number format/.cd, fixed,precision=5}
  }
}
\definecolor{darkgreen}{rgb}{0,0.4,0} 
\definecolor{darkblue}{rgb}{0,0,0.6} 
\usepackage[colorlinks=true,linkcolor=darkblue,citecolor=darkgreen]{hyperref}

\newcommand{\as}{\alpha_{\mathrm{s}}}
\newcommand{\LA}{\mathrm{A}}
\newcommand{\LB}{\mathrm{B}}
\newcommand{\LF}{\mathrm{F}}
\newcommand{\scF}{\textsc{f}}

\newcommand{\scH}{\textsc{h}}
\newcommand{\LI}{\mathrm{I}}

\newcommand{\LL}{\mathrm{L}}

\newcommand{\LR}{\mathrm{R}}
\newcommand{\scR}{\textsc{r}}

\newcommand{\LT}{\mathrm{T}}
\newcommand{\La}{\mathrm{a}}
\newcommand{\Lb}{\mathrm{b}}
\newcommand{\Lc}{\mathrm{c}}
\newcommand{\Lf}{\mathrm{f}}

\newcommand{\Ls}{\mathrm{s}}

\newcommand{\LZ}{\mathrm{Z}}

\newcommand{\cD}{\mathcal{D}}
\newcommand{\cF}{\mathcal{F}}

\newcommand{\cH}{\mathcal{H}}

\newcommand{\cM}{\mathcal{M}}

\newcommand{\cS}{\mathcal{S}}

\newcommand{\cU}{\mathcal{U}}
\newcommand{\cV}{\mathcal{V}}

\newcommand{\mur}{\mu^2}
\newcommand{\mus}{\mu_{\mathrm{s}}^2}

\definecolor{darkgreen}{rgb}{0,0.4,0} 

\newcommand{\GeV}{\ \mathrm{GeV}}
\newcommand{\TeV}{\ \mathrm{TeV}}

\definecolor{red}{rgb}{1,0,0}

\def\mi{{\mathrm i}}
\newcommand{\MSbar}{\overline {\text{MS}}}

\def\ket#1{\big|{#1}\big\rangle}
\def\bra#1{\big\langle{#1}\big|}
\def\brax#1{\big\langle{#1}}   
\def\<>#1{\big\langle{#1}\big\rangle}
\def\[]#1{\big[{#1}\big]}

\def\sket#1{\big|{#1}\big)}
\def\sbra#1{\big({#1}\big|}

\newbox\charbox
\newbox\slabox
\def\s#1{{      
        \setbox\charbox=\hbox{$#1$}
        \setbox\slabox=\hbox{$/$}
        \dimen\charbox=\ht\slabox
        \advance\dimen\charbox by -\dp\slabox
        \advance\dimen\charbox by -\ht\charbox
        \advance\dimen\charbox by \dp\charbox
        \divide\dimen\charbox by 2
        \raise-\dimen\charbox\hbox to \wd\charbox{\hss/\hss}
        \llap{$#1$}
}}

\newif\ifusefigs
\usefigstrue

\begin{document}

\title{Jets and threshold summation in Deductor}

\author{Zolt\'an Nagy}

\affiliation{
DESY,
Notkestrasse 85,
22607 Hamburg, Germany
}
\email{Zoltan.Nagy@desy.de}

\author{Davison E.\ Soper}

\affiliation{
Institute of Theoretical Science,
University of Oregon,
Eugene, OR  97403-5203, USA
}

\email{soper@uoregon.edu}

\begin{abstract}
We explore jet physics in hadron collisions using the parton shower event generator \textsc{Deductor}. Of particular interest is the one jet inclusive cross section $d\sigma/dP_\LT$ for jets of very high $P_\LT$. Compared to the Born level, the cross section decreases substantially because of $P_\LT$ loss from the jet during showering. We compare to the same effect in \textsc{Pythia} and \textsc{Dire}. The cross section then increases substantially because of the summation of threshold logarithms included in \textsc{Deductor}. We also study the cross section to have a gap with no jets between two hard jets that are widely separated in rapidity. Here we compare Atlas data to \textsc{Deductor} with virtuality based ordering and to \textsc{Deductor} with $k_\LT$ ordering. We also compare with perturbation theory. In both cases, we check whether adding an underlying event and hadronization has a significant effect beyond that found with just a parton shower.
\end{abstract}

\keywords{perturbative QCD, parton shower}

\preprint{DESY 17-179}

\maketitle


\section{Introduction}
\label{sec:intro}

Parton shower Monte Carlo event generators, such as \textsc{Herwig}  \cite{Herwig}, \textsc{Pythia} \cite{Pythia}, and \textsc{Sherpa} \cite{Sherpa}, perform calculations of cross sections according to an approximation to the standard model or its possible extensions. In most such programs, the shower develops with decreasing values of a parameter that measures of the hardness of interactions: smaller hardness corresponds to a larger scale of space-time separations. Thus a parton shower is essentially an application of the renormalization group. It describes how the description of physics changes as one changes the resolution scale at which a scattering event is examined.

Following this view, we have recently presented a general formulation \cite{NSallorder} of how a parton shower can be defined at any order of QCD perturbation theory by using an evolution equation based on operators that characterize the infrared singular behavior of QCD with a variable resolution scale. The current version\footnote{Version 2.1.1 of the code, 
  used in this paper, is available at 
  \href{http://www.desy.de/~znagy/deductor/}
  {http://www.desy.de/$\sim$znagy/deductor/}
  and
  \href{http://pages.uoregon.edu/soper/deductor/}
  {http://pages.uoregon.edu/soper/deductor/}. Matching to a next-to-leading order  
  perturbative calculation of the hard scattering cross section in included in 
  Ref.~\cite{NSallorder} but not in \textsc{Deductor} v.\ 2.1.1.} 
of the parton shower program \textsc{Deductor} \cite{NSI, NSII, NSspin, NScolor, Deductor, ShowerTime, PartonDistFctns, ColorEffects, DuctThreshold} approximately follows this framework, although in its present version its splitting functions are available only to order $\as$.

In \textsc{Deductor} and other parton shower programs, a hard interaction, based on a new physics model or on the electroweak part of the standard model or on just QCD, initiates an event. At this stage, there are just a few partons. Then, as the hardness decreases, the partons that carry QCD color split, making more partons in a parton shower. Thus the program describes the development of QCD jets.

In this paper, we use \textsc{Deductor} to explore the description of QCD jets created in a $2 \to 2$ QCD hard scattering. We study two problems. 

First, we look at the inclusive cross section to produce a jet with a large transverse momentum, $P_\LT > 0.3 \TeV$ with $\sqrt s = 13 \TeV$. Here the cross section is falling quickly as $P_\LT$ increases because the relevant parton distribution functions are falling quickly. This creates two important effects. One effect is related to the jet definition: any momentum lost from the jet or gained by the jet changes the cross section dramatically. The other effect arises from threshold logarithms, which, in a parton shower, arise from the relation between initial state radiation and the evolution of the parton distribution functions.

Second, we examine the gap survival probability: when two jets are produced with a wide separation $\Delta y$ in rapidity, this is the probability that there are no other jets between these two with transverse momenta greater than some specified value $p_\LT^{\rm cut}$. This problem is of interest because its theoretical description involves the summation of logarithms of the ratio of the jet $P_\LT$ to $p_\LT^{\rm cut}$ and the summation of the large logarithmic factor $\Delta y$. It is not initially evident whether a parton shower approach can provide a good description of this physics when $\Delta y$ is very large. We compare \textsc{Deductor} results to NLO perturbation theory and to data from Atlas for $\sqrt s = 7 \TeV$.

For these studies, we make use of some features of the current version of \textsc{Deductor}, version 2.1.1, that were not available in earlier versions. First, we can add a simple model for a nonperturbative underlying event and we can include the string model of hadronization provided by \textsc{Pythia}. Second, we can compare results obtained with the default hardness measure of \textsc{Deductor}, which is based on the virtuality in parton splittings, with results obtained with a transverse momentum hardness measure. Here we change only the hardness measure that orders splittings, leaving everything else unchanged. Finally, we include factors that sum threshold logarithms \cite{Sterman1987, AppellStermanMackenzie, Catani32, CataniWebberMarchesini, Magnea1990, Korchemsky1993, CataniManganoNason32, SudakovFactorization, KidonakisSterman, KidonakisOderdaSterman, KidonakisOderdaStermanColor, LaenenOderdaSterman, CataniManganoNason, LiJointR, KidonakisOwensPhotons, StermanVogelsang2001, KidonakisOwensJets, LSVJointR, KSVJointR, Kidonakis2005, MukherjeeVogelsang, Bolzoni2006, deFlorianVogelsang, RavindranSmithvanNeerven, RavindranSmith, BonviniForteRidolfi, BFRSaddlePt, deFlorianJets, Catani:2014uta, Ahmed:2014uya, BonviniMarzani, Bonvini:2015ira, ManoharSCET, IdilbiJiSCET, BecherNeubert, BecherNeubertPecjak, BecherNeubertXu, BecherSchwartz, Stewart:2009yx, Beneke:2010da, Bauer:2011uc, Wang:2014mqt, Lustermans:2016nvk, KrkNLO, Moch1, Moch2, Bonvini:2012az, StermanZeng, Bonvinietalcompare}, which are important when the scale of the hard interaction is large. We included a summation of threshold logs in an earlier paper \cite{DuctThreshold}. In that paper, the infrared behavior of the threshold contributions was not sufficiently well controlled, requiring an infrared cutoff. In this paper, we use an improved version based on the general formulation of Ref.~\cite{NSallorder}. Then no infrared cutoff is needed. We provide details of the threshold factor in Sec.~\ref{sec:threshold} and the Appendix.

\section{New features in \textsc{Deductor}}
\label{sec:DuctNew}

Our analysis is based on the parton shower event generator, \textsc{Deductor}.  In this section, we describe new features of \textsc{Deductor} that are not described in our previous papers \cite{NSI, NSII, NSspin, NScolor, Deductor, ShowerTime, PartonDistFctns, ColorEffects, DuctThreshold}.

\subsection{Non-perturbative effects}
\label{sec:nonperturbative}

One can use a parton shower event generator to make predictions for observables that are sensitive to non-perturbative physics. For example, one can ask what is the number of charged particles in an event. For this kind of prediction, we need to include a model of non-perturbative physics in the parton shower code and we  need to carefully fit this model to data. 

There is, however, a class of observables for which the only non-perturbative input needed, in principle, is the parton distribution functions. These observables are generally known as {\em infrared safe}. However, even with an infrared safe observable, one should be careful. The cross section for the calculated observable $J$ will have the form
\begin{equation}
\sigma[J] = \sigma[J, \textrm{pert}]\times(1 + \Delta[J])
\;.
\end{equation}
Here $\sigma[J, \textrm{pert}]$ has an expansion in powers of $\as$, and our calculation of it may include a summation of large logarithms.
An non-perturbative error term, $\Delta[J]$ is necessarily present, no matter how many orders of perturbation theory we use for $\sigma[J, \textrm{pert}]$. We may be able to neglect it because it is suppressed by a power of $(1 \GeV)/Q[J]$, where $Q[J]$ is a large scale characteristic of the observable. (We discuss this in some detail in Ref.~\cite{NSallorder}.) It can be that we know that the observable of interest has the mathematical property of being infrared safe, but we don't know how big $Q[J]$ is. In that case, it is worthwhile to be able to estimate $\Delta[J]$ in the context of the parton shower used. To do that, we need to add a non-perturbative model to the parton shower. For this purpose, we may not need a complicated model that is carefully tuned to a variety of data. A simple model may do. Suppose that the simple non-perturbative model confirms that $\Delta[J]$ is indeed small for the infrared safe observable $J$ of interest. Then we can be reasonably confident in using the results from our model as a rough estimate of $\Delta[J]$. If the simple model surprises us and tells us that $\sigma[J, \textrm{pert}]$ has, say, 50\% corrections from non-perturbative effects, then we will need to think again.

In \textsc{Deductor}, we can use a model to estimate a non-perturbative correction $\Delta[J]$. The \textsc{Deductor} shower stops when the splittings become too soft: splittings with transverse momenta $k_\LT$ smaller than $k_\LT^{\rm min} \approx 1 \GeV$ are not allowed.\footnote{In this paper, the precise value is  $k_\LT^{\rm min} = 1.0 \GeV$ for final state splittings and $k_\LT^{\rm min} = 1.295 \GeV$, set by the starting scale of the parton distributions that we use, for initial state splittings.} To estimate the contribution $\Delta[J]$ from physics at scales below $k_\LT^{\rm min}$, we use a non-perturbative model that includes two features. First, we supply a model for the ``underlying event,'' which we take to include soft scatterings of partons not involved in the primary hard scattering together with any radiation from the two initial state partons that involves transverse momentum scales smaller than $k_\LT^{\rm min}$. Second, we need a model for how partons turn into hadrons, for which we use the model in \textsc{Pythia} by sending the partonic final state at the end of the shower to \textsc{Pythia}.

At the end of the \textsc{Deductor} shower, we have a number $N$ of final state partons plus one initial state parton from each beam. These partons carry quantum color, which means that their color configuration is described by a color density matrix element $\ket{\{c\}_N}\bra{\{c'\}_N}$ \cite{NSI, NScolor}. Here $\ket{\{c\}_N}$ is a color basis state, which, for example, can consist of a quark and an antiquark and several gluons, all linked to each other, with a quark having one $\bm 3$ link, an antiquark having one $\bar{\bm 3}$ link, and a gluon having two links, one $\bm 3$ and one $\bar{\bm 3}$. The total probability associated with this color density operator basis element is the corresponding trace of the color density operator, $\brax{\{c'\}_N} \ket{\{c\}_N}$. We need to connect this with the color string model used in \textsc{Pythia}, which is based on classical string states $\{c_\Lf\}_N$.  If we use the leading color approximation, then the quantum color state has $\{c\}_N = \{c'\}_N$. Given the way that the quantum color states are defined, it is then evident that the corresponding classical string state should be $\{c_\Lf\}_N = \{c\}_N$ to leading power in $1/N_\Lc^2$, where $N_\Lc = 3$ is the number of colors. If we use the LC+ approximation for color that is available in \textsc{Deductor}, then we can have $\{c\}_N \ne \{c'\}_N$. In this case the corresponding total probability $\brax{\{c'\}_N} \ket{\{c\}_N}$ is proportional to $1/N_\Lc^I$ with $I \ge 1$. Then our hadronization model can be to use \textsc{Pythia} hadronization with $\{c_\Lf\}_N = \{c\}_N$ with probability 1/2 and with $\{c_\Lf\}_N = \{c'\}_N$ with probability 1/2. This is in fact the lowest order version of a more general algorithm that is specified in Sec.~8 of Ref.~\cite{NScolor}. One could improve this by using a higher order version of this algorithm, but for this paper we use just the lowest order version because of its simplicity.

The first step of linking to the Pythia hadronization model is to replace each of the two incoming partons by an outgoing parton hole with the same color links as the incoming parton, the opposite flavor, and a substantial fraction of the momentum of the hadron. Assuming that the incoming parton flavors were massless, we let the parton hole in hadron A have momentum $p_\La$ with components\footnote{We use components $p^\pm = (p^0 \pm p^3)/\sqrt 2$.} $p_\La^+ = \frac{1}{2}(p_\LA^+ - P_{\rm FS}^+)$, $p_\La^- = 0$ and $\bm p_\La^\perp = 0$. Here $p_\LA$ is the momentum of the beam hadron A and $P_{\rm FS}$ is the sum of the momenta of the final state partons. Similarly we let the parton hole in hadron B have momentum $p_\Lb^- = \frac{1}{2}(p_\LB^- - P_{\rm FS}^-)$, $p_\Lb^+ = 0$ and $\bm p_\Lb^\perp = 0$. This is adjusted slightly for parton mass in the case that either of the incoming partons is a charm or bottom quark or antiquark. Note that the parton holes make the final state plus the holes a color singlet and give it zero transverse momentum. When \textsc{Pythia} acts on this collection of partons, the color strings attached to the parton holes will create partons that do have a small amount of transverse momentum. 

The amount of transverse momentum associated with the parton holes will be too small to make realistic events, but we have left half of the total momentum $p_\LA + p_\LB - P_{\rm FS}$ to be available for the model of underlying event scattering. In this model, we generate a u-quark and a (ud)-diquark with large positive rapidity associated with proton A and a u-quark and a (ud)-diquark with large negative rapidity associated with proton B. The A quark is color-connected to the B diquark and the B quark is color-connected to the A diquark. Each of the quarks carries a random transverse momentum generated with a gaussian distribution with a typical transverse momentum around 5 GeV but adjusted  with a formula that we have tuned to obtain a reasonable result.  The B diquark carries the opposite transverse momentum to the A quark and the A diquark the opposite transverse momentum to the B quark. We now have a color string from the A quark to the B diquark and we populate this string with several gluons carrying gaussian distributed transverse momenta and uniformly distributed rapidities, with about one gluon per unit rapidity.  The color string from the B quark to the A diquark is populated with the same number of gluons carrying the opposite transverse momenta and uniformly distributed rapidities. We recognize that collisions can be fairly central, with many soft parton-parton interactions, or quite peripheral, with most of the partons missing each other. To model a peripheral component, we produce with probability 1/4 a softer event with typical transverse momenta around 1 GeV and fewer gluons. This procedure for producing partons with a distribution of momenta will generally not exactly conserve the plus and minus components of momenta, so we adjust the momenta of all of the produced partons by applying a rescaling and a Lorentz boost in the z-direction so that momentum is conserved. We now have an underlying event consisting of several partons that are widely dispersed in rapidity and carry small transverse momenta. The partons are color-connected to each other along two color strings. This configuration is sent to Pythia, which turns it into hadrons.

This model is, evidently, much less sophisticated than the dynamical model in \textsc{Pythia} \cite{SjostrandSkands}. Code for the model described above is quite simple and is separate from the main \textsc{Deductor} code. It is included in the \textsc{Deductor} 2.1.1 distribution as part of the suggested user routines that analyze \textsc{Deductor} events. It contains a number of parameters, which a user can easily adjust.\footnote{In fact, we encourage interested readers to tune this model to data with the aim of making it good enough to use with observables that are {\em not} infrared safe.}

\begin{figure}
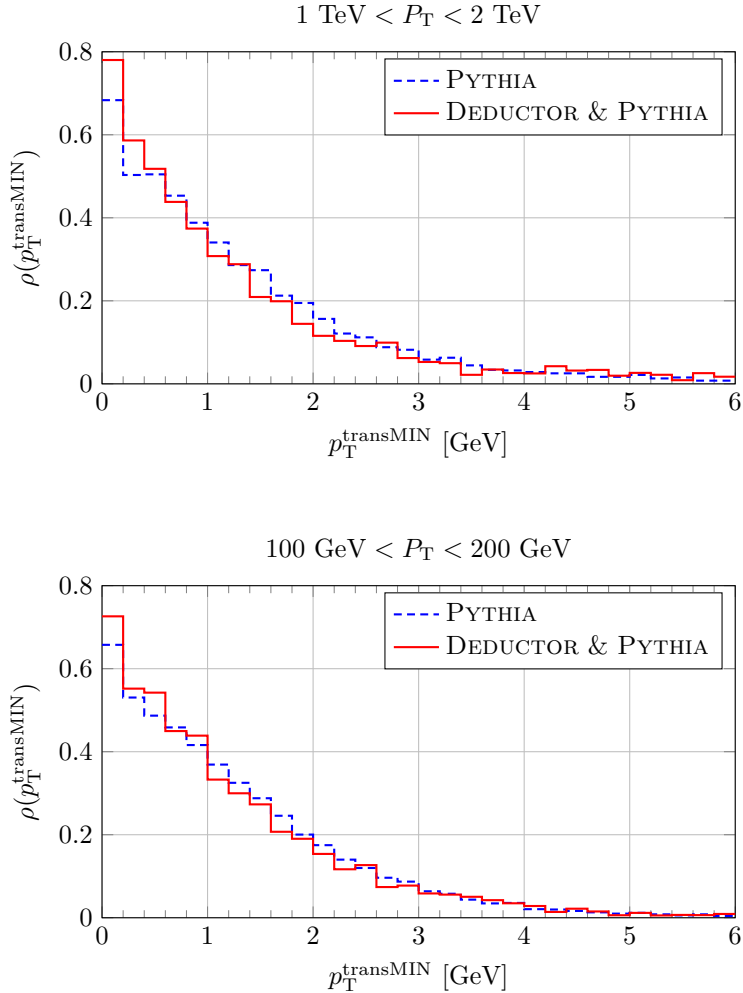

\begin{center}
\ifusefigs 
\begin{tikzpicture}
\begin{axis}[
   title = {$1 \TeV < P_\LT < 2 \TeV$},
   xlabel={$p_\LT^{\rm transMIN}$\ [GeV]}, 
   ylabel={$\rho(p_\LT^{\rm transMIN})$},
   xmin=0.0, xmax=6.0,
   ymin=0.0, ymax=0.8,
  legend cell align=left,
  width=10.0cm,	
  height=6.0cm,
 xminorgrids=false,
 yminorgrids=false,
 minor x tick num=4,
]
\pgfplotstableread{
  0.0000e-01  6.8339e-01
  1.0000e-01  6.8339e-01
  3.0000e-01  5.0314e-01
  5.0000e-01  5.0468e-01
  7.0000e-01  4.5329e-01
  9.0000e-01  3.8810e-01
  1.1000e+00  3.4054e-01
  1.3000e+00  2.8609e-01
  1.5000e+00  2.7382e-01
  1.7000e+00  2.1246e-01
  1.9000e+00  1.9482e-01
  2.1000e+00  1.5647e-01
  2.3000e+00  1.2118e-01
  2.5000e+00  1.1198e-01
  2.7000e+00  8.8204e-02
  2.9000e+00  8.2068e-02
  3.1000e+00  5.8291e-02
  3.3000e+00  6.2893e-02
  3.5000e+00  4.4485e-02
  3.7000e+00  3.3748e-02
  3.9000e+00  3.2214e-02
  4.1000e+00  2.8379e-02
  4.3000e+00  2.5311e-02
  4.5000e+00  2.5311e-02
  4.7000e+00  1.6874e-02
  4.9000e+00  1.6874e-02
  5.1000e+00  2.1476e-02
  5.3000e+00  1.3039e-02
  5.5000e+00  1.5340e-02
  5.7000e+00  7.6699e-03
  5.9000e+00  7.6699e-03
  6.0000e+00  7.6699e-03
}\Pythia

\pgfplotstableread{
  0.0000e-01  7.8013e-01
  1.0000e-01  7.8013e-01
  3.0000e-01  5.8638e-01
  5.0000e-01  5.1799e-01
  7.0000e-01  4.3829e-01
  9.0000e-01  3.7400e-01
  1.1000e+00  3.0776e-01
  1.3000e+00  2.8822e-01
  1.5000e+00  2.0922e-01
  1.7000e+00  1.9894e-01
  1.9000e+00  1.4453e-01
  2.1000e+00  1.1564e-01
  2.3000e+00  1.0348e-01
  2.5000e+00  9.1086e-02
  2.7000e+00  9.9249e-02
  2.9000e+00  6.2100e-02
  3.1000e+00  5.2679e-02
  3.3000e+00  4.9753e-02
  3.5000e+00  2.1732e-02
  3.7000e+00  3.4530e-02
  3.9000e+00  2.5963e-02
  4.1000e+00  2.5090e-02
  4.3000e+00  4.2453e-02
  4.5000e+00  3.1988e-02
  4.7000e+00  3.3578e-02
  4.9000e+00  1.9598e-02
  5.1000e+00  2.6416e-02
  5.3000e+00  2.1841e-02
  5.5000e+00  8.8044e-03
  5.7000e+00  2.5756e-02
  5.9000e+00  1.7160e-02
  6.0000e+00  1.7160e-02
  }\Deductor

\addplot [const plot mark mid, blue, thick, densely dashed] table [x={0},y={1}]{\Pythia};

\addplot [const plot mark mid, red, thick] table [x={0},y={1}]{\Deductor};

\legend{ \textsc{Pythia}, \textsc{Deductor} \& \textsc{Pythia}}

\end{axis}
\end{tikzpicture}

\vskip 0.8cm

\begin{tikzpicture}
\begin{axis}[
   title = {$100 \GeV < P_\LT < 200 \GeV$},
   xlabel={$p_\LT^{\rm transMIN}$\ [GeV]}, 
   ylabel={$\rho(p_\LT^{\rm transMIN})$},
   xmin=0.0, xmax=6.0,
   ymin=0.0, ymax=0.8,
   legend cell align=left,
   width=10.0cm,	
   height=6.0cm,
 xminorgrids=false,
 yminorgrids=false,
 minor x tick num=4,
]
\pgfplotstableread{
  0.0000e-01  6.5748e-01
  1.0000e-01  6.5748e-01
  3.0000e-01  5.3052e-01
  5.0000e-01  4.8688e-01
  7.0000e-01  4.5854e-01
  9.0000e-01  4.1603e-01
  1.1000e+00  3.6898e-01
  1.3000e+00  3.2506e-01
  1.5000e+00  2.8793e-01
  1.7000e+00  2.4571e-01
  1.9000e+00  2.0036e-01
  2.1000e+00  1.7486e-01
  2.3000e+00  1.4000e-01
  2.5000e+00  1.1988e-01
  2.7000e+00  9.6355e-02
  2.9000e+00  8.7003e-02
  3.1000e+00  6.3765e-02
  3.3000e+00  5.7813e-02
  3.5000e+00  4.3643e-02
  3.7000e+00  3.4575e-02
  3.9000e+00  3.5708e-02
  4.1000e+00  2.0688e-02
  4.3000e+00  1.9838e-02
  4.5000e+00  1.6437e-02
  4.7000e+00  1.3036e-02
  4.9000e+00  9.9189e-03
  5.1000e+00  1.1903e-02
  5.3000e+00  8.2186e-03
  5.5000e+00  5.6680e-03
  5.7000e+00  8.2186e-03
  5.9000e+00  4.2510e-03
  6.0000e+00  4.2510e-03
  }\Pythia

\pgfplotstableread{
  0.0000e-01  7.2613e-01
  1.0000e-01  7.2613e-01
  3.0000e-01  5.5197e-01
  5.0000e-01  5.4230e-01
  7.0000e-01  4.4973e-01
  9.0000e-01  4.3872e-01
  1.1000e+00  3.3283e-01
  1.3000e+00  2.9985e-01
  1.5000e+00  2.7315e-01
  1.7000e+00  2.0712e-01
  1.9000e+00  1.9025e-01
  2.1000e+00  1.5375e-01
  2.3000e+00  1.1677e-01
  2.5000e+00  1.2683e-01
  2.7000e+00  7.3939e-02
  2.9000e+00  7.7492e-02
  3.1000e+00  5.8540e-02
  3.3000e+00  5.5759e-02
  3.5000e+00  5.0468e-02
  3.7000e+00  4.2372e-02
  3.9000e+00  3.4853e-02
  4.1000e+00  2.8281e-02
  4.3000e+00  1.3978e-02
  4.5000e+00  2.1812e-02
  4.7000e+00  1.5166e-02
  4.9000e+00  6.0212e-03
  5.1000e+00  1.1721e-02
  5.3000e+00  5.5577e-03
  5.5000e+00  6.8007e-03
  5.7000e+00  6.3470e-03
  5.9000e+00  9.1400e-03
  6.0000e+00  9.1400e-03
  }\Deductor

\addplot [const plot mark mid, blue, thick, densely dashed] table [x={0},y={1}]{\Pythia};

\addplot [const plot mark mid, red, thick] table [x={0},y={1}]{\Deductor};

\legend{ \textsc{Pythia}, \textsc{Deductor} \& \textsc{Pythia}}

\end{axis}
\end{tikzpicture}
\else 
\includegraphics[width = 10 cm]{figures/fig01A.eps}
\vskip 0.8cm
\includegraphics[width = 10 cm]{figures/fig01B.eps}
\fi
\end{center}
\caption{
Distribution of $p_\LT^{\rm transMIN}$ in jet events. The upper figure is for events with a jet in the range $1 \TeV < P_\LT < 2 \TeV$ while the lower figure is for events with a jet in the range $100 \GeV < P_\LT < 200 \GeV$. The red histogram is from \textsc{Deductor} with an underlying event added, with the partons then turned into hadrons by \textsc{Pythia}. The blue, dashed histogram is the same distribution calculated with \textsc{Pythia} alone.
}
\label{fig:transMIN}
\end{figure}

To see how the model works, we use \textsc{Deductor} with parameters as described in   Sec.~\ref{sec:PythiaDireDeductor} below to generate  p-p scattering events with $\sqrt s = 13 \TeV$, add the parton holes and the underlying event, and send the resulting partonic event to \textsc{Pythia} to be hadronized. For this, we use \textsc{Pythia} \cite{Pythia} version 8.226 with its default parameters but with everything except hadronization turned off. We select events such that, after hadronization, the jet with the highest $P_\LT$ has $P_\LT$ in the range $1 \TeV < P_\LT < 2 \TeV$ and rapidity in the range $|y| < 2$. For these events, we examine a distribution that is sensitive to the underlying event, the $p_\LT^{\rm transMIN}$ distribution. A review can be found in Ref.~\cite{Field}. For each event, let $\bm P_\LT$ be the transverse momentum of the leading jet. Define a transverse unit vector $\bm n$ that is orthogonal to $\bm P_\LT$. Look at all final state charged particles with transverse momenta $\bm p_i$, with $|\bm p_i| > 0.5 \GeV$. Let the particle pseudorapidities be $\eta_i$. Define two angular regions. For region $R_1$, $\bm p_i \cdot \bm n/|\bm p_i| > \sqrt{3}/2$ and $|\eta_i|< 2.5$. For region $R_2$, $\bm p_i \cdot \bm n/|\bm p_i| < - \sqrt{3}/2$ and $|\eta_i|< 2.5$. Sum the absolute values of transverse momenta of the hadrons in the two regions and divide by the solid angle, $5\pi/3$, contained in each region:
\begin{equation}
\begin{split}
p_\LT^{(1)} ={}& \frac{3}{5\pi} \sum_{i\in R_1} |\bm p_i|
\;,
\\
p_\LT^{(2)} ={}& \frac{3}{5\pi} \sum_{i\in R_2} |\bm p_i|
\;.
\end{split}
\end{equation}
The leading jet is well separated in angle from regions $R_1$ and $R_2$. So is any second jet that recoils against the leading jet with the opposite transverse momentum. However, a third jet could be directed into one of regions $R_1$ or $R_2$. In that case, one of $p_\LT^{(1)}$ and $p_\LT^{(2)}$ would be large, while the other would be small. We would like to examine the nature of the underlying event without being confused by perturbatively produced jets. For this reason, we look at
\begin{equation}
p_\LT^{\rm transMIN} = \min(p_\LT^{(1)},p_\LT^{(2)})
\end{equation}
for each event. One often looks at the average, $\langle p_\LT^{\rm transMIN} \rangle$, but we can look at the whole distribution, $\rho(p_\LT^{\rm transMIN})$, normalized to 
\begin{equation}
\int_0^{20 \GeV}\! dp_\LT^{\rm transMIN}\
\rho(p_\LT^{\rm transMIN})
= 1
\;.
\end{equation}
We can examine the distribution $\rho(p_\LT^{\rm transMIN})$ in events produced by \textsc{Deductor} and hadronized by \textsc{Pythia}. We can also examine $\rho(p_\LT^{\rm transMIN})$ in events produced entirely by \textsc{Pythia}, which we regard as being a reasonably good model for nature. For this purpose we use \textsc{Pythia} \cite{Pythia} version 8.226 for p-p collisions with $\sqrt s = 13 \TeV$ and otherwise with its default parameters.

We show the results of this comparison in the top panel of Fig.~\ref{fig:transMIN}. We see that the \textsc{Deductor} shower and underlying event model with hadronization by \textsc{Pythia} gives a $p_\LT^{\rm transMIN}$ distribution that matches reasonably well the distribution from \textsc{Pythia}. In the lower panel of Fig.~\ref{fig:transMIN} we show the same comparison in which we choose the $P_\LT$ range of the lead jet to be $100 \GeV < P_\LT < 200 \GeV$. Again, the distributions match reasonably well.

The results in Fig.~\ref{fig:transMIN} show that, using either the simple model for the underlying event available with \textsc{Deductor} or the much more sophisticated model in \textsc{Pythia}, particles that carry roughly 1.5 GeV per unit solid angle are added to a jet event by non-perturbative sources. One expects that this is not enough to much affect most infrared safe measurements of the jets. However, the effect can depend on exactly what the measurement is, so it is worthwhile to check quantitatively how big the nonperturbative effect is. Based on Fig.~\ref{fig:transMIN}, we judge that the underlying event model available in \textsc{Deductor} together with hadronization by \textsc{Pythia} is adequate for this purpose.

\subsection{Shower ordering variable}
\label{sec:ordering}

In \textsc{Deductor}, we order splittings according to decreasing values of a hardness parameter. The default choice of the hardness, $\Lambda$, is based on virtuality. For massless partons, the definition is
\begin{equation}
\begin{split}
\label{eq:Lambdadef}
\Lambda^2 ={}& \frac{(\hat p_l + \hat p_{m+1})^2}{2 p_l\cdot Q_0}\ Q_0^2
\hskip 1 cm {\rm final\ state},
\\
\Lambda^2 ={}&  \frac{|(\hat p_\La - \hat p_{m+1})^2|}{2 p_\La \cdot Q_0}\ Q_0^2
\hskip  1 cm {\rm initial\ state}.
\end{split}
\end{equation}
Here the mother parton in a final state splitting has momentum $p_l$ and the daughters have momenta $\hat p_l$ and $\hat p_{m+1}$. For an initial state splitting in hadron A, the mother parton has momentum $p_\La$, the new (in backward evolution) initial state parton has momentum $\hat p_\La$ and the final state parton created in the splitting has momentum $\hat p_{m+1}$. We denote by $Q_0$ a fixed vector equal to the total momentum of all of the final state partons just after the hard scattering that initiates the shower. The motivation for this choice is described in Ref.~\cite{ShowerTime}.

One can make other choices for the hardness variable. For instance, one can use the transverse momentum $k_\LT =|\bm k_\LT|$ in the splitting. Transverse momentum is not a Lorentz invariant concept, so there are various definitions available. The definition used in \textsc{Deductor} is given in Eqs.~(B.10) and (A.8) of Ref.~\cite{DuctThreshold}:
\begin{equation}
\begin{split}
\label{eq:kTdef}
\bm k_\LT^2 ={}& z(1-z)\, (\hat p_l + \hat p_{m+1})^2
\hskip 1 cm {\rm final\ state},
\\
\bm k_\LT^2 ={}&  (1-z)\, |(\hat p_\La - \hat p_{m+1})^2|
\hskip 1 cm {\rm initial\ state}.
\end{split}
\end{equation}
We define the momentum fraction for a final state splitting by
\begin{equation}
\frac{\hat p_{m+1}\cdot \tilde n_l}{\hat p_{l}\cdot \tilde n_l}
= \frac{1-z}{z}
\;,
\end{equation}
where the auxiliary lightlike vector $\tilde n_l$ is defined using the total momentum $Q$ of all of the final state partons:
\begin{equation}
\label{tildenldef}
\tilde n_l = \frac{2 p_l\cdot Q}{Q^2}\, Q - p_l
\;.
\end{equation}
For an initial state splitting, $z$ is the ratio of momentum fractions before and after the splitting: 
\begin{equation}
z = \frac{\eta_\La}{\hat \eta_\La}  
= \frac{p_\La\cdot p_\Lb}{\hat p_\La\cdot p_\Lb}
\;.
\end{equation}
Here $p_\Lb$ is the momentum of the initial state parton from hadron B.

We can change from $\Lambda$ ordering to $k_\LT$ ordering in \textsc{Deductor}. Doing that allows us to investigate the extent to which the choice of ordering variable really matters. In this paper, when we show results obtained with $k_\LT$ ordering, we specify that we have made this choice. The default ordering choice for \textsc{Deductor} is $\Lambda$ ordering, so we do not always specify the ordering explicitly when $\Lambda$ ordering is used.

With $k_\LT$ ordering, the $\bm k_\LT^2$ of each splitting is required to be smaller than the $\bm k_\LT^2$ of the previous splitting. With $\Lambda$ ordering, the $\Lambda^2$ of each splitting is required to be smaller than the $\Lambda^2$ of the previous splitting. In our numerical investigations in this paper, we look at jet production, which begins with a $2 \to 2$ scattering that produces two partons with equal absolute values of transverse momenta, $P_\LT^{\rm Born}$. The Born cross section is calculated with a factorization scale $\mu_\scF^2$ and renormalization scale $\mu_\scR^2$. We set $\mu_\scR^2 = \mu_\scF^2$ and
\begin{equation}
\label{eq:muFduct}
\mu_\scF  = \frac{1}{\sqrt 2}\, P_\LT^{\rm Born} 
\;.
\end{equation}
The motivation for this choice is that the next-to-leading order (NLO) one jet inclusive cross section is quite stable with respect changes in $\mu_\scF^2$ near this point. We then need to choose the scale $\mu_\Ls^{2}$ at which the parton shower starts. With $k_\LT$ ordering, the first splitting is required to satisfy $\bm k_\LT^2 <  \mu_\Ls^2$ where
\begin{equation}
\label{eq:musforkT}
\mu_\Ls = P_\LT^{\rm Born}
\hskip 1 cm k_\LT\ {\rm ordering}.
\end{equation}
With $\Lambda$ ordering, the first splitting is required to satisfy $\Lambda^2 <  \mu_\Ls^2$, where we make a different choice for $\mu_\Ls$. We recognize that the transverse momentum for the first splitting is intrinsically smaller than $\Lambda^2$: $k_\LT^2 = (1-z) \Lambda^2$ for an initial state splitting or $k_\LT^2 = z(1-z) \Lambda^2$ for a final state splitting. In either case, $0 < (1-z) < 1$. Thus, $\Lambda^2$ for the first splitting is on average quite a lot larger than $k_\LT^2$ for this splitting.  Accordingly, we choose 
\begin{equation}
\label{eq:musforLambda}
\mu_\Ls = \frac{3}{2}\,P_\LT^{\rm Born}
\hskip 1 cm \Lambda\ {\rm ordering}.
\end{equation}

There is an additional requirement in the case of initial state splittings with $\Lambda$ ordering. In successive initial state splittings, the factor $2p_\La\cdot Q_0$ in Eq.~(\ref{eq:Lambdadef}) can grow, so that successive virtualities $|(\hat p_\La - \hat p_{m+1})^2|$ can grow even while successive $\Lambda^2$ values get smaller. We require that for each splitting, $|\bm k_\LT| < P_\LT^{\rm Born}$, where, as above, $P_\LT^{\rm Born}$ is the transverse momentum in the hard $2 \to 2$ scattering that initiates the shower. In this way, we ensure that the hard scattering is indeed the hardest scattering in the whole event. This is discussed in Sec.\ 5.4 of Ref.~\cite{ShowerTime}, although we now replace Eq.~(5.30) of Ref.~\cite{ShowerTime} by $|\bm k_\LT| < P_\LT^{\rm Born}$.

\subsection{Threshold logarithms}
\label{sec:threshold}

We have presented a general analysis \cite{NSallorder} of the structure of parton showers at any order of perturbation theory. In this treatment, there is a factor associated with the summation of threshold logarithms,
\begin{equation}
\label{eq:UVexponential00}
\cU_\cV(\mu_\Lf^2,\mu_\Ls^{2})
=\mathbb{T} \exp\!\left(
\int_{\mu_\Lf^2}^{\mu_\Ls^{2}}\!\frac{d\mu^2}{\mu^2}\,\cS_\cV(\mu^2)
\right)
\;.
\end{equation}
Here $\mu^2$ is the hardness scale, which runs between the scale $\mu_\Ls^{2}$ at which the shower starts and the scale $\mu_\Lf^2$ at which the shower is stopped.  The scale $\mu_\Ls^2$ is of the order of the scale of the hard interaction, while  $\mu_\Lf^2$ is much smaller, typically of order $1 \GeV^2$. The integrand is well behaved in the infrared, so that the integral is not sensitive to the value of $\mu_\Lf^2$.

We included a summation of threshold logarithms in an earlier paper \cite{DuctThreshold}. The important change between that treatment and the treatment of threshold logarithms in Ref.~\cite{NSallorder} lies in the fact that $\cU_\cV(\mu_\Lf^2,\mu_\Ls^{2})$ in Eq.~(\ref{eq:UVexponential01}) is a single operator that acts on the state at the start of the shower. In Ref.~\cite{DuctThreshold}, it was divided into smaller factors that acted on the states at intermediate stages of shower development. This led to an unphysical sensitivity to small scales, all the way down to the shower cutoff $\mu_\Lf^2$. To avoid this, we inserted an {\it ad hoc} infrared cutoff into the threshold factors. The general analysis of Ref.~\cite{NSallorder} indicates that there should be a single threshold factor as in Eq.~(\ref{eq:UVexponential00}).

The operator $\cS_\cV(\mu^2)$ has an expansion to any order in $\as$. The notation of Ref.~\cite{NSallorder} is adapted to working to arbitrary perturbative order. For this paper, we need only its first order contribution, $\cS_\cV^{(1)}(\mu^2)$. Since we want to work only to first order, it is most convenient to use the notation of Ref.~\cite{DuctThreshold} and previous \textsc{Deductor} papers \cite{NSI, NSII, NSspin, NScolor, Deductor, ShowerTime, PartonDistFctns, ColorEffects} instead of the notation of Ref.~\cite{NSallorder}. In the notation of Ref.~\cite{DuctThreshold}, the first order splitting operator $\cH_\LI(\mu^2)$ acts on a state $\sket{\rho}$ with $m$ final state partons and creates a state with $m+1$ final state partons. The inclusive probability associated with $\cH_\LI(\mu^2)\sket{\rho}$ is denoted $\sbra{1} \cH_\LI(\mu^2) \sket{\rho}$. There is another operator, $\cV(\mu^2)$, that leaves the number of partons, their momenta and flavors unchanged. It is defined by integrating the parton spitting functions over the splitting variables $z$ and $\phi$, so that 
\begin{equation}
\sbra{1} \cV(\mu^2) \sket{\rho} = \sbra{1} \cH_\LI(\mu^2) \sket{\rho}
\end{equation}
for any $\sket{\rho}$. This is the operator that, after approximations for color and spin, appears in the Sudakov exponent that comes between parton splittings.

When we evaluate $\cS_\cV(\mu^2)$ at order $\as$ and use the notation of Ref.~\cite{DuctThreshold}, the threshold operator in Eq.~(\ref{eq:UVexponential00}) is
\begin{equation}
\label{eq:UVexponential01}
\cU_\cV(\mu_\Lf^2,\mu_\Ls^{2})
=\mathbb{T} \exp\!\left(
\int_{\mu_\Lf^2}^{\mu_\Ls^{2}}\!\frac{d\mu^2}{\mu^2}\,
\left[
\cV(\mu^2) - \left\{\cS(\mu^2) - \cS_{\mi\pi}(\mu^2)\right\}
\right]
\right)
\;.
\end{equation}
Here the operator $\cS(\mu^2)$ has two contributions. One comes from the operator $\cF(\mu^2)$ that multiplies by the proper parton distribution functions to make a cross section. The derivative of $\cF(\mu^2)$ with respect to the scale contributes to $\cS(\mu^2)$. There is also a contribution $\cS^{\rm pert}(\mu^2)$ that comes from virtual graphs. The total is 
\begin{equation}
\begin{split}
\label{eq:VfromVpertandF}
\cS(\mu^2)
 ={}& 
\cS^{\rm pert}(\mu^2)
- {\cal F}(\mu^2)^{-1}\left[\mu^2\,\frac{d}{d\mu^2}\,{\cal F}(\mu^2)\right]
\;.
\end{split}
\end{equation}
The operator $\cS^{\rm pert}(\mu^2)$ has a contribution $\cS_{\mi\pi}(\mu^2)$ from the imaginary part of the one loop graphs. We remove this term. It belongs in the Sudakov exponent rather than the threshold factor because, although it changes parton colors, it preserves probabilities. We have, in fact, not included $\cS_{\mi\pi}(\mu^2)$ in the \textsc{Deductor} code used in this paper.

An exact treatment of leading threshold logarithms requires an exact treatment of color, which is available in the general formalism of Ref.~\cite{NSI}. The exact color treatment is not implemented in the code of \textsc{Deductor}. Rather, we are able to use only an approximation, the leading-color-plus (LC+) approximation \cite{NScolor}. The LC+ approximation consists of simply dropping some terms that appear in the exact color formulas. 

We could simply use $\cS(\mu^2)$ and $\cV(\mu^2)$ as given in Ref.~\cite{DuctThreshold} to construct the threshold factor (\ref{eq:UVexponential01}). However, we have found that some of the integrations that go into these operators can be performed so that they are accurate in a wider range of the kinematic variables compared to Ref.~\cite{DuctThreshold}. Thus we use the improved versions of $\cS(\mu^2)$ and $\cV(\mu^2)$ in \textsc{Deductor} v.~2.1.1. We explain the changes relative to Ref.~\cite{DuctThreshold} in Appendix \ref{sec:thresholdmods}.

There is a factor associated with parton distribution functions that is related to the summation of threshold logarithms. \textsc{Deductor} begins with a Born color density matrix consisting of a sum of products of matrix elements $\ket{\cM}\bra{\cM}$ and parton factors
\begin{equation}
{\rm pdf}[{\rm LO}] = 
f_{a/A}^{\MSbar,{\rm NLO}}(\eta_\La,\mu_\scF^{2})\,
f_{b/B}^{\MSbar,{\rm NLO}}(\eta_\Lb,\mu_\scF^{2})
\;.
\end{equation}
For this paper, we use the CT14 NLO parton distributions \cite{CT14}. Then with a ``standard'' shower (std.) we apply a probability preserving shower, still starting with the same parton factor:
\begin{equation}
{\rm pdf}[{\rm std.}] = 
f_{a/A}^{\MSbar,{\rm NLO}}(\eta_\La,\mu_\scF^{2})\,
f_{b/B}^{\MSbar,{\rm NLO}}(\eta_\Lb,\mu_\scF^{2})
\;.
\end{equation}
The shower splitting functions use parton distribution functions $f_{a/A}(\eta_\La,\mu^{2})$ and $f_{b/B}(\eta_\Lb,\mu^{2})$ that are adapted to the definition of the parton shower. These are described in Ref.~\cite{NSallorder} and in Ref.~\cite{DuctThreshold}. With $k_\LT$ ordering, $f_{a/A}(\eta_\La,\mu^{2})$ and $f_{b/B}(\eta_\Lb,\mu^{2})$ are just $\MSbar$ parton distribution functions with leading order evolution, $f_{a/A}^{\MSbar, {\rm LO}}(\eta_\La,\mu^{2})$.\footnote{There is a small difference that arises from the fact that in the dimensionally regulated definition of the $\MSbar$ parton distribution functions, a gluon has $2 - 2\epsilon$ polarization states instead of $2$ polarization states. See Appendix B of Ref.~\cite{NSallorder}. We ignore this difference in this paper.} With $\Lambda$ ordering, there is a substantial difference between $f_{a/A}(\eta_\La,\mu^{2})$ and $f_{a/A}^{\MSbar, {\rm LO}}(\eta_\La,\mu^{2})$. We define all of these parton distribution functions to agree at a low scale $\mu_\Lf^2$. (See Appendix B of Ref.~\cite{NSallorder} and Sec.\ 4.1 of Ref.~\cite{DuctThreshold}.) They differ in their evolution equations.

The full result (full) given by \textsc{Deductor} includes the factor $\cU_\cV(\mu_\Lf^2,\mu_\Ls^{2})$ from Eq.~(\ref{eq:UVexponential01}) and a modified  parton factor,
\begin{equation}
{\rm pdf}[{\rm full}] = Z_\La Z_\Lb\,
f_{a/A}^{\MSbar,{\rm NLO}}(\eta_\La,\mu_\scF^{2})\,
f_{b/B}^{\MSbar,{\rm NLO}}(\eta_\Lb,\mu_\scF^{2})
\;,
\end{equation}
where
\begin{equation}
\label{eq:ZaZb}
Z_\La  = 
\frac{f_{a/A}(\eta_\La,\mu_\Ls^{2})}
{f_{a/A}^{\MSbar, {\rm LO}}(\eta_\La,\mu_\Ls^{2})}\,,
\hskip 1 cm
Z_\Lb =
\frac{f_{b/B}(\eta_\Lb,\mu_\Ls^{2})}
{f_{b/B}^{\MSbar, {\rm LO}}(\eta_\Lb,\mu_\Ls^{2})}
\;.
\end{equation}
To a good approximation, the $\MSbar$ parton distribution functions cancel in ${\rm pdf}[{\rm full}]$, leaving just $f_{a/A}(\eta_\La,\mu_\Ls^{2})\,f_{b/B}(\eta_\Lb,\mu_\Ls^{2})$. However, this cancellation is not exact because we define the denominators in $Z_\La$ and $Z_\Lb$ at a scale $\mu_\Ls^2$ and with just lowest order evolution so that it matches the numerator except for the change of evolution kernels.

The factor $Z_\La Z_\Lb$ equals 1 with $k_\LT$ ordering, but it can be substantially larger than 1 with $\Lambda$ ordering. It is part of the summation of threshold logarithms, as described in Secs.\ 4.2 and 9.4 of Ref.~\cite{DuctThreshold}. 

\section{Pythia, Dire, and Deductor}
\label{sec:PythiaDireDeductor}

In the next section, we use \textsc{Deductor} to examine some questions in jet physics. We compare to results from NLO perturbation theory, results from \textsc{Pythia}, results from \textsc{Deductor} with hadronization by \textsc{Pythia}, and results from \textsc{Dire} \cite{Dire}, which is available as an add-on to \textsc{Pythia}. Before beginning, we briefly sketch some of the ways in which \textsc{Pythia} and \textsc{Dire} differ from \textsc{Deductor} and from each other.

In \textsc{Deductor} version 2.1.1, used in this paper, all partons are massless. In \textsc{Pythia} and \textsc{Dire}, final state charm and bottom quarks have non-zero masses, while initial state partons are massless.

The three parton shower algorithms share the feature that the shower evolution starts from a hard scattering. In the cases examined in this paper the hard scattering is $2 \to 2$ parton-parton scattering. There are then splittings of either initial state partons or final state partons. As the shower develops the splittings become softer and softer, according to a hardness ordering variable that is defined in the algorithm. \textsc{Deductor} is a dipole shower, meaning that it includes quantum interference between the amplitude to emit a soft gluon from one parton and the amplitude to emit the same gluon from another parton. \textsc{Pythia} is a dipole shower for final state splittings, but not for initial state splittings. We have included some comparisons to \textsc{Dire} because it is similar to \textsc{Deductor} in being a dipole shower for both initial and final state splittings.  

The splitting probabilities in all three algorithms reflect the behavior of order $\as$ QCD matrix elements in the limit of the daughter partons becoming collinear or one of them becoming soft. In the strict collinear limit, this gives the Dokshitzer-Gribov-Lipatov-Altarelli-Parisi (DGLAP) parton evolution kernels. However, away from the strict soft or collinear limits there is freedom to choose the splitting probabilities and the three algorithms make different choices. In particular, for graphs in a physical gauge that to not involve interference, \textsc{Deductor} makes only minimal approximations to the matrix elements \cite{NSI}, rather than taking the collinear limit and obtaining the DGLAP kernel.

The partons in all three shower algorithms are approximated as being on shell and, for initial state partons, as having zero transverse momentum relative to the beam axis. It is not kinematically possible to have, for instance, a massless final state parton split into two massless daughter partons that are not exactly collinear with each other. Thus in order to conserve momentum in the splitting, the algorithm must take some momentum from the other partons in the event. This is evidently an approximation. In \textsc{Deductor}, the algorithm takes a small fraction of its momentum from each of the other final state partons in the event. This global recoil strategy is shared by \textsc{Pythia} initial state splittings. The final state splittings in \textsc{Pythia} use a local recoil strategy, in which all of the needed momentum comes from the single parton that forms a color dipole with the mother parton in the splitting. Splittings in \textsc{Dire} mostly follow this local recoil strategy.

All of the shower algorithms make some approximation with respect to color. \textsc{Pythia} and \textsc{Dire} use the leading color approximation, in which only the terms that are leading in an expansion in powers of $1/N_\Lc^2$ are retained, where $N_\Lc = 3$ is the number of colors. \textsc{Deductor} uses an approximation in which some contributions suppressed by a power of $1/N_\Lc^2$ are retained. This ``LC+'' approximation is described in Ref.~\cite{NScolor}. In this paper, we set the LC+ approximation to omit contributions suppressed by more than $1/N_\Lc^4$. However, we emphasize that only some but not all of the contributions suppressed by $1/N_\Lc^2$ and $1/N_\Lc^4$ are retained.

All three algorithms average over the spins of partons in intermediate states. This is an approximation that potentially affects angular distributions of the partons in the shower \cite{NSspin}, but we do not know the size of the effect on the physical observables examined in this paper.

The choices for the hardness variable that orders successive splittings are different in all three algorithms. The default choice for \textsc{Deductor} is $\Lambda$, Eq.~(\ref{eq:Lambdadef}) but we can also use $k_\LT$, Eq.~(\ref{eq:kTdef}). The choice for \textsc{Pythia} is described in Ref.~\cite{SjostrandSkands}. For a final state splitting with daughter partons $i$ and $j$ and a final state dipole partner parton $k$, the \textsc{Pythia} definition (with massless partons) is
\begin{equation}
\label{eq:PythiaOrdering}
p_{\perp, \rm evol}^2 = 2 p_i\cdot p_j\
\frac{[p_i\cdot(p_j+p_k)][p_j\cdot(p_i+p_k)]}
{[p_k\cdot(p_i + p_j)+2p_i\cdot p_j]^2}
\;.
\end{equation}
The choice for \textsc{Dire} is described in Ref.~\cite{Dire}. For a final state splitting with daughter partons $i$ and $j$ and a final state dipole partner parton $k$, the \textsc{Dire} evolution variable (with massless partons) is
\begin{equation}
\label{DireOrdering}
t = 2 p_i\!\cdot\!p_j\ 
\frac{  p_j\!\cdot\!p_k}{p_k\cdot (p_i + p_j) + p_i\cdot p_j}
\;.
\end{equation}
We have discussed the issues related to the choice of ordering variable at some length in Ref.~\cite{ShowerTime}. Perhaps the most important conclusion of that paper is that algorithms that use very different ordering variables can give approximately the same results: in one algorithm two splittings can occur in one order while in another algorithm the splittings can occur in the opposite order, but after accounting for the quantum interference in a dipole shower, one gets the same result in the appropriate limit.

In all three algorithms, the splitting function for an initial state splitting is proportional to a ratio of parton distributions. \textsc{Pythia} and \textsc{Dire} use $\MSbar$ parton distributions.  The evolution of the parton distributions needs to be matched with the evolution of the shower \cite{PartonDistFctns, DuctThreshold, NSallorder}. \textsc{Deductor} uses $\MSbar$ parton distributions if one chooses $k_\LT$, Eq.~(\ref{eq:kTdef}), as the ordering variable.  However, with the default virtuality based ordering variable $\Lambda$,  Eq.~(\ref{eq:Lambdadef}), \textsc{Deductor} uses modified parton distribution functions internally in the shower, as described in Ref.~\cite{DuctThreshold}. The scale parameter in the parton distributions is $\lambda_\LR k_\LT^2$ for \textsc{Deductor} with $k_\LT$ ordering, and $\lambda_\LR$ times the virtuality in the splitting for \textsc{Deductor} with $\Lambda$ ordering, where  $\lambda_\LR \approx 0.4$ is the group theory factor defined in Ref.~\cite{lambdaR}.

The splitting functions in each shower algorithm are proportional to $\as$. In \textsc{Pythia} and \textsc{Dire}, $\as$ is evaluated at a scale $\mu^2$ equal to the hardness ordering variable for that shower. In \textsc{Deductor} with $k_\LT$ ordering, $\as$ is evaluated at $\lambda_\LR k_\LT^2$. In \textsc{Deductor} with $\Lambda$ ordering, there are terms in the splitting functions that are singular when a gluon is emitted and that gluon becomes soft. For those terms, \textsc{Deductor} uses $\mu^2 = \lambda_\LR k_\LT^2$ where $k_\LT^2$ is given in Eq.~(\ref{eq:kTdef}). For the remaining splitting terms \textsc{Deductor} uses $\lambda_\LR$ times the virtuality in the splitting. \textsc{Deductor} and \textsc{Dire} use a standard value \cite{PDG} for the strong coupling, $\as(M_\LZ^2) = 0.118$. \textsc{Pythia} uses $\as(M_\LZ^2) = 0.1365$ in its shower. This, by itself, would lead to more splittings in a \textsc{Pythia} shower, but one cannot look just at $\as(M_\LZ^2)$ without also accounting for the way the splitting functions are constructed. 

We conclude from the considerations in this section that there are multiple differences between \textsc{Deductor} and the two other programs that we use, \textsc{Pythia} and \textsc{Dire}. For this reason, we expect to see some differences in results. We also expect that it may not always be possible to attribute any difference in results to a specific combination of the differences among the programs.

\section{The physics of a single jet}
\label{sec:onejetphysics} 

In this section, we use the new version of \textsc{Deductor} to address questions in related to the one jet inclusive cross section. We will be concerned with the interplay of the internal structure of the jet and the jet definition in calculating the jet cross section. We also examine the effect of the summation of threshold logarithms on the predicted jet cross section at high jet transverse momentum.

\subsection{One jet inclusive cross section}
\label{sec:onejet}

We begin with the one jet inclusive cross section $d\sigma/dP_\LT$ as calculated in \textsc{Deductor} with $\Lambda$ ordering for jets with $|y_{\rm jet}| < 2$ for the LHC at 13 TeV. We use CT14 NLO parton distributions \cite{CT14}. The jets are defined with the anti-$k_\LT$ algorithm \cite{antikt, FastJet} with $R = 0.4$. In \textsc{Deductor}, we use the LC+ approximation for color, with the maximum color suppression index, as defined in Ref.~\cite{NScolor}, set to 4.

\textsc{Deductor} starts with the Born matrix elements for two jet production. It would, of course, be best to put in the right NLO matrix elements and subtractions to make the shower cross section match the NLO jet cross section \cite{MCatNLO, PowHeg1, PowHeg2, MENLOOPS, PowHegBox, FxFx, MINLO, MINLO2, MEPSatNLO, UNLOPS, PlatzerMatching, NNLOPS, MG5aMCNLO, GENEVA, Czakon, KrkNLO, MINLO3, HocheMatching1, HocheMatching2, review2011, SjostrandReview}. In principle, this is straightforward in the \textsc{Deductor} framework \cite{NSallorder}. However, we have not developed the code to do this.  Without matching, it is important to calculate the Born cross matrix elements using scales $\mu_{\scR}$ and $\mu_{\scF}$ such that the NLO cross section is fairly stable with respect to scale variations about these choices and such that the LO cross section with these scale choices is fairly close to the NLO cross section. Based on these criteria, we choose $\mu_{\scR} = \mu_{\scF} = P_\LT^{\textrm{Born}}/\sqrt 2$. We will show the effect of other choices later, in Fig.~\ref{fig:JetKfactorsaltmuRmuF}.

Without matching, but with reasonable choices for $\mu_{\scR}$ and $\mu_{\scF}$, the parton shower needs to do as good a job as possible to calculate the cross section accurately. For this purpose, it is important to include the threshold factors $\cU_\cV(\mu_\Lf^2,\mu_\Ls^{2})$, Eq.~(\ref{eq:UVexponential00}), and $Z_\La Z_\Lb$, Eq.~(\ref{eq:ZaZb}). It is also important to make a sensible choice for the starting scale $\mu_\Ls$ of the shower. We use Eq.~(\ref{eq:musforLambda}) for $\mu_\Ls$ in the default $\Lambda$ ordered shower. We will show the effect of an alternative choice later, in Fig.~\ref{fig:JetKfactorsR04altmus}.

\begin{figure}
\begin{center}
\ifusefigs 
\begin{tikzpicture}
\begin{semilogyaxis}[title = {One jet inclusive cross section},
   xlabel={$P_\LT\,\mathrm{[TeV]}$}, ylabel={$d\sigma/dP_\LT\,\mathrm{[nb/GeV]}$},
   xmin=0, xmax=3800,
   ymin=1.01e-10, ymax=1.0,
  legend cell align=left,
  width=14cm,	
  height=11cm,
  x coord trafo/.code={
    \pgflibraryfpuifactive{
    \pgfmathparse{(#1)/(1000)}
  }{
     \pgfkeys{/pgf/fpu=true}
     \pgfmathparse{(#1)/(1000)}
     \pgfkeys{/pgf/fpu=false}
   }
 },
 xminorgrids=false,
 yminorgrids=false,
 minor x tick num=4,
]
 
\addplot[red, domain=300:3500] 
{exp(-1.3146599761746292 + 4.2130715526917095/exp(x/200.) - 
0.009179792449765788*x + (0.04589891310873323*x)/exp(x/200.) + 
1.6218405497595238e-6*x^2 - (0.0001329788971422548*x^2)/exp(x/200.) - 
1.883388150288521e-10*x^3 + (2.1371632222697485e-7*x^3)/exp(x/200.) - 
(1.2947116753916147e-10*x^4)/exp(x/200.))};

\addplot[black, dashed, domain=300:3500] 
{exp(-89.43435686639582 + 97.74159682256081/exp(x/400.) + 
0.06373702674018425*x + (0.12366319025063298*x)/exp(x/400.) - 
0.00002184162264881441*x^2 + (0.00015681748444247627*x^2)/exp(x/400.) 
+ 3.26427371209164e-9*x^3 + (1.3879593301841329e-8*x^3)/exp(x/400.) - 
1.9513040776090802e-13*x^4 + (3.959183680137702e-11*x^4)/exp(x/400.))};

\legend{ \textsc{Deductor}(full),  NLO}
\end{semilogyaxis}
\end{tikzpicture}
\else 
\includegraphics[width = 14 cm]{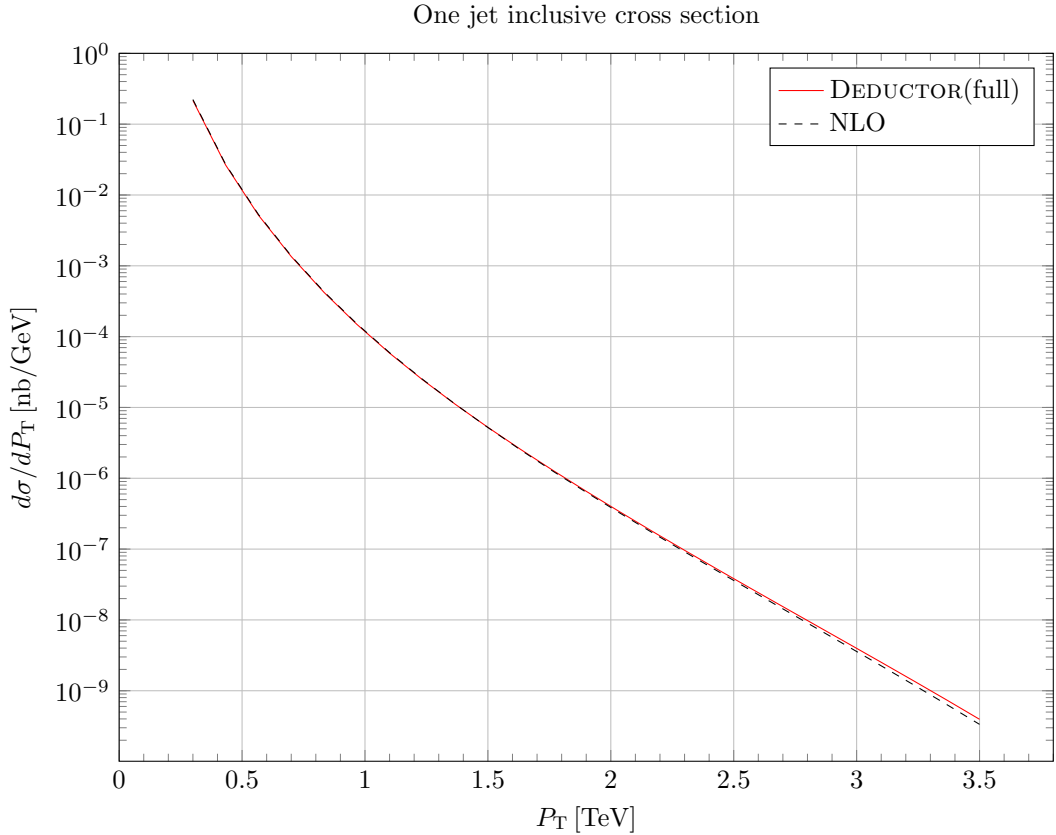}
\fi
\end{center}
\caption{
One jet inclusive cross section $d\sigma/dP_\LT$ for the production of a jet with transverse momentum $P_\LT$ and rapidity in the range $-2 < y < 2$. The cross section is for the LHC at 13 TeV. We use the anti-$k_\LT$ algorithm \cite{antikt} with $R = 0.4$.  The solid red curve is $d\sigma({\rm full})/dP_\LT$, obtained with threshold effects. The dashed black curve is an NLO calculation \cite{EKS}.
}
\label{fig:jetcrosssection}

\end{figure}

We show the \textsc{Deductor} result for $d\sigma({\rm full})/dP_\LT$ as the solid red curve in Fig.\ \ref{fig:jetcrosssection}. Here ``full'' refers the calculation including all contributions. We also show, as a black dashed curve, the purely perturbative NLO cross section \cite{EKS} with the same parton distribution functions and scale choices. We see that the parton shower calculation matches the NLO calculation reasonably well.

In the following subsections, we will examine the jet cross section in more detail. Since it is not easy to see details in a plot like that in Fig.\ \ref{fig:jetcrosssection} in which the cross section falls by nine orders of magnitude, we will show ratios of calculated cross sections as functions of $P_\LT$.

\subsection{Underlying event and hadronization}
\label{sec:nonperturbativeII}

\begin{figure}
\begin{center}
\ifusefigs 
\begin{tikzpicture}
\begin{semilogyaxis}[
   log ticks with fixed point,
   ytick = {0.7, 0.8, 0.9, 1.0},
   title = { },
   xlabel={$P_\LT\,\mathrm{[TeV]}$}, ylabel={Ratio of cross sections},
   xmin=0, xmax=3800,
   ymin=0.6, ymax=1.1,
  legend cell align=left,
  width=14cm,	
  height=11cm,
  every axis legend/.append style={
  at={(0.02,0.02)},
  anchor= south west,
  },
  x coord trafo/.code={
    \pgflibraryfpuifactive{
    \pgfmathparse{(#1)/(1000)}
  }{
     \pgfkeys{/pgf/fpu=true}
     \pgfmathparse{(#1)/(1000)}
     \pgfkeys{/pgf/fpu=false}
   }
 },
 xminorgrids=false,
 yminorgrids=false,
 minor x tick num=4,
]

\addplot[white, opacity=0, forget plot, name path=A07, domain=300:3500] 
{1.0089275584234483 - 0.000013711910125072416*x + 
1.43471044648449e-9*x^2 + 
1.992102154002228*sqrt(0.000051407534332086624 - 
1.0379021298767348e-7*x + 8.651800381355856e-11*x^2 - 
2.6664070438411357e-14*x^3 + 2.7659422894301317e-18*x^4)};

\addplot[white, opacity=0, forget plot, name path=B07, domain=300:3500] 
{1.0089275584234483 - 0.000013711910125072416*x + 
1.43471044648449e-9*x^2 - 
1.992102154002228*sqrt(0.000051407534332086624 - 
1.0379021298767348e-7*x + 8.651800381355856e-11*x^2 - 
2.6664070438411357e-14*x^3 + 2.7659422894301317e-18*x^4)};

\addplot[darkgreen, very thick, dashdotted, domain=300:3500] 
{1.0089275584234483 - 0.000013711910125072416*x + 
1.43471044648449e-9*x^2};

\addlegendentry{$R = 0.7$}

\addplot[white, opacity=0, forget plot, name path=A04, domain=300:3500] 
{0.9747944479847453 + 8.116277419136401e-6*x - 
1.9369846434832953e-9*x^2 + 
1.992102154002228*sqrt(0.000033813790856239197 - 
6.826899209400889e-8*x + 5.6908033506384057e-11*x^2 - 
1.753854396820816e-14*x^3 + 1.8193246439528595e-18*x^4)};

\addplot[white, opacity=0, forget plot, name path=B04, domain=300:3500] 
{0.9747944479847453 + 8.116277419136401e-6*x - 
1.9369846434832953e-9*x^2 -  
1.992102154002228*sqrt(0.000033813790856239197 - 
6.826899209400889e-8*x + 5.6908033506384057e-11*x^2 - 
1.753854396820816e-14*x^3 + 1.8193246439528595e-18*x^4)};

\addplot[blue, very thick, domain=300:3500] 
{0.9747944479847453 + 8.116277419136401e-6*x - 
1.9369846434832953e-9*x^2};

\addlegendentry{$R = 0.4$}

\addplot[white, opacity=0, forget plot, name path=A02, domain=300:3500] 
{0.9522775602007918 + 0.00002141105318920893*x - 
3.834370133535018e-9*x^2 + 
1.992102154002228*sqrt(0.00004403534227731489 - 
8.890598651207675e-8*x + 7.411073027678984e-11*x^2 - 
2.2840260353218517e-14*x^3 + 2.3692872458645306e-18*x^4)};

\addplot[white, opacity=0, forget plot, name path=B02, domain=300:3500] 
{0.9522775602007918 + 0.00002141105318920893*x - 
3.834370133535018e-9*x^2 - 
1.992102154002228*sqrt(0.00004403534227731489 - 
8.890598651207675e-8*x + 7.411073027678984e-11*x^2 - 
2.2840260353218517e-14*x^3 + 2.3692872458645306e-18*x^4)};

\addplot[red!30,opacity=0.5, forget plot] fill between[of=A02 and B02];
\addplot[blue!30,opacity=0.5, forget plot] fill between[of=A04 and B04];
\addplot[green!30,opacity=0.5, forget plot] fill between[of=A07 and B07];

\addplot[red, very thick, densely dashed, domain=300:3500] 
{0.9522775602007918 + 0.00002141105318920893*x - 
3.834370133535018e-9*x^2};

\addlegendentry{$R = 0.2$}

\end{semilogyaxis}
\end{tikzpicture}
\else 
\includegraphics[width = 14 cm]{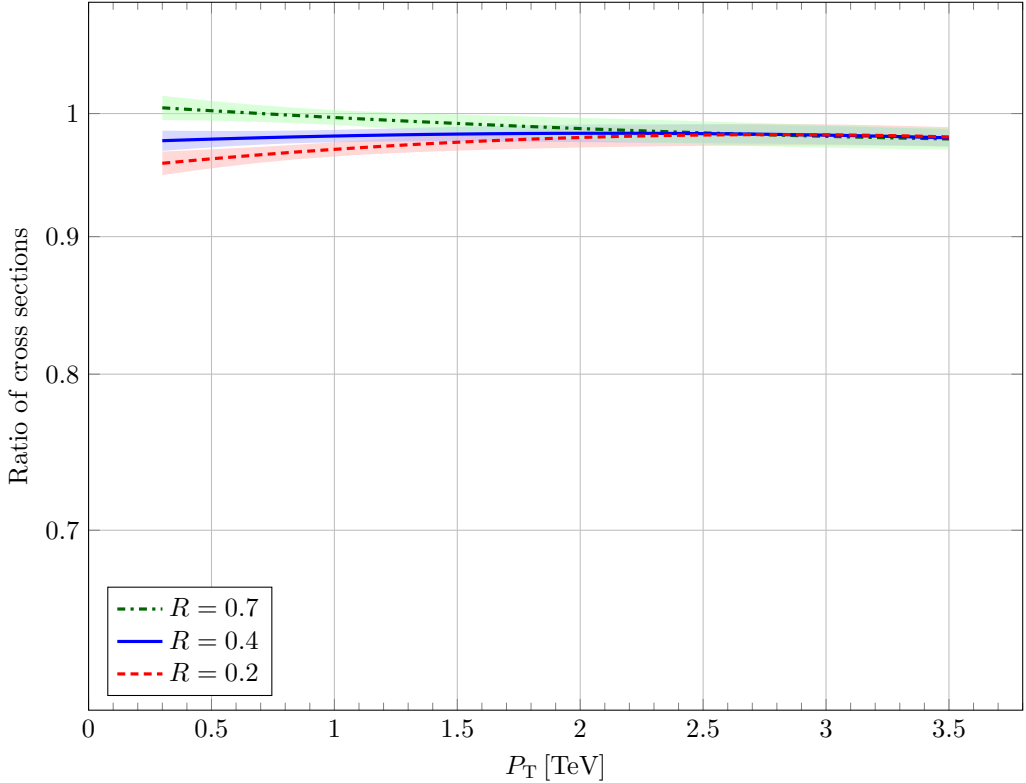}
\fi
\end{center}
\caption{
Cone size dependence of the ratio of the one jet inclusive cross section $d\sigma/dP_\LT$ calculated by \textsc{Deductor} with an underlying event and with hadronization by \textsc{Pythia} to the same cross section calculated by \textsc{Deductor} with no non-perturbative input. The solid blue curve is for a jet with $R = 0.4$, while the dash-dotted green curve is for a jet with $R = 0.7$ and the dashed red curve is for a jet with $R = 0.2$. The bands represent the Monte Carlo statistical error. They indicate that the differences among the curves are not statistically significant for $P_\LT$ greater than about 1.5 TeV.
}
\label{fig:HadronizationEfffect}
\end{figure}

The \textsc{Deductor} results presented above represent a purely perturbative parton shower, with an infrared cutoff at around a 1 GeV scale for the transverse momentum of splittings. One expects that nonperturbative effects from an underlying event or hadronization do not significantly affect the jet cross section. However, it is worthwhile to check this expectation numerically in the context of the \textsc{Deductor} shower. Accordingly, we include an underlying event using the simple model described in Sec.\ \ref{sec:nonperturbative} and then let \textsc{Pythia} hadronize the event as also described in Sec.\ \ref{sec:nonperturbative}. In Fig.\ \ref{fig:HadronizationEfffect}, we show the ratio of the jet cross section $d\sigma/dP_\LT$ calculated with an underlying event and hadronization to the same cross section calculated at the parton level. We show three curves, corresponding to the anti-$k_\LT$ jet algorithm with $R = 0.2$, $R = 0.4$, and $R = 0.7$. We see that the non-perturbative effect is small.

Although the non-perturbative effect is small, it is of interest to understand its dependence on the jet cone size $R$. We expect that there are two effects. First, some hadrons from the underlying event can get into the jet, raising its $P_\LT$. Since the jet cross section falls quickly with increasing $P_\LT$, this has the effect of raising the cross section at the measured value of $P_\LT$. This effect increases with increasing $R$. Thus this effect increases the cross section as $R$ increases. Second, the result of hadronization is that a parton near the edge of the jet cone effectively becomes a set of hadrons with a range in angles from the jet center. This smears the distribution of $P_\LT$ as a function of angle from the jet center. Since the $P_\LT$ distribution is peaked near the jet center, the effect is for the jet to lose some $P_\LT$ as a result of hadronization. This decreases the jet cross section at a measured $P_\LT$ value. If we increase $R$, there are fewer partons near the edge of the cone, so there is less $P_\LT$ loss. Thus this effect increases the cross section as $R$ increases. We conclude that there are two competing non-perturbative effects \cite{SnowmassJetDef, EllisJets, DasguptaJets} and that the result from either of them is to give a jet cross section that increases with $R$. This is what we see.

\begin{figure}
\begin{center}
\ifusefigs 
\begin{tikzpicture}
\begin{semilogyaxis}[
   log ticks with fixed point,
   ytick = {0.7, 0.8, 0.9, 1.0},
   title = { },
   xlabel={$P_\LT\,\mathrm{[TeV]}$}, ylabel={Ratio of cross sections},
   xmin=0, xmax=3800,
   ymin=0.6, ymax=1.1,
  legend cell align=left,
  width=14cm,	
  height=11cm,
  every axis legend/.append style={
  at={(0.02,0.02)},
  anchor= south west,
  },
  x coord trafo/.code={
    \pgflibraryfpuifactive{
    \pgfmathparse{(#1)/(1000)}
  }{
     \pgfkeys{/pgf/fpu=true}
     \pgfmathparse{(#1)/(1000)}
     \pgfkeys{/pgf/fpu=false}
   }
 },
 xminorgrids=false,
 yminorgrids=false,
 minor x tick num=4,
]

\addplot[white, opacity=0, forget plot, name path=A07, domain=300:3500] 
{1.0332380959817058 - 0.00002231076101672148*x + 
5.031132073722046e-9*x^2 + 
1.9971379083919978*sqrt(9.126832229265987e-6 - 2.288157318396274e-8*x 
+ 2.2527686048345593e-11*x^2 - 8.376953642760105e-15*x^3 + 
1.0606411071714367e-18*x^4};

\addplot[white, opacity=0, forget plot, name path=B07, domain=300:3500] 
{1.0332380959817058 - 0.00002231076101672148*x + 
5.031132073722046e-9*x^2 -
1.9971379083919978*sqrt(9.126832229265987e-6 - 2.288157318396274e-8*x 
+ 2.2527686048345593e-11*x^2 - 8.376953642760105e-15*x^3 + 
1.0606411071714367e-18*x^4};

\addplot[darkgreen, very thick, dashdotted, domain=300:3500] 
{1.0332380959817058 - 0.00002231076101672148*x + 
5.031132073722046e-9*x^2};

\addlegendentry{$R = 0.7$}

\addplot[white, opacity=0, forget plot, name path=A04, domain=300:3500] 
{1.0081234617790016 - 5.538669282275593e-6*x + 
1.9415686490636302e-9*x^2 + 
1.9971379083919978*sqrt(9.194173800373756e-6 - 
2.3050402965087066e-8*x + 2.2693904702727518e-11*x^2 - 
8.438762297201185e-15*x^3 + 1.0684669592024816e-18*x^4)};

\addplot[white, opacity=0, forget plot, name path=B04, domain=300:3500] 
{1.0081234617790016 - 5.538669282275593e-6*x + 
1.9415686490636302e-9*x^2 - 
1.9971379083919978*sqrt(9.194173800373756e-6 - 
2.3050402965087066e-8*x + 2.2693904702727518e-11*x^2 - 
8.438762297201185e-15*x^3 + 1.0684669592024816e-18*x^4)};

\addplot[blue, very thick, domain=300:3500] 
{1.0081234617790016 - 5.538669282275593e-6*x + 
1.9415686490636302e-9*x^2};

\addlegendentry{$R = 0.4$}

\addplot[white, opacity=0, forget plot, name path=A02, domain=300:3500] 
{0.98170985260079 + 0.000015379815439908834*x - 
2.623072442218689e-9*x^2 + 
1.9971379083919978*sqrt(9.281562517576654e-6 - 
2.3269492269886474e-8*x + 2.2909605565399432e-11*x^2 - 
8.518970984566068e-15*x^3 + 1.078622516293929e-18*x^4)};

\addplot[white, opacity=0, forget plot, name path=B02, domain=300:3500] 
{0.98170985260079 + 0.000015379815439908834*x - 
2.623072442218689e-9*x^2 - 
1.9971379083919978*sqrt(9.281562517576654e-6 - 
2.3269492269886474e-8*x + 2.2909605565399432e-11*x^2 - 
8.518970984566068e-15*x^3 + 1.078622516293929e-18*x^4)};

\addplot[red!30,opacity=0.5, forget plot] fill between[of=A02 and B02];
\addplot[blue!30,opacity=0.5, forget plot] fill between[of=A04 and B04];
\addplot[green!30,opacity=0.5, forget plot] fill between[of=A07 and B07];

\addplot[red, very thick, densely dashed, domain=300:3500] 
{0.98170985260079 + 0.000015379815439908834*x - 
2.623072442218689e-9*x^2};

\addlegendentry{$R = 0.2$}

\end{semilogyaxis}
\end{tikzpicture}
\else 
\includegraphics[width = 14 cm]{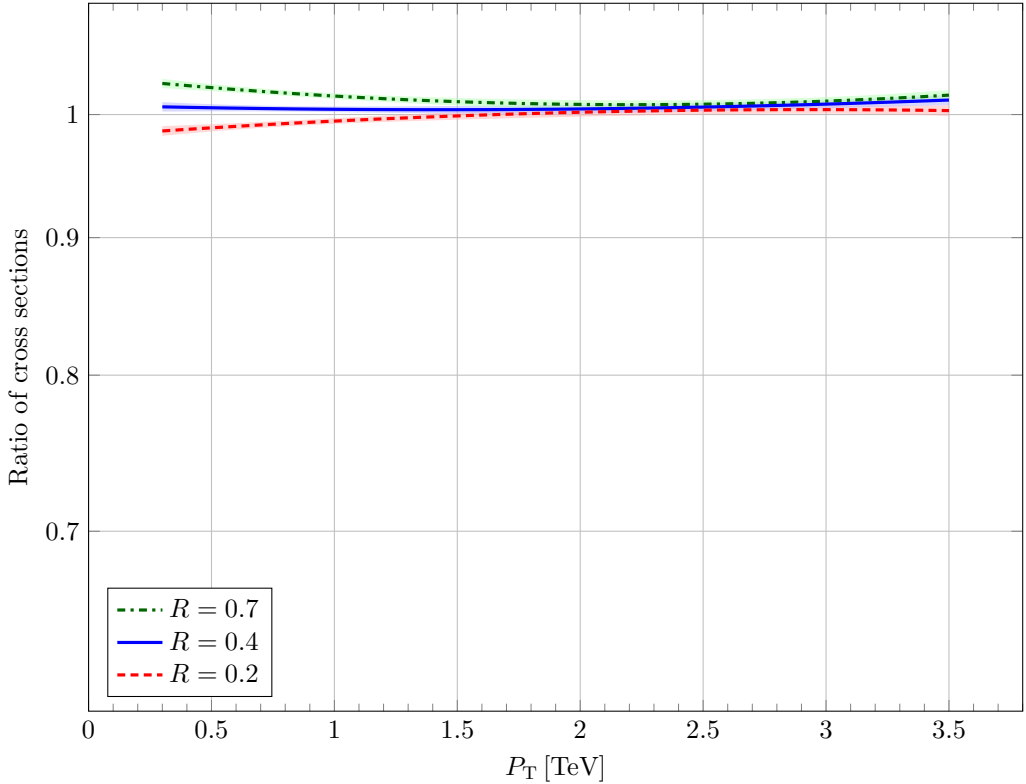}
\fi
\end{center}
\caption{
Cone size dependence of the ratio of the one jet inclusive cross section $d\sigma/dP_\LT$ calculated by full \textsc{Pythia} to the same cross section calculated by \textsc{Pythia} with no non-perturbative input. The curves are labeled as in Fig.~\ref{fig:HadronizationEfffect}.
}
\label{fig:HadronizationEfffectPythia}
\end{figure}

We can also ask what \textsc{Pythia} says about the non-perturbative effects. For this, we calculate a purely perturbative \textsc{Pythia} jet cross section with no underlying event from multiple parton interactions, no hadronization,  \textsc{Pythia}'s primordial $P_\LT$ parameter set to zero, and with the shower stopped at a 1 GeV $k_\LT$ value (compared to its default value of 0.5 GeV). We compare this to full \textsc{Pythia} with default settings, so that all non-perturbative effects are included. We plot the ratio of the full \textsc{Pythia} jet cross section to the perturbative \textsc{Pythia} jet cross section in Fig.~\ref{fig:HadronizationEfffectPythia}.

Our expectation is as follows. This ratio should increase with increasing $R$, as before. Also, we will find later in this paper that in a perturbative \textsc{Pythia} jet it is somewhat less likely than in a \textsc{Deductor} jet to find a parton carrying a rather substantial $P_\LT$ at a fairly large angle to the jet axis. Thus the $P_\LT$ loss effect should be diminished in a \textsc{Pythia} jet, leading to an increase in the cross section ratio that we calculate compared to a \textsc{Deductor} jet. We show the results of this calculation in Fig.~\ref{fig:HadronizationEfffectPythia}. We see that these expectations are confirmed.

We see that the underlying event and hadronization changes the cross section by less than 5\%. This is a small effect, so in the following subsections, we work only at the partonic level.

\subsection{Effect of jet finding}
\label{sec:jetfinding}

The intuitive picture of a jet is that it is a parton created in a hard interaction. However, this picture does not closely match reality, as we can see by simulating reality with a parton shower. At the Born level, one has just two final state partons, with $\bm p_{\LT, 1} = - \bm p_{\LT, 2}$. After showering, we have many final state partons, which are grouped into jets. We use the anti-$k_\LT$ jet algorithm with $R = 0.4$. The jet $P_\LT$ is typically quiet close to $|\bm p_{\LT, 1}|$. However, it can be larger if a parton from an initial state splitting enters the jet. It can be smaller if a parton from a final state splitting leaves the jet. Additionally, $P_\LT$ can be either larger or smaller than $|\bm p_{\LT, 1}|$ if the jet recoils against an initial state emission. These effects have the potential to substantially change the cross section because the cross section falls very steeply as a function of $P_\LT$.

We can examine this effect in \textsc{Deductor} by turning off the threshold factor. We call the resulting cross section $d\sigma({\rm std.})/dP_\LT$ because it is the result of a standard probability preserving shower. Then we look at the ratio to the Born cross section, $d\sigma({\rm LO})/dP_\LT$:
\begin{equation}
\label{eq:ratiostdtoLO}
r(P_\LT) = \frac{d\sigma({\rm std.})/dP_\LT}{d\sigma({\rm LO})/dP_\LT}
\;.
\end{equation}
For $d\sigma({\rm std.})/dP_\LT$ calculated with \textsc{Deductor}, we use the default $\Lambda$ ordering, but we also try $k_\LT$ ordering.

We can do the same thing using the parton level showering produced by \textsc{Pythia} \cite{Pythia} and \textsc{Dire} \cite{Dire}. For $d\sigma({\rm LO})/dP_\LT$, we use the Born cross section produced by these programs. For $d\sigma({\rm std.})/dP_\LT$, we use \textsc{Pythia} and \textsc{Dire} with no underlying event, no hadronization, and no primordial $k_\LT$. In \textsc{Pythia} and \textsc{Dire}, there is no threshold factor to turn off. We set the minimum $p_\LT$ in shower splittings in \textsc{Pythia} to 1 GeV, which is a change from the default value 0.5 GeV. In \textsc{Dire}, we retain the default, $p_\LT^{\rm min} = 1.732 \GeV$. We retain the default factorization scale setting in \textsc{Pythia} and \textsc{Dire}, $\mu_\LF = P_\LT^{\textrm{Born}}$, where $P_\LT^{\textrm{Born}}$ is the transverse momentum of the hard scattering. This is then the starting scale for the shower. For $\as$ at the hard interaction, we set $\as(M_\LZ^2) = 0.118$ and use the renormalization scale $\mu_\LR = P_\LT/\sqrt 2$, although this $\as$ cancels in the ratio $r$ in Eq.~(\ref{eq:ratiostdtoLO}). We use CT14 NLO parton distributions in each case. For other parameters, we use the default choices in \textsc{Pythia} and \textsc{Dire}. For \textsc{Deductor}, we use the parameters from Sec.\ \ref{sec:onejet}.

\begin{figure}
\begin{center}
\ifusefigs 
\begin{tikzpicture}
\begin{semilogyaxis}[
   log ticks with fixed point,
   ytick = {0.5,0.6,0.7, 0.8, 0.9, 1.0, 1.1},
   title = {Effect of jet finding, $R = 0.4$},
   xlabel={$P_\LT\,\mathrm{[TeV]}$}, ylabel={$r(P_\LT)$},
   xmin=0, xmax=3800,
   ymin=0.42, ymax=1.2,
  legend cell align=left,
  width=14cm,	
  height=11cm,
  every axis legend/.append style={
  at={(0.02,0.02)},
  anchor= south west,
  },
  x coord trafo/.code={
    \pgflibraryfpuifactive{
    \pgfmathparse{(#1)/(1000)}
  }{
     \pgfkeys{/pgf/fpu=true}
     \pgfmathparse{(#1)/(1000)}
     \pgfkeys{/pgf/fpu=false}
   }
 },
 xminorgrids=false,
 yminorgrids=false,
 minor x tick num=4,
]


\addplot[white, forget plot, name path=PythiaPlus, domain=300:3500] 
{0.8643408710260655 + 0.13435291232546342/exp(x/300.) + 
0.00005810615081302507*x - 7.344942816886196e-9*x^2 + 
1.9944371117711854*sqrt((0.0004229756493378255 + 
exp(x/300.)*(-0.00023122963457877376 + 1.871556483738557e-7*x - 
3.248583227929018e-11*x^2) + exp(x/150.)*(0.000039945039033334126 - 
6.948925046485878e-8*x + 4.588217105357603e-11*x^2 - 
1.2582245894809113e-14*x^3 + 1.2270784224347636e-18*x^4))/exp(x/150.))};

\addplot[white, forget plot, name path=PythiaMinus, domain=300:3500] 
{0.8643408710260655 + 0.13435291232546342/exp(x/300.) + 
0.00005810615081302507*x - 7.344942816886196e-9*x^2 - 
1.9944371117711854*sqrt((0.0004229756493378255 + 
exp(x/300.)*(-0.00023122963457877376 + 1.871556483738557e-7*x - 
3.248583227929018e-11*x^2) + exp(x/150.)*(0.000039945039033334126 - 
6.948925046485878e-8*x + 4.588217105357603e-11*x^2 - 
1.2582245894809113e-14*x^3 + 1.2270784224347636e-18*x^4))/exp(x/150.))};

\addplot[blue!20, forget plot, ] fill between[of=PythiaPlus and PythiaMinus];

\addplot[blue, dashdotted, thick, domain=300:3500] 
{0.8643408710260655 + 0.13435291232546342/exp(x/300.) + 
0.00005810615081302507*x - 7.344942816886196e-9*x^2};

\addlegendentry{\textsc{Pythia}}


\addplot[white, forget plot, name path=DirePlus, domain=300:3500] 
{0.7985554160840983 + 0.000030995668888619*x - 
8.995968988658967e-9*x^2 + 
1.9939433678456226*sqrt(0.0001377223030574595 - 
3.026187293583954e-7*x + 2.6906006425544546e-10*x^2 - 
8.90583966277746e-14*x^3 + 9.959109315245145e-18*x^4)};

\addplot[white, forget plot, name path=DireMinus, domain=300:3500] 
{0.7985554160840983 + 0.000030995668888619*x - 
8.995968988658967e-9*x^2 - 
1.9939433678456226*sqrt(0.0001377223030574595 - 
3.026187293583954e-7*x + 2.6906006425544546e-10*x^2 - 
8.90583966277746e-14*x^3 + 9.959109315245145e-18*x^4)};

\addplot[black!20, forget plot] fill between[of=DirePlus and DireMinus];

\addplot[black, densely dashed, very thick, domain=300:3500] 
{0.7985554160840983 + 0.000030995668888619*x - 
8.995968988658967e-9*x^2};

\addlegendentry{\textsc{Dire}}


\addplot[white, forget plot, name path=DuctPlus, domain=300:3500] 
{0.7019122439594058 - 3.6979677734060272e-6*x - 
5.567257820047609e-9*x^2 + 
1.992102154002228*sqrt(6.669345402326024e-6 - 1.3465200943584765e-8*x 
+ 1.1224394603812561e-11*x^2 - 3.4592574395212752e-15*x^3 + 
3.588389276160253e-19*x^4)};

\addplot[white, forget plot, name path=DuctMinus, domain=300:3500] 
{0.7019122439594058 - 3.6979677734060272e-6*x - 
5.567257820047609e-9*x^2 - 
1.992102154002228*sqrt(6.669345402326024e-6 - 1.3465200943584765e-8*x 
+ 1.1224394603812561e-11*x^2 - 3.4592574395212752e-15*x^3 + 
3.588389276160253e-19*x^4)};

\addplot[red!20, forget plot] fill between[of=DuctPlus and DuctMinus ];

\addplot[red, thick, domain=300:3500] 
{0.7019122439594058 - 3.6979677734060272e-6*x - 
5.567257820047609e-9*x^2};

\addlegendentry{\textsc{Deductor}-$\Lambda$}


\addplot[white, forget plot, name path=DuctKTPlus, domain=300:3500] 
{0.6459850957620701 + 2.7903819974961875e-6*x - 
7.925854313056382e-9*x^2 + 
1.992102154002228*sqrt(0.00003471397527282324 - 
7.008643643440253e-8*x + 5.842302853189557e-11*x^2 - 
1.8005451805808448e-14*x^3 + 1.867758364388304e-18*x^4)};

\addplot[white, forget plot, name path=DuctKTMinus, domain=300:3500] 
{0.6459850957620701 + 2.7903819974961875e-6*x - 
7.925854313056382e-9*x^2 - 
1.992102154002228*sqrt(0.00003471397527282324 - 
7.008643643440253e-8*x + 5.842302853189557e-11*x^2 - 
1.8005451805808448e-14*x^3 + 1.867758364388304e-18*x^4)};

\addplot[red!20, forget plot] fill between[of=DuctKTPlus and DuctKTMinus ];

\addplot[red, dashed, domain=300:3500] 
{0.6459850957620701 + 2.7903819974961875e-6*x - 
7.925854313056382e-9*x^2
};

\addlegendentry{\textsc{Deductor}-$k_\LT$}

\end{semilogyaxis}
\end{tikzpicture}
\else 
\includegraphics[width = 14 cm]{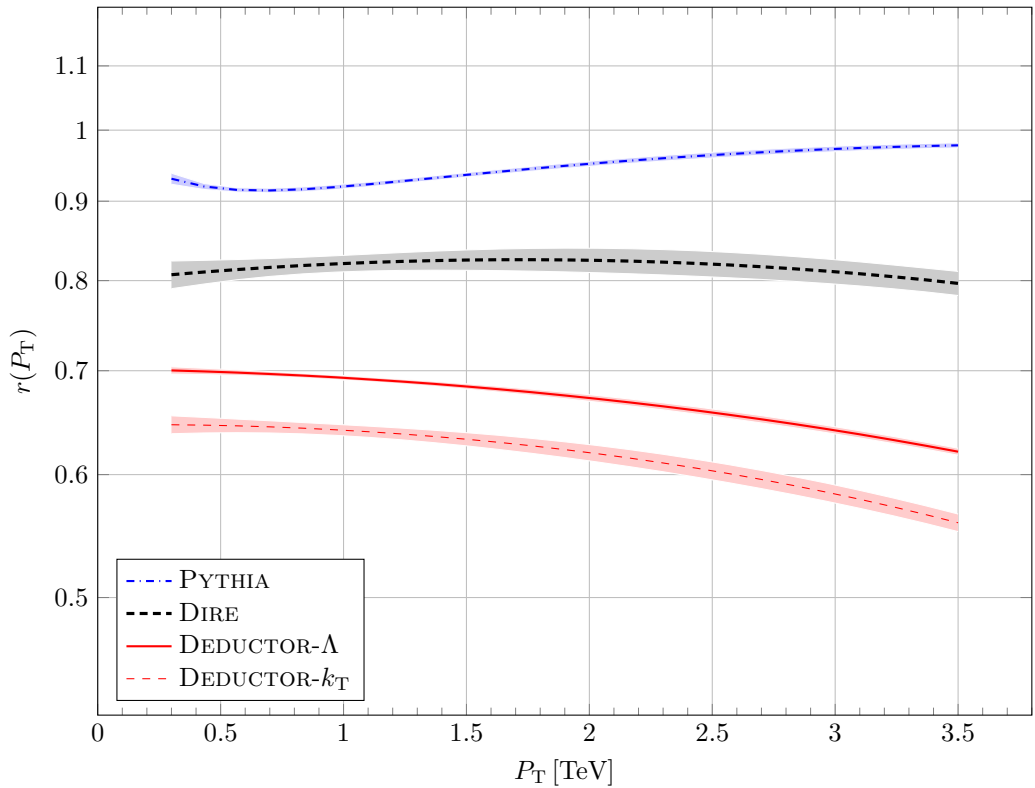}
\fi
\end{center}
\caption{
Ratio $r(P_\LT)$ of the one jet inclusive cross section $d\sigma/dP_\LT$ calculated after showering to the cross section at the Born level according to (from top to bottom) \textsc{Pythia}, \textsc{Dire}, \textsc{Deductor} with $\Lambda$ ordering, and also \textsc{Deductor} with $k_\LT$ ordering. The bands represent the Monte Carlo statistical error.
}
\label{fig:JetFindingEffectR04}
\end{figure}

We exhibit $r(P_\LT)$ for \textsc{Deductor} with $\Lambda$ ordering, \textsc{Deductor} with $k_\LT$ ordering, \textsc{Pythia}, and \textsc{Dire} in Fig.\ \ref{fig:JetFindingEffectR04}. We can make three observations. First, \textsc{Pythia}, and \textsc{Dire} and \textsc{Deductor} give substantially different results for $r(P_\LT)$, while the two orderings for \textsc{Deductor} give  results that differ by about 5\%. Second, $r(P_\LT)$ for \textsc{Deductor} and, to a lesser extent for \textsc{Dire}, is substantially smaller than 1. Third, there is some dependence on $P_\LT$, but it is not large. 

The calculated results depend on the starting scale for the shower, $\mu_\Ls$. For \textsc{Pythia}, \textsc{Dire}, and \textsc{Deductor} with $k_\LT$ ordering, this is $\mu_\Ls = P_\LT^{\textrm{Born}}$. For \textsc{Deductor} with $\Lambda$ ordering, the starting scale for $\Lambda$ is $\mu_\Ls = (3/2)\,P_\LT^{\textrm{Born}}$, Eq.~(\ref{eq:musforLambda}). If we were to use $\mu_\Ls = P_\LT^{\textrm{Born}}$ for \textsc{Deductor} with $\Lambda$ ordering, the result would change slightly, as we will see later in Fig.~\ref{fig:JetKfactorsR04altmus}.

\begin{figure}
\begin{center}
\ifusefigs 
\begin{tikzpicture}
\begin{semilogyaxis}[
   log ticks with fixed point,
   ytick = {0.5,0.6,0.7, 0.8, 0.9, 1.0, 1.1},
   title = {$R = 0.2$},
   xlabel={$P_\LT\,\mathrm{[TeV]}$}, ylabel={$r(P_\LT)$},
   xmin=0, xmax=3800,
   ymin=0.42, ymax=1.2,
  width=8.12cm,	
  height=6.38cm,
  x coord trafo/.code={
    \pgflibraryfpuifactive{
    \pgfmathparse{(#1)/(1000)}
  }{
     \pgfkeys{/pgf/fpu=true}
     \pgfmathparse{(#1)/(1000)}
     \pgfkeys{/pgf/fpu=false}
   }
 },
 xminorgrids=false,
 yminorgrids=false,
 minor x tick num=4,
]

\addplot[white, forget plot, name path=PythiaPlus, domain=300:3500] 
{0.6679179498901145 + 0.019416318158412615/exp(x/300.) + 
0.00006819957967449873*x - 1.0389485413906096e-8*x^2 + 
1.9944371117711854*sqrt((0.000787698706295838 + 
exp(x/300.)*(-0.00043061411289302626 + 3.4853605007956513e-7*x - 
6.049768609469712e-11*x^2) + exp(x/150.)*(0.00007438881084230834 - 
1.2940837794877233e-7*x + 8.544540764350373e-11*x^2 - 
2.34316534040508e-14*x^3 + 2.285162484858354e-18*x^4))/exp(x/150.))};

\addplot[white, forget plot, name path=PythiaMinus, domain=300:3500] 
{0.6679179498901145 + 0.019416318158412615/exp(x/300.) + 
0.00006819957967449873*x - 1.0389485413906096e-8*x^2 - 
1.9944371117711854*sqrt((0.000787698706295838 + 
exp(x/300.)*(-0.00043061411289302626 + 3.4853605007956513e-7*x - 
6.049768609469712e-11*x^2) + exp(x/150.)*(0.00007438881084230834 - 
1.2940837794877233e-7*x + 8.544540764350373e-11*x^2 - 
2.34316534040508e-14*x^3 + 2.285162484858354e-18*x^4))/exp(x/150.))};

\addplot[blue!20, forget plot, ] fill between[of=PythiaPlus and PythiaMinus];

\addplot[blue, dashdotted, thick, domain=300:3500] 
{0.6679179498901145 + 0.019416318158412615/exp(x/300.) + 
0.00006819957967449873*x - 1.0389485413906096e-8*x^2};


\addplot[white, forget plot, name path=DirePlus, domain=300:3500] 
{0.6136763672043848 + 0.000029786134866743298*x -
6.20264156453471e-9*x^2 +
1.9939433678456226*sqrt(0.00038371871962553734 -
8.431493577018061e-7*x + 7.496489752672073e-10*x^2 -
2.481324605184343e-13*x^3 + 2.7747841781750094e-17*x^4)};

\addplot[white, forget plot, name path=DireMinus, domain=300:3500] 
{0.6136763672043848 + 0.000029786134866743298*x -
6.20264156453471e-9*x^2 -
1.9939433678456226*sqrt(0.00038371871962553734 -
8.431493577018061e-7*x + 7.496489752672073e-10*x^2 -
2.481324605184343e-13*x^3 + 2.7747841781750094e-17*x^4)};

\addplot[black!20, forget plot] fill between[of=DirePlus and DireMinus];

\addplot[black, densely dashed, very thick, domain=300:3500] 
{0.6136763672043848 + 0.000029786134866743298*x - 
6.20264156453471e-9*x^2};


\addplot[white, forget plot, name path=DuctPlus, domain=300:3500] 
{0.5293170695250425 + 0.000025945877040418022*x - 
1.0200935352386659e-8*x^2 + 
1.992102154002228*sqrt(3.6614916226344072e-6 - 7.392437709828245e-9*x 
+ 6.1622279746778525e-12*x^2 - 1.89914322489962e-15*x^3 + 
1.9700370097535443e-19*x^4)};

\addplot[white, forget plot, name path=DuctMinus, domain=300:3500] 
{0.5293170695250425 + 0.000025945877040418022*x - 
1.0200935352386659e-8*x^2 - 
1.992102154002228*sqrt(3.6614916226344072e-6 - 7.392437709828245e-9*x 
+ 6.1622279746778525e-12*x^2 - 1.89914322489962e-15*x^3 + 
1.9700370097535443e-19*x^4)};

\addplot[red!20, forget plot] fill between[of=DuctPlus and DuctMinus ];

\addplot[red, thick, domain=300:3500] 
{0.5293170695250425 + 0.000025945877040418022*x - 
1.0200935352386659e-8*x^2};


\addplot[white, forget plot, name path=DuctKTPlus, domain=300:3500] 
{0.4706691275883809 + 0.00003138086938405285*x - 
1.1764225255422875e-8*x^2 + 
1.992102154002228*sqrt(0.000022955027834878136 - 
4.63454872729211e-8*x + 3.8632920476770057e-11*x^2 - 
1.1906318539826376e-14*x^3 + 1.2350773688810391e-18*x^4)};

\addplot[white, forget plot, name path=DuctKTMinus, domain=300:3500] 
{0.4706691275883809 + 0.00003138086938405285*x - 
1.1764225255422875e-8*x^2 - 
1.992102154002228*sqrt(0.000022955027834878136 - 
4.63454872729211e-8*x + 3.8632920476770057e-11*x^2 - 
1.1906318539826376e-14*x^3 + 1.2350773688810391e-18*x^4)};

\addplot[red!20, forget plot] fill between[of=DuctKTPlus and DuctKTMinus ];

\addplot[red, dashed, domain=300:3500] 
{0.4706691275883809 + 0.00003138086938405285*x - 
1.1764225255422875e-8*x^2
};

\end{semilogyaxis}
\end{tikzpicture}
\hskip -0.4cm
\begin{tikzpicture}
\begin{semilogyaxis}[
   log ticks with fixed point,
   ytick = {0.5,0.6,0.7, 0.8, 0.9, 1.0, 1.1},
   title = {$R = 0.7$},
   xlabel={$P_\LT\,\mathrm{[TeV]}$}, 
   xmin=2, xmax=3800,
   ymin=0.42, ymax=1.2,
  width=8.12cm,	
  height=6.38cm,
  x coord trafo/.code={
    \pgflibraryfpuifactive{
    \pgfmathparse{(#1)/(1000)}
  }{
     \pgfkeys{/pgf/fpu=true}
     \pgfmathparse{(#1)/(1000)}
     \pgfkeys{/pgf/fpu=false}
   }
 },
 xminorgrids=false,
 yminorgrids=false,
 minor x tick num=4,
 yticklabel = \empty,
]

\addplot[white, forget plot, name path=PythiaPlus, domain=300:3500] 
{1.0528387786719522 + 0.29666494726314846/exp(x/300.) + 
0.00005725787161999672*x - 7.247711040392513e-9*x^2 + 
1.9944371117711854*sqrt((0.002217631404226557 + 
exp(x/300.)*(-0.0012123206147504886 + 9.812438232839173e-7*x - 
1.7032092028884607e-10*x^2) + exp(x/150.)*(0.00020942901356627963 - 
3.6432722386804987e-7*x + 2.40556976700185e-10*x^2 - 
6.596782504078327e-14*x^3 + 6.433485353826403e-18*x^4))/exp(x/150.))};

\addplot[white, forget plot, name path=PythiaMinus, domain=300:3500] 
{1.0528387786719522 + 0.29666494726314846/exp(x/300.) + 
0.00005725787161999672*x - 7.247711040392513e-9*x^2 - 
1.9944371117711854*sqrt((0.002217631404226557 + 
exp(x/300.)*(-0.0012123206147504886 + 9.812438232839173e-7*x - 
1.7032092028884607e-10*x^2) + exp(x/150.)*(0.00020942901356627963 - 
3.6432722386804987e-7*x + 2.40556976700185e-10*x^2 - 
6.596782504078327e-14*x^3 + 6.433485353826403e-18*x^4))/exp(x/150.))};

\addplot[blue!20, forget plot, ] fill between[of=PythiaPlus and PythiaMinus];

\addplot[blue, thick, dashdotted, domain=300:3500] 
{1.0528387786719522 + 0.29666494726314846/exp(x/300.) + 
0.00005725787161999672*x - 7.247711040392513e-9*x^2};


\addplot[white, forget plot, name path=DirePlus, domain=300:3500] 
{1.0076356925331056 + 6.551980662055589e-6*x - 
6.0893979815879235e-9*x^2 + 
1.9939433678456226*sqrt(0.0008784559537477036 - 
1.930240916821851e-6*x + 1.7161883740959838e-9*x^2 - 
5.680552605648254e-13*x^3 + 6.352376251180806e-17*x^4)};

\addplot[white, forget plot, name path=DireMinus, domain=300:3500] 
{1.0076356925331056 + 6.551980662055589e-6*x - 
6.0893979815879235e-9*x^2 - 
1.9939433678456226*sqrt(0.0008784559537477036 - 
1.930240916821851e-6*x + 1.7161883740959838e-9*x^2 - 
5.680552605648254e-13*x^3 + 6.352376251180806e-17*x^4)};

\addplot[black!20, forget plot] fill between[of=DirePlus and DireMinus];

\addplot[black, densely dashed, very thick, domain=300:3500] 
{1.0076356925331056 + 6.551980662055589e-6*x - 
6.0893979815879235e-9*x^2};


\addplot[white, forget plot, name path=DuctPlus, domain=300:3500] 
{0.8839042938140268 - 0.000037437203447146006*x - 
8.515064867027739e-10*x^2 + 
1.992102154002228*sqrt(0.000017380436139025302 - 
3.5090560014705007e-8*x + 2.925097769606384e-11*x^2 - 
9.014888177043412e-15*x^3 + 9.35140810588621e-19*x^4)};

\addplot[white, forget plot, name path=DuctMinus, domain=300:3500] 
{0.8839042938140268 - 0.000037437203447146006*x - 
8.515064867027739e-10*x^2 - 
1.992102154002228*sqrt(0.000017380436139025302 - 
3.5090560014705007e-8*x + 2.925097769606384e-11*x^2 - 
9.014888177043412e-15*x^3 + 9.35140810588621e-19*x^4)};

\addplot[red!20, forget plot] fill between[of=DuctPlus and DuctMinus ];

\addplot[red, thick, domain=300:3500] 
{0.8839042938140268 - 0.000037437203447146006*x - 
8.515064867027739e-10*x^2};


\addplot[white, forget plot, name path=DuctKTPlus, domain=300:3500] 
{0.840351218986353 - 0.00003102142651114759*x - 
4.56520266993207e-9*x^2 + 
1.992102154002228*sqrt(0.00007767893014573237 - 
1.5683134406716692e-7*x + 1.3073231505652786e-10*x^2 - 
4.0290523400836525e-14*x^3 + 4.179454250807467e-18*x^4)};

\addplot[white, forget plot, name path=DuctKTMinus, domain=300:3500] 
{0.840351218986353 - 0.00003102142651114759*x - 
4.56520266993207e-9*x^2 - 
1.992102154002228*sqrt(0.00007767893014573237 - 
1.5683134406716692e-7*x + 1.3073231505652786e-10*x^2 - 
4.0290523400836525e-14*x^3 + 4.179454250807467e-18*x^4)};

\addplot[red!20, forget plot] fill between[of=DuctKTPlus and DuctKTMinus ];

\addplot[red, dashed, domain=300:3500] 
{0.840351218986353 - 0.00003102142651114759*x - 
4.56520266993207e-9*x^2
};

\end{semilogyaxis}
\end{tikzpicture}
\else 
\includegraphics[width = 14.3 cm]{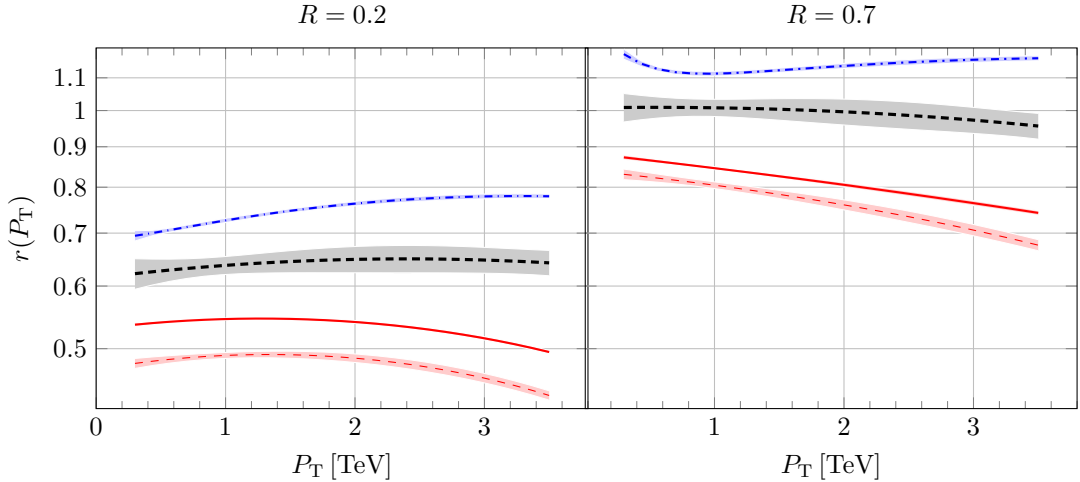}
\fi
\end{center}
\caption{
Ratio $r(P_\LT)$ of the one jet inclusive cross section $d\sigma/dP_\LT$ calculated after showering to the cross section at the Born level for $R = 0.2$ and for $R = 0.7$. The labeling is the same as in Fig.\ \ref{fig:JetFindingEffectR04}.
}
\label{fig:JetFindingEffectR0207}
\end{figure}

One would expect that the effect of the jet algorithm would depend on the value of $R$: for smaller $R$, it should be much easier for partons to be radiated out of the jet, so that $r(P_\LT)$ should be smaller. To investigate this, we exhibit in Fig.\ \ref{fig:JetFindingEffectR0207} the same plot for $R = 0.2$ and $R = 0.7$. We see that for smaller $R$, $r(P_\LT)$ is indeed smaller, while  $r(P_\LT)$ is larger for larger $R$. In fact, according to  \textsc{Dire}, $r(P_\LT)$ is close to 1 for $R = 0.7$ while according to  \textsc{Pythia}, $r(P_\LT)$ is greater than 1 for $R = 0.7$.

\begin{figure}
\begin{center}
\ifusefigs 
\begin{tikzpicture}
\begin{loglogaxis}[
   log ticks with fixed point,
   ytick = {0.5, 0.6, 0.7, 0.8, 0.9, 1.0, 1.1, 1.2},
   xtick = {0.2, 0.3, 0.4, 0.5, 0.6, 0.7, 0.8},
   title = {Dependence of $r(R)$ on $R$},
   xlabel={$R$}, ylabel={$r(R)$},
   xmin=0.15, xmax=0.9,
   ymin=0.45, ymax=1.2,
  legend cell align=left,
  width=14cm,	
  height=11cm,
 xminorgrids=false,
 yminorgrids=false,
 minor x tick num=4,
 every axis legend/.append style={
  at={(0.02,0.98)},
  anchor= north west,
  },
]
\pgfplotstableread{
0.2  0.762784
0.4  0.951344
0.7  1.13874
}\Pythia

\pgfplotstableread{
0.2  0.648438
0.4  0.824563
0.7  0.996382
}\Dire

\pgfplotstableread{
0.2  0.540405
0.4  0.672247
0.7  0.805624
}\Deductor

\pgfplotstableread{
0.2  0.486374
0.4  0.619862
0.7  0.760048
}\DeductorKT

\addplot[blue, dashdotted, thick, domain=0.18:0.78] 
{exp(0.242137 + 0.318693*ln(x))};
\addlegendentry{\textsc{Pythia}}
\addplot[black, densely dashed, very thick, domain=0.18:0.78] 
{exp(0.12474 + 0.346661*ln(x))};
\addlegendentry{\textsc{Dire}}
\addplot[red, thick, domain=0.18:0.78] 
{exp(-0.108543 + 0.314951*ln(x))};
\addlegendentry{\textsc{Deductor-$\Lambda$}}
\addplot[red, densely dashed, very thick, domain=0.18:0.78] 
{exp(-0.157664 + 0.349882*ln(x))};
\addlegendentry{\textsc{Deductor-$k_\LT$}}

\addplot [only marks, mark size=3pt, blue] table [x={0},y={1}]{\Pythia};
\addplot [only marks, mark size=3pt, black] table [x={0},y={1}]{\Dire};
\addplot [only marks, mark size=3pt, red] table [x={0},y={1}]{\Deductor};
\addplot [only marks, mark size=3pt, red] table [x={0},y={1}]{\DeductorKT};

\end{loglogaxis}
\end{tikzpicture}
\else 
\includegraphics[width = 14 cm]{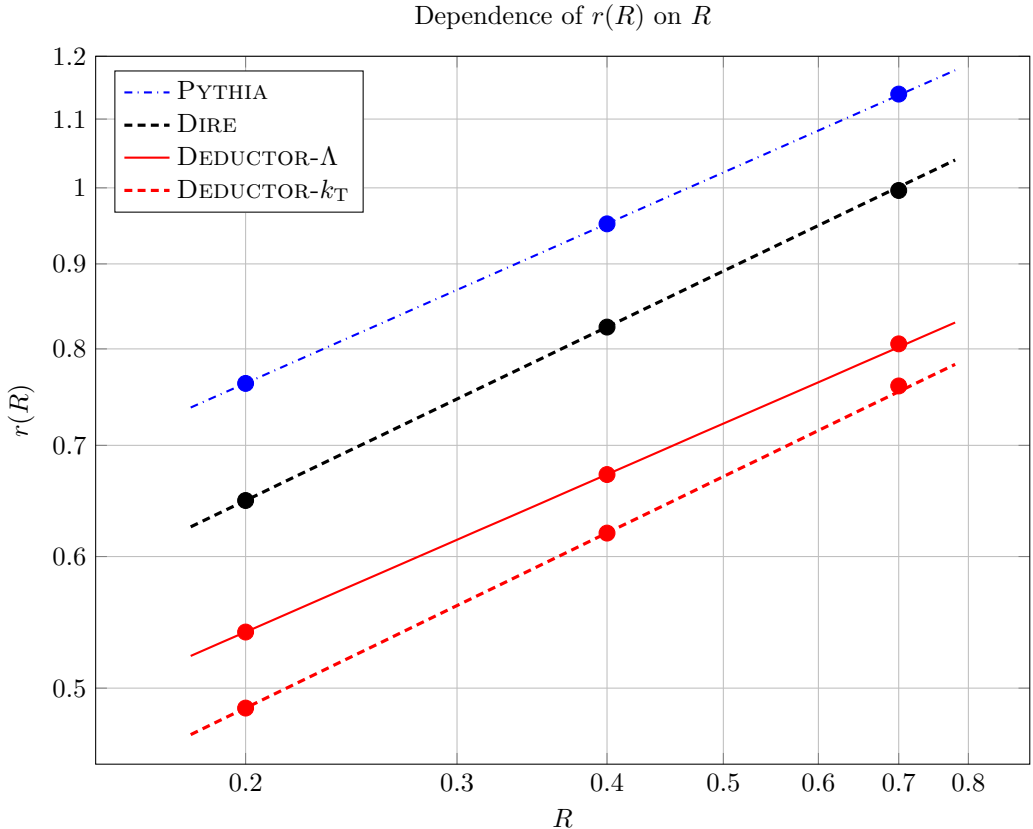}
\fi
\end{center}
\caption{
Ratio of the one jet inclusive cross section $d\sigma/dP_\LT$ calculated after showering to the cross section at the Born level at $P_\LT = 2 \TeV$ versus $R$ for (from top to bottom) \textsc{Pythia}, \textsc{Dire}, \textsc{Deductor} with its default $\Lambda$ ordering, and \textsc{Deductor} with $k_\LT$ ordering. 
}
\label{fig:Rdependence}
\end{figure}

We explore the difference among the parton showers by plotting $\log(r(2 \TeV))$ versus $\log(R)$ in Fig.~\ref{fig:Rdependence}. We see that \textsc{Pythia}, \textsc{Dire}, \textsc{Deductor} with $\Lambda$ ordering, and \textsc{Deductor} with $k_\LT$ ordering all give approximately straight lines in this plot. We see also that the slopes as determined from the $R = 0.4$ and $R = 0.2$ points are not very different among the shower algorithms. This is to be expected since the slopes $d\log(r(P_\LT))/d\log(R)$ are given rather directly by perturbation theory \cite{DasguptaSmallRjets1, DasguptaSmallRjets2, KangJetinParton, Dai:2016hzf, Dai:2017dpc}. We also note that the four algorithms give very different results at fixed $R$, for instance at $R = 0.4$.

\begin{figure}
\begin{center}
\ifusefigs 
\begin{tikzpicture}
\begin{axis}[
   title = {Distribution of jets in a parton},
   xlabel={$z$}, ylabel={$f(z)$},
   xmin=0.8, xmax=1.2,
   ymin=0.0, ymax=30,
  legend cell align=left,
  width=14cm,	
  height=11cm,
 xminorgrids=false,
 yminorgrids=false,
 minor x tick num=4,
]
\pgfplotstableread{
0.795 0.34407
0.805 0.319725
0.815 0.35463
0.825 0.39661
0.835 0.42877
0.845 0.475935
0.855 0.48261
0.865 0.54905
0.875 0.66265
0.885 0.73355
0.895 0.8341
0.905 0.9958
0.915 1.15505
0.925 1.40865
0.935 1.7316
0.945 2.1671
0.955 2.9138
0.965 4.12475
0.975 6.2045
0.985 10.8765
0.995 25.7795
1.005 17.3385
1.015 5.437
1.025 2.78685
1.035 1.632
1.045 1.10135
1.055 0.7606
1.065 0.5503
1.075 0.40896
1.085 0.29314
1.095 0.233755
1.105 0.17272
1.115 0.126695
1.125 0.098385
1.135 0.083765
1.145 0.08126
1.155 0.05571
1.165 0.047482
1.175 0.0368215
1.185 0.032066
1.195 0.0270775
1.205 0.0241325
}\Pythia

\pgfplotstableread{
0.795  0.52225
0.805  0.548
0.815  0.5887
0.825  0.633
0.835  0.68535
0.845  0.7366
0.855  0.81825
0.865  0.88415
0.875  0.9907
0.885  1.10685
0.895  1.22895
0.905  1.3941
0.915  1.5897
0.925  1.84665
0.935  2.255
0.945  2.6916
0.955  3.41735
0.965  4.5218
0.975  6.4175
0.985  10.5755
0.995  26.7125
1.005  14.3955
1.015  2.5475
1.025  1.20105
1.035  0.69575
1.045  0.436215
1.055  0.285765
1.065  0.213
1.075  0.14804
1.085  0.110195
1.095  0.08348
1.105  0.062705
1.115  0.049806
1.125  0.0394065
1.135  0.0322725
1.145  0.026718
1.155  0.0202625
1.165  0.01441
1.175  0.0148605
1.185  0.0106815
1.195  0.009589
1.205  0.0064475
}\Dire

\pgfplotstableread{
0.795   0.791472
0.805   0.840024
0.815   0.882459
0.825   0.933391
0.835   1.00318
0.845   1.07089
0.855   1.11878
0.865   1.31603
0.875   1.38606
0.885   1.56218
0.895   1.69454
0.905   1.90151
0.915   2.11822
0.925   2.50301
0.935   2.95114
0.945   3.57089
0.955   4.33463
0.965   5.59687
0.975   7.58524
0.985   11.0829
0.995   16.8991
1.005   6.80874
1.015   2.29147
1.025   1.15135
1.035   0.710988
1.045   0.464648
1.055   0.327455
1.065   0.228982
1.075   0.170262
1.085   0.130909
1.095   0.098121
1.105   0.0796263
1.115   0.063852
1.125   0.0454354
1.135   0.0413118
1.145   0.0355901
1.155   0.0204591
1.165   0.025282
1.175   0.0189362
1.185   0.0141743
1.195   0.012118
1.205   0.0110698
}\Deductor

\addplot [const plot mark mid, blue, thick, dashdotted] table [x={0},y={1}]{\Pythia};
\addplot [const plot mark mid, black, very thick, densely dashed] table [x={0},y={1}]{\Dire};
\addplot [const plot mark mid, red, thick] table [x={0},y={1}]{\Deductor};

\legend{ \textsc{Pythia}, \textsc{Dire}, \textsc{Deductor}}

\end{axis}
\end{tikzpicture}
\else 
\includegraphics[width = 14 cm]{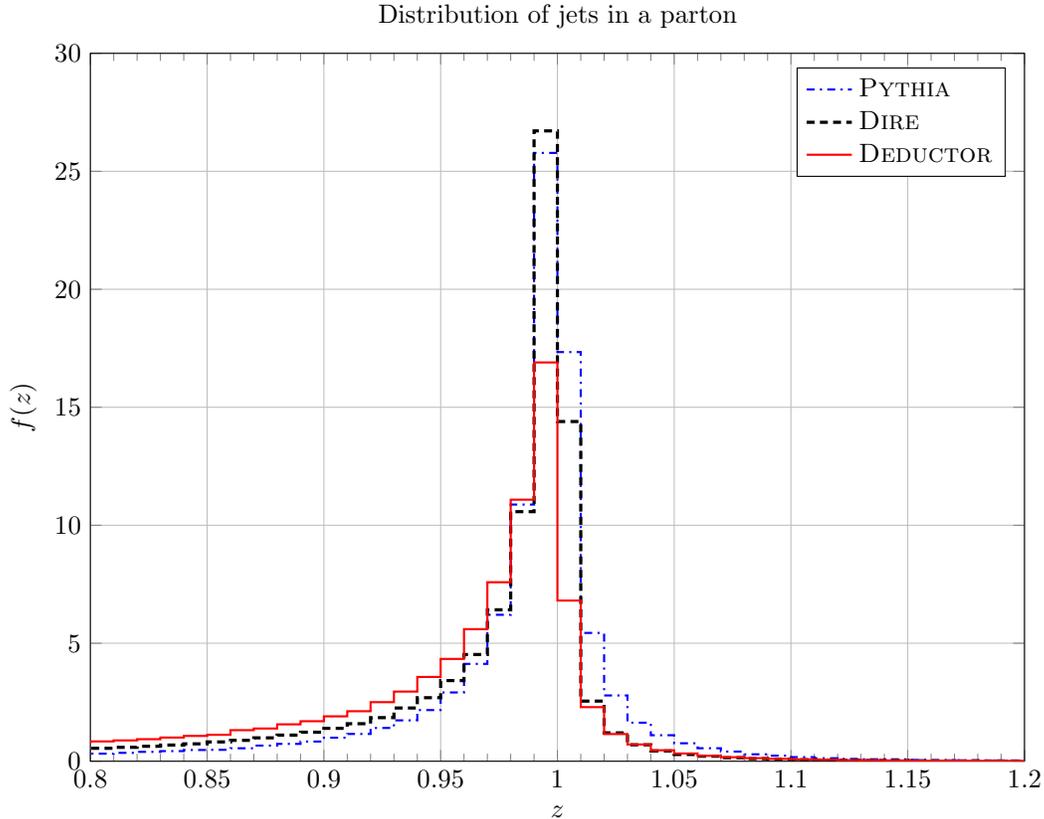}
\fi
\end{center}
\caption{
Effective distribution of jets in a parton, $f(z)$, for \textsc{Pythia}, \textsc{Dire}, and \textsc{Deductor} for $R = 0.4$ jets with $P_\LT \approx 3 \TeV$.
}
\label{fig:jetsinaparton}
\end{figure}

\begin{figure}
\begin{center}
\ifusefigs 
\begin{tikzpicture}
\begin{axis}[
   title = {Distribution of jets in a parton},
   xlabel={$z$}, ylabel={$f(z)$},
   xmin=0.8, xmax=1.2,
   ymin=0.0, ymax=30,
  legend cell align=left,
  width=14cm,	
  height=11cm,
 xminorgrids=false,
 yminorgrids=false,
 minor x tick num=4,
]
\pgfplotstableread{
0.795 0.34407
0.805 0.319725
0.815 0.35463
0.825 0.39661
0.835 0.42877
0.845 0.475935
0.855 0.48261
0.865 0.54905
0.875 0.66265
0.885 0.73355
0.895 0.8341
0.905 0.9958
0.915 1.15505
0.925 1.40865
0.935 1.7316
0.945 2.1671
0.955 2.9138
0.965 4.12475
0.975 6.2045
0.985 10.8765
0.995 25.7795
1.005 17.3385
1.015 5.437
1.025 2.78685
1.035 1.632
1.045 1.10135
1.055 0.7606
1.065 0.5503
1.075 0.40896
1.085 0.29314
1.095 0.233755
1.105 0.17272
1.115 0.126695
1.125 0.098385
1.135 0.083765
1.145 0.08126
1.155 0.05571
1.165 0.047482
1.175 0.0368215
1.185 0.032066
1.195 0.0270775
1.205 0.0241325
}\Pythia

\pgfplotstableread{
0.795   0.301
0.805   0.33262
0.815   0.334485
0.825   0.345805
0.835   0.39392
0.845   0.44628
0.855   0.5013
0.865   0.5461
0.875   0.62805
0.885   0.70525
0.895   0.79995
0.905   0.9246
0.915   1.1017
0.925   1.26515
0.935   1.6261
0.945   2.07235
0.955   2.7509
0.965   3.9094
0.975   5.9235
0.985   10.542
0.995   27.0325
1.005   18.2975
1.015   5.474
1.025   2.7587
1.035   1.6094
1.045   1.0441
1.055   0.741
1.065   0.51355
1.075   0.373995
1.085   0.277765
1.095   0.233455
1.105   0.17038
1.115   0.14036
1.125   0.09927
1.135   0.08509
1.145   0.067325
1.155   0.05357
1.165   0.045806
1.175   0.0361
1.185   0.033939
1.195   0.0240655
1.205   0.0166675
}\PythiaMod

\addplot [const plot mark mid, blue, very thick, dashdotted] table [x={0},y={1}]{\Pythia};
\addplot [const plot mark mid, blue] table [x={0},y={1}]{\PythiaMod};

\legend{ \textsc{Pythia}, \textsc{Pythia}-modified}

\end{axis}
\end{tikzpicture}
\else 
\includegraphics[width = 14 cm]{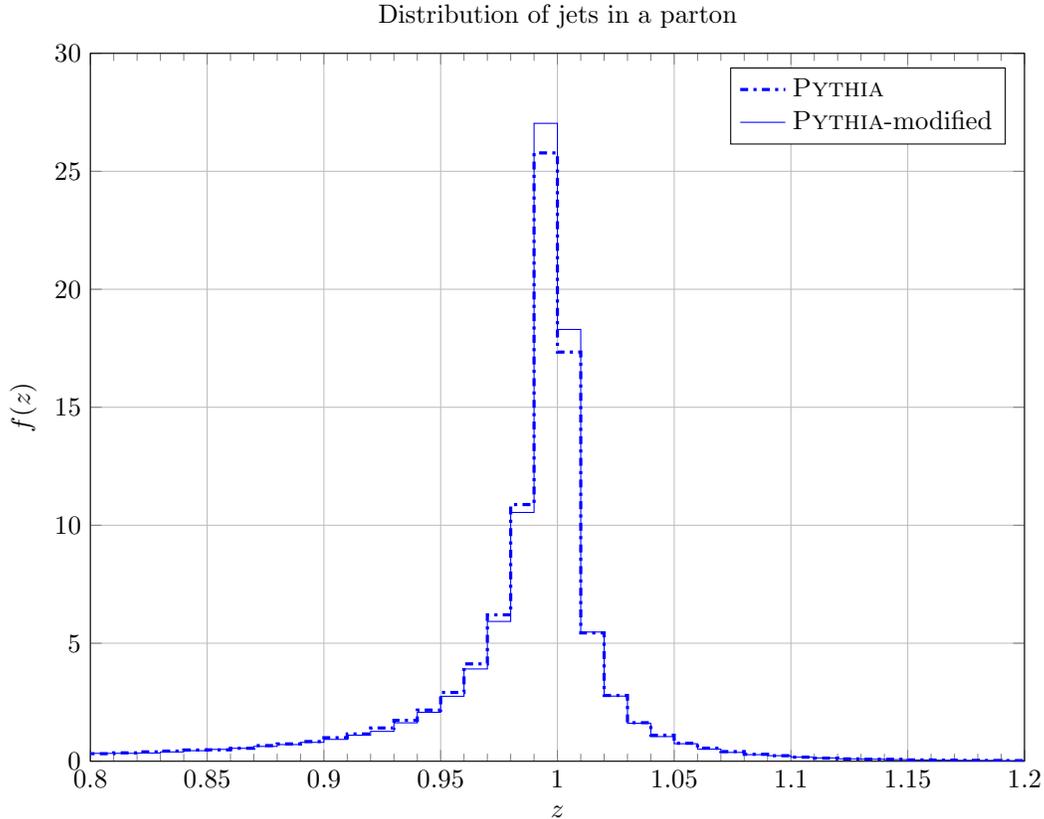}
\fi
\end{center}
\caption{
Effective distribution of jets in a parton, $f(z)$, as in Fig.~\ref{fig:jetsinaparton}, for \textsc{Pythia} compared with \textsc{Pythia} with $\as(M_\LZ^2) = 0.118$ and with $\as$ evaluated at a scale $\mu^2 = \lambda_\LR k_\LT^2$ where $\lambda_\LR \approx 0.4$ is the group theory factor defined in Ref.~\cite{lambdaR}.
}
\label{fig:jetsinapartonPytia}
\end{figure}

Why do the shower programs give different results at $R = 0.4$? We can get some idea by calculating a function $f(z)$ that gives an effective distribution of jets in a parton as a function of the ratio $z$ of the jet $P_\LT$ to the parton $p_\LT$. For theoretical analyses of $f(z)$ see Refs.~\cite{KangJetinParton, Dai:2016hzf, Dai:2017dpc}. In nature, of course, we do not have access to the parton that initiates a jet, but in a parton shower program, we do. Each event starts with a Born level $2 \to 2$ scattering that produces two final state partons with $\bm p_{\LT,1} = -\bm p_{\LT,2}$. We define $P_\LT^{\rm Born} = |\bm p_{\LT,1}|$. We select events in which the two initiating partons have $P_\LT^{\rm Born}$ close to 3 TeV, $2.8 \TeV < P_\LT^{\rm Born} < 3.3 \TeV$, and for which $|y_1| < 2$ and $|y_2| < 2$. For these events, we look at the final state after showering. We look at all final state jets with rapidity $|y| < 2$ and transverse momentum $P_\LT$ and define
\begin{equation}
z = \frac{P_\LT}{P_\LT^{\rm Born}}
\,.
\end{equation}
Then we define $f(z)\,dz$ to be the ratio of the number of final state jets in a range $dz$ to the number of Born level partons in the sample, which is twice the number of events in the sample. We display $f(z)$ in Fig.\ \ref{fig:jetsinaparton} for \textsc{Pythia}, \textsc{Dire}, and \textsc{Deductor} with $\Lambda$ ordering.  For $z > 1.01$, we see that $f(z)$ for \textsc{Dire} (in black) and \textsc{Deductor} (in red) are close to each other while $f(z)$ for \textsc{Pythia} (in blue) is larger. With a steeply falling $P_\LT$ spectrum, this increases the \textsc{Pythia} after-showering cross section. It may be surprising to have jets with more transverse momentum than the initiating parton, but it can easily happen. When there is a parton emitted from the initial state, the two original partons recoil in the opposite direction in order for transverse momentum to be conserved. Thus an initial state emission in the direction opposite to the observed jet increases the transverse momentum of the jet.\footnote{We thank Gavin Salam for this observation.} It appears that this effect is bigger in \textsc{Pythia} than in \textsc{Dire} or \textsc{Deductor}. Perhaps this is due to the lack of quantum interference for initial state emissions in \textsc{Pythia}. We will see in Fig.~\ref{fig:jetsinapartonPytia} below that it is not due to the different treatment of $\as$ in \textsc{Pythia}.

For $z < 0.99$, $f(z)$ for \textsc{Pythia} and for \textsc{Dire} are close to each other while $f(z)$ for \textsc{Deductor} is larger. It seems plausible that this happens because \textsc{Deductor} has more final state radiation out of the jet. More radiation out of the jet makes $f(z)$ larger for $z < 1$. Here the cone size, $R = 0.4$ is not particularly small, so this is wide angle radiation. To see how this happens in more detail, we fit $f(z)$ according to \textsc{Pythia}, \textsc{Dire}, and \textsc{Deductor} in the range $0.8 < z < 0.97$. We estimate errors in the coefficients of the fits that arise from the Monte Carlo statistical errors. We find
\begin{equation}
\begin{split}
\label{eq:jetinpartonfits}
f(z;\textsc{Pythia}) \approx{}& 
(2.1 \pm 0.1)\,\frac{1-0.9}{1-z}
- (1.5 \pm 0.2)\,\frac{\log(1-z)}{\log(1-0.9)}
+ (0.3 \pm 0.1)
\;,
\\
f(z;\textsc{Dire}) \approx{}& 
(1.8 \pm 0.1)\,\frac{1-0.9}{1-z}
- (0.4 \pm 0.1)\,\frac{\log(1-z)}{\log(1-0.9)}
- (0.1 \pm 0.1)
\;,
\\
f(z;\textsc{Deductor}) \approx{}& 
(2.1 \pm 0.1)\,\frac{1-0.9}{1-z}
- (0.1 \pm 0.4)\,\frac{\log(1-z)}{\log(1-0.9)}
- (0.2 \pm 0.2)
\;.
\end{split}
\end{equation}
In each case, we find that around $z = 0.9$, $f(z)$ is dominated by a term proportional to $1/(1-z)$. This is expected from the $1/(1-z)$ singularity from soft gluon emissions. In some of the cases, there is also a term with a weak $\log(1-z)$ singularity and a constant term. The coefficient of the $1/(1-z)$ leading singularity for \textsc{Deductor} lies close to the corresponding coefficients for \textsc{Pythia} and \textsc{Dire}. To a good approximation, the differences between $f(z)$ for \textsc{Deductor} and \textsc{Dire} or \textsc{Pythia} have only a constant term and a $\log(1-z)$ term. The sum of these terms is consistent with zero for \textsc{Deductor}, slightly negative for \textsc{Dire}, and negative for \textsc{Pythia}.

It is not a surprise that there are differences between \textsc{Deductor} splitting and \textsc{Dire} or \textsc{Pythia} splitting that are only weakly singular. The \textsc{Deductor} splitting kernel has contributions corresponding to soft interference diagrams in an eikonal approximation and then ``direct'' terms whose collinear limit is the DGLAP evolution kernel. However, the direct terms are not obtained from the DGLAP kernel but are rather designed to approximate as closely as possible the Feynman diagrams from which the DGLAP kernel is derived \cite{NSI}. On the other hand, \textsc{Dire} and \textsc{Pythia} are based more closely on the DGLAP kernel. Additionally, \textsc{Deductor}, in contrast to \textsc{Dire} and \textsc{Pythia}, uses a global recoil strategy for each splitting.

Although one can expect differences for the function $f(z)$ among \textsc{Deductor}, \textsc{Dire}, and \textsc{Pythia} simply based on the many structural differences among these programs, it is reasonable to ask if a substantial part of the differences could arise from different treatments of $\as$ at the splitting vertices. In particular, \textsc{Deductor} sets $\as(M_\LZ^2) = 0.118$ \cite{PDG} and inserts a factor $\lambda_\LR \approx 0.4$ defined in Ref.~\cite{lambdaR} in the scale argument of $\as$, whereas \textsc{Pythia} sets $\as(M_\LZ^2) = 0.1365$ and does not include a factor $\lambda_\LR$. Could these differences matter? To find out, we set the parameters in \textsc{Pythia} so that $\as(M_\LZ^2) = 0.118$ and a factor $\lambda_\LR$ is included in the argument of $\as$. The resulting $f(z)$ is compared to the $f(z)$ in \textsc{Pythia} with default parameters in Fig.~\ref{fig:jetsinapartonPytia}. We see that adjusting $\as$ in this way makes hardly any difference.

\begin{figure}
\begin{center}
\ifusefigs 
\begin{tikzpicture}
\begin{axis}[
   title = {Distribution of jets in a parton},
   xlabel={$z$}, ylabel={$f(z)$},
   xmin=0.8, xmax=1.2,
   ymin=0.0, ymax=30,
  legend cell align=left,
  width=14cm,	
  height=11cm,
 xminorgrids=false,
 yminorgrids=false,
 minor x tick num=4,
]
\pgfplotstableread{
0.795   0.791472
0.805   0.840024
0.815   0.882459
0.825   0.933391
0.835   1.00318
0.845   1.07089
0.855   1.11878
0.865   1.31603
0.875   1.38606
0.885   1.56218
0.895   1.69454
0.905   1.90151
0.915   2.11822
0.925   2.50301
0.935   2.95114
0.945   3.57089
0.955   4.33463
0.965   5.59687
0.975   7.58524
0.985   11.0829
0.995   16.8991
1.005   6.80874
1.015   2.29147
1.025   1.15135
1.035   0.710988
1.045   0.464648
1.055   0.327455
1.065   0.228982
1.075   0.170262
1.085   0.130909
1.095   0.098121
1.105   0.0796263
1.115   0.063852
1.125   0.0454354
1.135   0.0413118
1.145   0.0355901
1.155   0.0204591
1.165   0.025282
1.175   0.0189362
1.185   0.0141743
1.195   0.012118
1.205   0.0110698
}\Deductor

\pgfplotstableread{
0.795   0.937086
0.805   0.847872
0.815   0.822452
0.825   1.18401
0.835   1.31256
0.845   1.21729
0.855   1.36904
0.865   1.34787
0.875   1.61659
0.885   1.68949
0.895   1.84316
0.905   2.1214
0.915   2.40983
0.925   2.65162
0.935   3.14099
0.945   3.64969
0.955   4.58154
0.965   5.85659
0.975   7.72962
0.985   11.4233
0.995   14.6714
1.005   4.99777
1.015   1.82701
1.025   0.942673
1.035   0.539461
1.045   0.359072
1.055   0.234268
1.065   0.146448
1.075   0.11016
1.085   0.105865
1.095   0.0494794
1.105   0.0747068
1.115   0.0523943
1.125   0.0423252
1.135   0.0386926
1.145   0.0355359
1.155   0.00174571
1.165   0.0228977
1.175   0.0252178
1.185   0.0182999
1.195   -0.00607924
1.205   0.00640448
}\DeductorkT

\addplot [const plot mark mid, red] table [x={0},y={1}]{\Deductor};
\addplot [const plot mark mid, red, very thick, dashdotted] table [x={0},y={1}]{\DeductorkT};

\legend{ \textsc{Deductor}, \textsc{Deductor}-$k_\LT$}

\end{axis}
\end{tikzpicture}
\else 
\includegraphics[width = 14 cm]{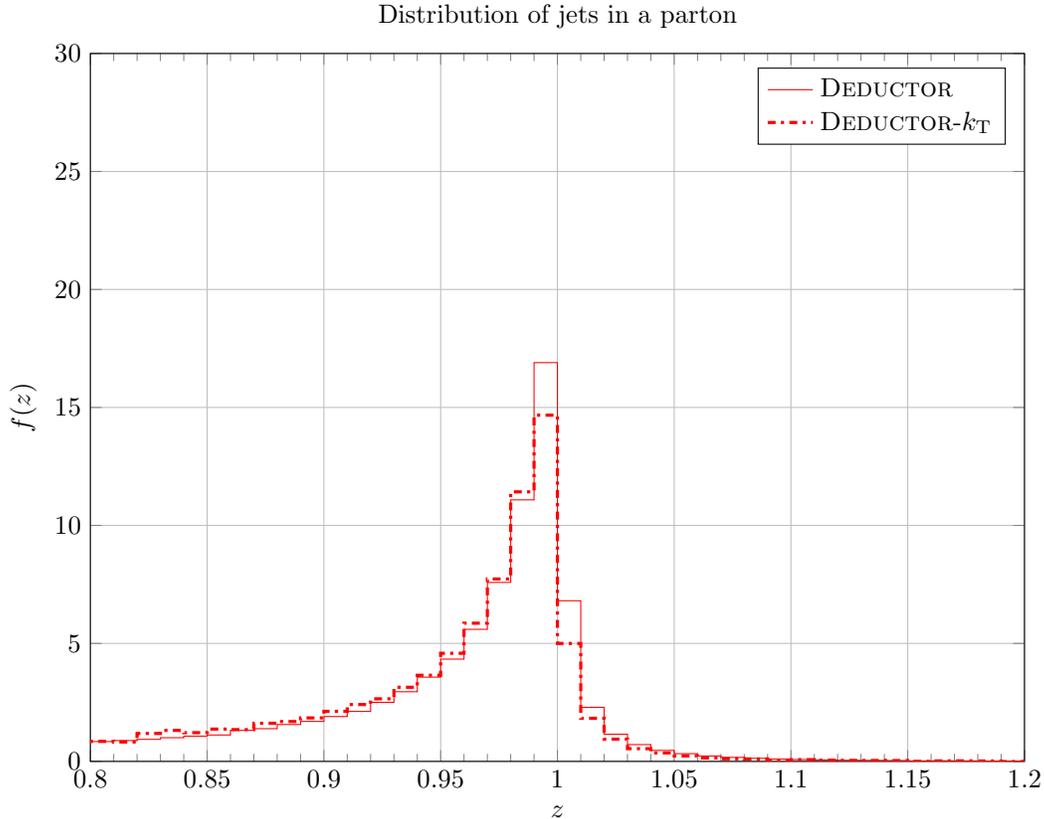}
\fi
\end{center}
\caption{
Effective distribution of jets in a parton, $f(z)$, as in Fig.~\ref{fig:jetsinaparton}, for \textsc{Deductor} with $\Lambda$ ordering compared with \textsc{Deductor} with $k_\LT$ ordering.
}
\label{fig:jetsinapartonkT}
\end{figure}

One can also ask if the choice of shower ordering parameter matters for $f(z)$. In Fig.~\ref{fig:jetsinapartonkT}, we compare $f(z)$ obtained using \textsc{Deductor} with $\Lambda$ ordering to $f(z)$ obtained using \textsc{Deductor} with $k_\LT$ ordering. We see that for $z < 0.99$, $f(z)$ obtained with $k_\LT$ ordering is slightly larger than with $\Lambda$ ordering. This makes $r(P_\LT)$ with $k_\LT$ ordering somewhat smaller than $r(P_\LT)$ with $\Lambda$ ordering, as we have seen see in Fig.~\ref{fig:JetFindingEffectR04}.

We conclude that there are significant differences among the three results for $f(z)$ in Fig.\ \ref{fig:jetsinaparton}. There are many structural differences among the three shower algorithms, so it is not a surprise to see differences in the results. However, it is important to appreciate also the similarities among the results for $f(z)$. In each case, we have a function that is strongly peaked near $z = 1$. The rather small differences in $f(z)$ lead to substantial differences in $d\sigma/dP_\LT$ because the cross section falls so steeply with $P_\LT$.

\subsection{Effect of the threshold factors}
\label{sec:thresholdeffects}

The threshold factors $\cU_\cV(\mu_\Lf^2,\mu_\Ls^{2})$ from Eq.~(\ref{eq:UVexponential01}) and $Z_\La Z_\Lb$ from Eq.~(\ref{eq:ZaZb}) are included in \textsc{Deductor}. A factor similar to $\cU_\cV(\mu_\Lf^2,\mu_\Ls^{2})$ was included in an earlier version of \textsc{Deductor} that we used for Ref.~\cite{DuctThreshold}. However, in this version we did not have the general results of Ref.~\cite{NSallorder} and consequently did not have the proper organization for the threshold effect. As a result, the threshold effects were contaminated by soft scale physics. To eliminate this contamination, we had to use an {\em ad hoc} infrared regulator. With the organization from Ref.~\cite{NSallorder}, we have a single factor in which the integral over $\mu^2$ in Eq.~(\ref{eq:UVexponential01}) is well behaved in the infrared, so that no regulator is needed.

\begin{figure}
\begin{center}
\ifusefigs 
\begin{tikzpicture}
\begin{semilogyaxis}[
   log ticks with fixed point,
   ytick = {0.4, 0.5, 0.6, 0.7, 0.8, 0.9, 1.0, 1.1, 1.2, 1.3, 1.4, 1.5},
   title = {Ratios to NLO jet cross section, $R = 0.4$},
   xlabel={$P_\LT\,\mathrm{[TeV]}$}, ylabel={$K(P_\LT)$},
   xmin=0, xmax=3800,
   ymin=0.38, ymax=1.6,
  legend cell align=left,
  every axis legend/.append style={
  at={(0.02,0.02)},
  anchor= south west,
  },
  width=14cm,	
  height=11cm,
  x coord trafo/.code={
    \pgflibraryfpuifactive{
    \pgfmathparse{(#1)/(1000)}
  }{
     \pgfkeys{/pgf/fpu=true}
     \pgfmathparse{(#1)/(1000)}
     \pgfkeys{/pgf/fpu=false}
   }
 },
 xminorgrids=false,
 yminorgrids=false,
 minor x tick num=4,
]


\addplot[purple, very thick, dashdotted, domain=300:3500] 
{1.551224347974936 - 0.00031962651521130216*x + 
8.046065413106988e-8*x^2 - 6.5782530399931215e-12*x^3 + 
(18344.97800963932*x)/(1. + x)^3};

\addlegendentry{\textsc{FHMRV}(NNLO)}

\addplot[black, very thick, densely dashed, domain=300:3500] 
{1.10306 - 0.0000640375*x - 4.22713e-10*x^2};

\addlegendentry{\textsc{Deductor}(LO)}


\addplot[green!50!black, very thick, dashdotted, domain=300:3500] 
{0.943498088497596 + 0.000022899989715684918*x - 
2.202386768376987e-9*x^2};

\addlegendentry{\textsc{FHMRV}(NLO)}

\addplot[red, very thick, domain=300:3500] 
{0.9854211660494355 - 0.000018716423257807602*x + 
2.028986810514001e-8*x^2};

\addlegendentry{\textsc{Deductor}(full)}

\addplot[blue, very thick, domain=300:3500] 
{0.7780090775098257 - 0.000056369267929229*x - 3.068166458874673e-9*x^2};

\addlegendentry{\textsc{Deductor}(std.)}

\end{semilogyaxis}
\end{tikzpicture}
\else 
\includegraphics[width = 14 cm]{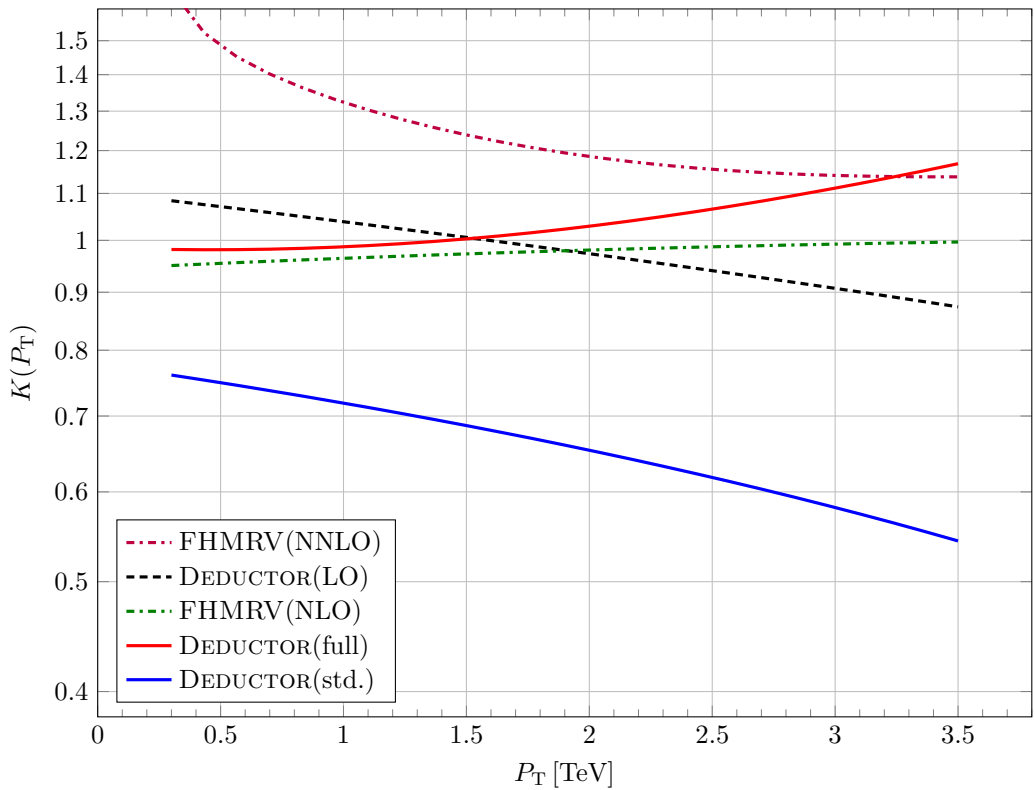}
\fi
\end{center}
\caption{
Ratios $K(P_\LT)$ of the one jet inclusive cross section $d\sigma/dP_\LT$ calculated in different approximations to the NLO cross section.
}
\label{fig:JetKfactorsR04}
\end{figure}

In this subsection, we examine the effect of the threshold factors. We look first at the inclusive jet cross section with $R = 0.4$ calculated with \textsc{Deductor} with $\Lambda$ ordering using the parameters from  Sec.\ \ref{sec:onejet}. In Fig.\ \ref{fig:JetKfactorsR04}, we plot ratios $K$ of the cross sections calculated with various approximations ``A'' to the purely perturbative NLO cross section,
\begin{equation}
\label{eq:Kdef}
K({\rm ``A"}) = 
\frac{d\sigma({\rm ``A"})/dP_\LT}{d\sigma({\rm NLO})/dP_\LT}
\;.
\end{equation}

First, in black (dashed), we plot $K({\rm LO})$ using the Born cross section in the numerator. We see that $K({\rm LO})$ is rather close to 1. That is, with the choice of $\mu_\LR^2$ and $\mu_\LF^2$ that we use, the LO cross section is fairly close to the NLO cross section.

Next, in blue, we plot $K({\rm std.})$ using \textsc{Deductor} with the threshold factors turned off, so that we have a standard probability preserving parton shower. Showering reduces the cross section substantially, multiplying it by a factor of roughly 0.6. This is the effect that we examined in section \ref{sec:jetfinding}.

Next, in red, we plot $K({\rm full})$ using \textsc{Deductor} including the threshold factors. We see that the threshold factors increase the cross section by a factor that ranges from about 1.3 at $P_\LT = 0.3 \TeV$ to more than 2 at $P_\LT = 3.5 \TeV$. With both showering and the threshold factors included, the full cross section is quite close to what one gets with an NLO calculation.

There is a calculation of de Florian, Hinderer, Mukherjee, Ringer and Vogelsang (FHMRV) \cite{deFlorianJets} that is relevant to this analysis. This calculation sums threshold logarithms and also the logarithms of $1/R$ that arise from the jet definition. These authors then expand the analytic result perturbatively to either NLO or NNLO. We show as a green dash-dotted line the ratio $K({\rm FHMRV\ NLO})$ from this calculation expanded to NLO. We note that this result is pretty close to the purely perturbative NLO result. We also show as a purple dashdotted line the ratio $K({\rm FHMRV\ NNLO})$ corresponding to the NNLO calculation. It is  puzzling to us that $K({\rm FHMRV\ NNLO})$ decreases with increasing $P_\LT$ since one normally expects the effects of threshold logarithms to grow strongly as one approaches the largest possible jet $P_\LT$, $P_\LT = \sqrt{s}/2$.

\begin{figure}
\begin{center}
\ifusefigs 
\begin{tikzpicture}
\begin{semilogyaxis}[
   log ticks with fixed point,
   ytick = {0.4, 0.6, 0.8, 1.0, 1.2, 1.4},
   title = {$R = 0.2$},
   xlabel={$P_\LT\,\mathrm{[TeV]}$}, ylabel={$K(P_\LT)$},
   xmin=0, xmax=3800,
   ymin=0.38, ymax=1.6,
  legend cell align=left,
  every axis legend/.append style={
  at={(0.02,0.02)},
  anchor= south west,
  },
  width=8.12cm,	
  height=6.38cm,
  x coord trafo/.code={
    \pgflibraryfpuifactive{
    \pgfmathparse{(#1)/(1000)}
  }{
     \pgfkeys{/pgf/fpu=true}
     \pgfmathparse{(#1)/(1000)}
     \pgfkeys{/pgf/fpu=false}
   }
 },
 xminorgrids=false,
 yminorgrids=false,
 minor x tick num=4,
]


\addplot[purple, very thick, dashdotted, domain=300:3500] 
{1.9819947811980345 - 0.0005507179806514999*x + 
1.3293844929397182e-7*x^2 - 1.116249185353589e-11*x^3 + 
(33330.28424437179*x)/(1. + x)^3};

\addplot[black, very thick, densely dashed, domain=300:3500] 
{1.36274 - 0.000128797*x + 6.76654e-9*x^2};


\addplot[green!50!black, very thick, dashdotted, domain=300:3500] 
{0.9220390545655046 + 0.000036414086746785855*x - 
4.207044744901027e-9*x^2};

\addplot[red, very thick, domain=300:3500] 
{0.9091947226813564 + 5.334769565626307e-6*x + 
1.1069905186422183e-8*x^2};

\addplot[blue, very thick, domain=300:3500] 
{0.7289123820249489 - 0.00004932264815051893*x -
5.045218651446636e-9*x^2};

\end{semilogyaxis}
\end{tikzpicture}
\hskip -0.4cm
\begin{tikzpicture}
\begin{semilogyaxis}[
   log ticks with fixed point,
   ytick = {0.4,  0.6,  0.8,  1.0,  1.2,  1.4},
   title = {$R = 0.7$},
   xlabel={$P_\LT\,\mathrm{[TeV]}$},
   xmin=2, xmax=3800,
   ymin=0.38, ymax=1.6,
  width=8.12cm,	
  height=6.38cm,
  x coord trafo/.code={
    \pgflibraryfpuifactive{
    \pgfmathparse{(#1)/(1000)}
  }{
     \pgfkeys{/pgf/fpu=true}
     \pgfmathparse{(#1)/(1000)}
     \pgfkeys{/pgf/fpu=false}
   }
 },
 xminorgrids=false,
 yminorgrids=false,
 minor x tick num=4,
 yticklabel = \empty,
]


\addplot[purple, very thick, dashdotted, domain=300:3500] 
{1.341813543581268 - 0.0002043408386846479*x + 
5.324672167936075e-8*x^2 - 3.907528703672166e-12*x^3 + 
(11063.057550856285*x)/(1. + x)^3};

\addplot[red, very thick, domain=300:3500] 
{1.0691789710609612 - 0.000043069941635524414*x + 
2.8274624423665577e-8*x^2};


\addplot[green!50!black, very thick, dashdotted, domain=300:3500] 
{0.9452264167279727 + 0.000014804301295256464*x - 
3.6276960262057674e-10*x^2};

\addplot[black, very thick, densely dashed, domain=300:3500] 
{0.94327 - 0.0000321198*x - 
3.16413e-9*x^2};

\addplot[blue, very thick, domain=300:3500] 
{0.8359908544166179 - 0.00006781114223609346*x - 
8.513100432725473e-10*x^2};

\end{semilogyaxis}
\end{tikzpicture}
\else 
\includegraphics[width = 14.3 cm]{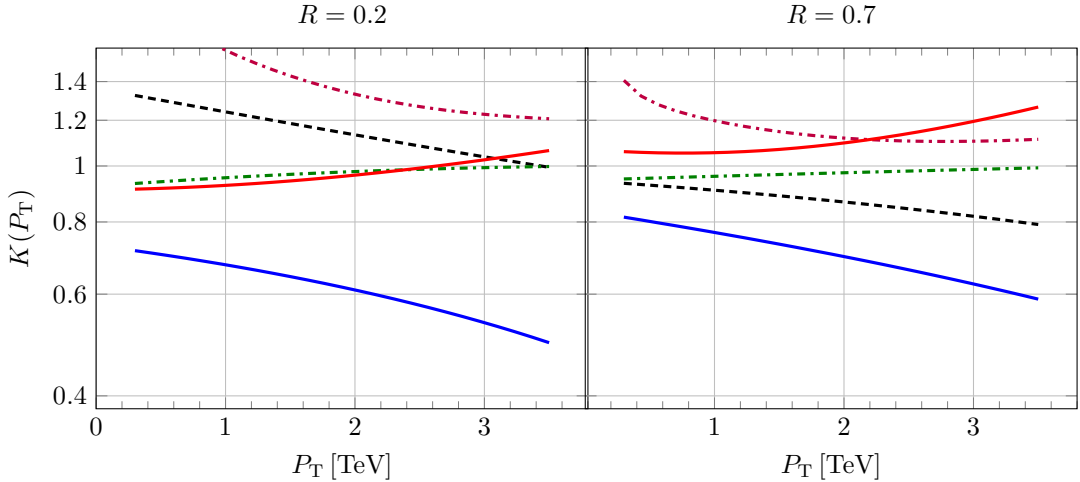}
\fi
\end{center}
\caption{
Ratios $K(P_\LT)$ of the one jet inclusive cross section $d\sigma/dP_\LT$ to the NLO cross section with $R = 0.2$ and $R = 0.7$. The labeling is the same as in Fig.\ \ref{fig:JetKfactorsR04}.
}
\label{fig:JetKfactorsR0207}
\end{figure}

\begin{figure}
\begin{center}
\ifusefigs 
\begin{tikzpicture}
\begin{semilogyaxis}[
   log ticks with fixed point,
   ytick = {0.4, 0.5, 0.6, 0.7, 0.8, 0.9, 1.0, 1.1, 1.2, 1.3, 1.4, 1.5},
   title = {Ratios to NLO jet cross section, $R = 0.4$, $k_\LT$ ordering},
   xlabel={$P_\LT\,\mathrm{[TeV]}$}, ylabel={$K(P_\LT)$},
   xmin=0, xmax=3800,
   ymin=0.38, ymax=1.6,
  legend cell align=left,
  every axis legend/.append style={
  at={(0.02,0.02)},
  anchor= south west,
  },
  width=14cm,	
  height=11cm,
  x coord trafo/.code={
    \pgflibraryfpuifactive{
    \pgfmathparse{(#1)/(1000)}
  }{
     \pgfkeys{/pgf/fpu=true}
     \pgfmathparse{(#1)/(1000)}
     \pgfkeys{/pgf/fpu=false}
   }
 },
 xminorgrids=false,
 yminorgrids=false,
 minor x tick num=4,
]


\addplot[purple, very thick, dashdotted, domain=300:3500] 
{1.551224347974936 - 0.00031962651521130216*x + 
8.046065413106988e-8*x^2 - 6.5782530399931215e-12*x^3 + 
(18344.97800963932*x)/(1. + x)^3};

\addlegendentry{\textsc{FHMRV}(NNLO)}

\addplot[black, very thick, densely dashed, domain=300:3500] 
{1.1031128474479892 - 0.00006435525316735763*x - 
3.562761732521691e-10*x^2};

\addlegendentry{\textsc{Deductor}(LO)}


\addplot[green!50!black, very thick, dashdotted, domain=300:3500] 
{0.943498088497596 + 0.000022899989715684918*x - 
2.202386768376987e-9*x^2};

\addlegendentry{\textsc{FHMRV}(NLO)}

\addplot[red, very thick, domain=300:3500] 
{1.007227856392672 - 0.0001046134549329782*x + 
1.3688291116372765e-8*x^2};

\addlegendentry{\textsc{Deductor}(full)}

\addplot[blue, very thick, domain=300:3500] 
{0.7175367858635915 - 0.00004835421992870293*x - 
4.829786460292932e-9*x^2};

\addlegendentry{\textsc{Deductor}(std.)}

\end{semilogyaxis}
\end{tikzpicture}
\else 
\includegraphics[width = 14 cm]{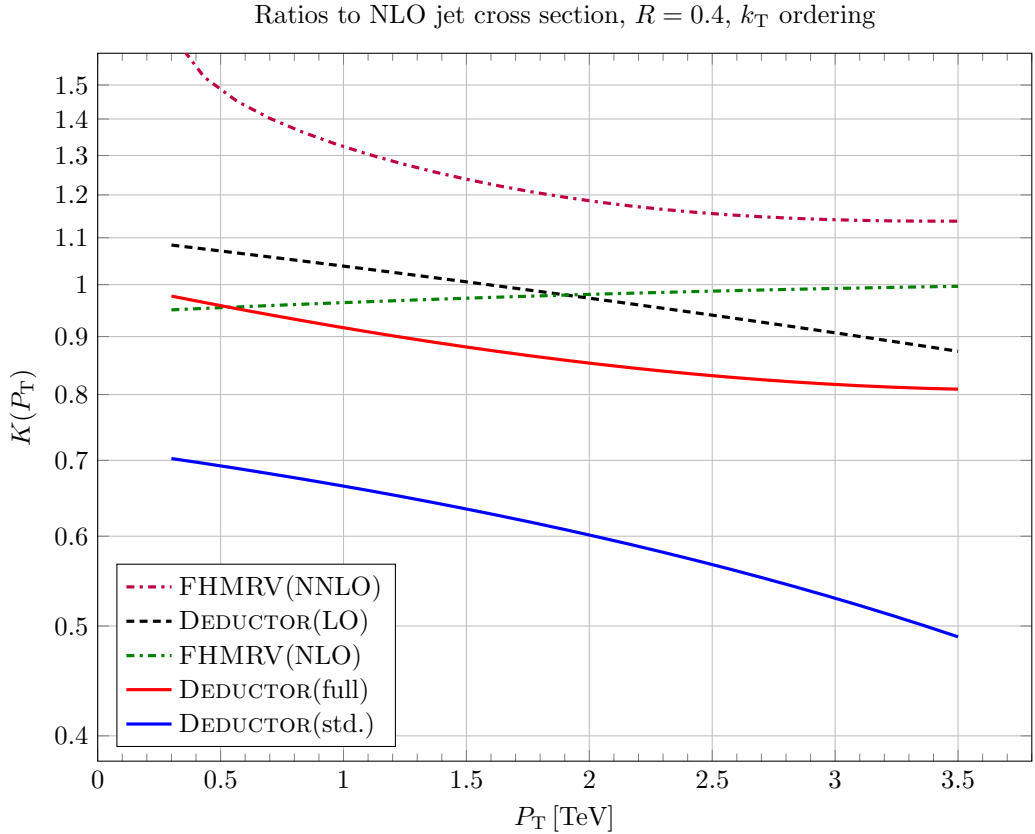}
\fi
\end{center}
\caption{
Ratios $K(P_\LT)$ of the one jet inclusive cross section $d\sigma/dP_\LT$ to the NLO cross section calculated using \textsc{Deductor} in different approximations with $k_\LT$ ordering. The FHMRV curves are the same as in Fig.~\ref{fig:JetKfactorsR04}.
}
\label{fig:JetKfactorsR04kT}
\end{figure}

How do these results depend on the cone size? In Fig.\ \ref{fig:JetKfactorsR0207}, we show the results for $R = 0.2$ on the left and for $R = 0.7$ on the right. We see that for $R = 0.2$ there is a bigger drop going from the LO cross section to the std.\ cross section, while for $R = 0.7$ the drop is smaller. This is the effect that we investigated in Sec.~\ref{sec:jetfinding}. On the other hand, the threshold factors that take us from the std.\ result to the full result are independent of $R$. In the end, the full \textsc{Deductor} cross section is fairly close to the NLO cross section.

How do the results depend on the choice of ordering variable for the shower? In the results presented above, we used $\Lambda$, Eq.~(\ref{eq:Lambdadef}), as the ordering variable. This is the default in \textsc{Deductor}. In this case, the threshold factor is $Z_\La Z_\Lb\, \cU_\cV(\mu_\Lf^2,\mu_\Ls^{2})$, where $\cU_\cV(\mu_\Lf^2,\mu_\Ls^{2})$ is an exponential of a certain integral, Eq.~(\ref{eq:UVexponential00}), and $Z_\La Z_\Lb$ is the factor in Eq.~(\ref{eq:ZaZb}) that results from the fact that the parton distributions in a $\Lambda$ ordered shower are not the $\MSbar$ parton distributions. However, we can choose $k_\LT$, Eq.~(\ref{eq:kTdef}), as the ordering variable. Then we use $\MSbar$ parton distributions in the shower, so $Z_\La Z_\Lb = 1$. In this case, the factor $\cU_\cV(\mu_\Lf^2,\mu_\Ls^{2})$ is modified by changing variables and integration limits as in Sec.~9.4 of Ref.~\cite{DuctThreshold}. With $k_\LT$ ordering, instead of Fig.\ \ref{fig:JetKfactorsR04} we get Fig.\ \ref{fig:JetKfactorsR04kT}. The \textsc{Deductor}(std.) result is modified a little. The main difference is that with $k_\LT$ ordering, the threshold effect is quite a lot smaller at large $P_\LT$ than with $\Lambda$ ordering. The result is that at $P_\LT = 2 \TeV$ the \textsc{Deductor}(full) result is 20\% smaller with $k_\LT$ ordering than with $\Lambda$ ordering. This is a larger effect than we anticipated based on the analytical results in Sec.~9.4 of Ref.~\cite{DuctThreshold}. It indicates to us that in this large $P_\LT$ range we should ascribe a 20\% uncertainty to a leading order hard scattering calculation coupled to a leading order shower without matching to an NLO hard scattering calculation.

We have seen that the choice of shower oriented parton distribution functions affects how the summation of threshold logarithms appears in the cross section: part of the summation of threshold logarithms appears as $\cU_\cV(\mu_\Lf^2,\mu_\Ls^{2})$ and part appears as the redefinition of the parton distributions, $Z_\La Z_\Lb$. If one uses $k_\LT$ ordering instead of $\Lambda$ ordering for the shower, then $Z_\La Z_\Lb = 1$ and all of the leading threshold logarithms appear in $\cU_\cV(\mu_\Lf^2,\mu_\Ls^{2})$. On the other hand, one could try to define a perturbative factorization scheme to make $\cU_\cV(\mu_\Lf^2,\mu_\Ls^{2}) = 1$ and put all of the threshold effect into $Z_\La Z_\Lb$. This idea has been worked out in Ref.~\cite{KrkNLO} for processes in which the hard scattering produces colorless partons. It will be of interest to see how it compares at the level of cross sections to other methods for treating threshold effects for this kind of process.

There is a new calculation of Liu, Moch, and Ringer \cite{Moch1, Moch2} that jointly sums threshold logarithms and logarithms of $1/R$. Although the results presented in that work use a narrower jet rapidity range than is used here, these results are largely consistent with ours. In in the range $1 \TeV < P_\LT < 2 \TeV$, they lie between the $K({\rm full})$ curves with $\Lambda$ ordering in Fig.\ \ref{fig:JetKfactorsR04} and with $k_\LT$ ordering in Fig.\ \ref{fig:JetKfactorsR04kT}.

\begin{figure}
\begin{center}
\ifusefigs 
\begin{tikzpicture}
\begin{semilogyaxis}[
   log ticks with fixed point,
   ytick = {0.4, 0.6, 0.8, 1.0, 1.2, 1.4},
   title = {$\mu_\LR = \mu_\LF = P_\LT^{\textrm{Born}}/(2\sqrt 2)$},
   xlabel={$P_\LT\,\mathrm{[TeV]}$}, ylabel={$K(P_\LT)$},
   xmin=0, xmax=3800,
   ymin=0.38, ymax=1.6,
  legend cell align=left,
  every axis legend/.append style={
  at={(0.02,0.02)},
  anchor= south west,
  },
  width=8.12cm,	
  height=6.38cm,
  x coord trafo/.code={
    \pgflibraryfpuifactive{
    \pgfmathparse{(#1)/(1000)}
  }{
     \pgfkeys{/pgf/fpu=true}
     \pgfmathparse{(#1)/(1000)}
     \pgfkeys{/pgf/fpu=false}
   }
 },
 xminorgrids=false,
 yminorgrids=false,
 minor x tick num=4,
]

\addplot[red, very thick, domain=300:3500] 
{1.3017390014465298 - 0.000015420140515062166*x + 
3.369752987551821e-8*x^2};

\addplot[black, very thick, densely dashed, domain=300:3500] 
{1.4404388069217624 - 0.00005178736312198219*x - 
1.2565542739712624e-9*x^2};

\addplot[blue, very thick, domain=300:3500] 
{1.0171191302055742 - 0.00005405004914257162*x - 
4.928124106652084e-9*x^2};

\end{semilogyaxis}
\end{tikzpicture}
\hskip -0.4cm
\begin{tikzpicture}
\begin{semilogyaxis}[
   log ticks with fixed point,
   ytick = {0.4,  0.6,  0.8,  1.0,  1.2,  1.4},
   title = {$\mu_\LR = \mu_\LF = \sqrt{2}\, P_\LT^{\textrm{Born}}$},
   xlabel={$P_\LT\,\mathrm{[TeV]}$},
   xmin=2, xmax=3800,
   ymin=0.38, ymax=1.6,
  width=8.12cm,	
  height=6.38cm,
  x coord trafo/.code={
    \pgflibraryfpuifactive{
    \pgfmathparse{(#1)/(1000)}
  }{
     \pgfkeys{/pgf/fpu=true}
     \pgfmathparse{(#1)/(1000)}
     \pgfkeys{/pgf/fpu=false}
   }
 },
 xminorgrids=false,
 yminorgrids=false,
 minor x tick num=4,
 yticklabel = \empty,
]

\addplot[red, very thick, domain=300:3500] 
{0.7578429423157969 - 0.000024813723191173675*x + 
1.3650932644630185e-8*x^2};

\addplot[black, very thick, densely dashed, domain=300:3500] 
{0.8623420175645086 - 0.00006962015941702349*x + 
6.874930176387708e-10*x^2};

\addplot[blue, very thick, domain=300:3500] 
{0.6051220924783136 - 0.00005927822100571617*x - 
5.304582384601176e-10*x^2};

\end{semilogyaxis}
\end{tikzpicture}
\else 
\includegraphics[width = 14.3 cm]{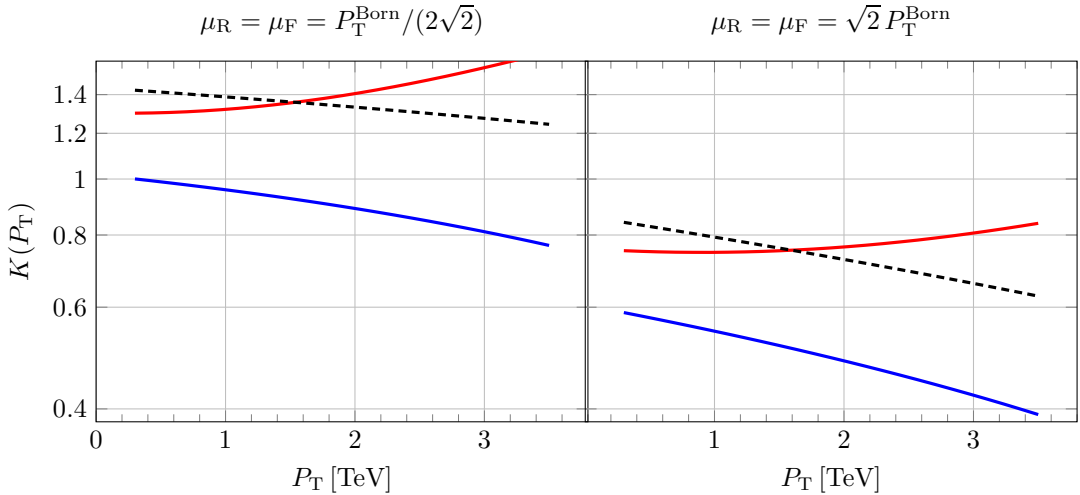}
\fi
\end{center}
\caption{
Ratios $K(P_\LT)$ using \textsc{Deductor} of the one jet inclusive cross section $d\sigma/dP_\LT$ to the NLO cross section with $R = 0.4$ for $\mu_\LR = \mu_\LF = P_\LT^{\textrm{Born}}/(2\sqrt 2)$ and $\mu_\LR = \mu_\LF = \sqrt{2}\, P_\LT^{\textrm{Born}}$. The labeling of the \textsc{Deductor} curves is the same as in Fig.\ \ref{fig:JetKfactorsR04}. Here we define ratios $K(P_\LT)$ in eq.~(\ref{eq:Kdef}) with denominators $d\sigma(\mathrm{NLO})/dP_\LT$ with $\mu_\LF = \mu_\LR = P_\LT^{\textrm{Born}}/\sqrt{2}$.
}
\label{fig:JetKfactorsaltmuRmuF}
\end{figure}

Before concluding this study of the one-jet-inclusive cross section, we pause to ask how the results depend on the choice of scale parameters. In calculating the Born matrix elements, we have set $\mu_\LF = \mu_\LR = P_\LT^{\textrm{Born}}/\sqrt{2}$. With this value, the NLO cross section is quite insensitive to variations in $\mu_\LF$ and $\mu_\LR$ and the Born cross section is fairly close to the NLO cross section. We could, however choose, say, $\mu_\LF = \mu_\LR = P_\LT^{\textrm{Born}}/(2\sqrt{2})$ or $\mu_\LF = \mu_\LR = \sqrt{2}\, P_\LT^{\textrm{Born}}$. The results from these choices are shown in Fig.~\ref{fig:JetKfactorsaltmuRmuF}. In these curves, we have kept unchanged the scales in the reference cross section in the denominators of $K(P_\LT)$, eq.~(\ref{eq:Kdef}). We observe that changing $\mu_\LF$ and $\mu_\LR$ changes the LO cross section, the dashed black curves, quite drastically. Then the \textsc{Deductor}(std.) and \textsc{Deductor}(full) cross sections change by the same factor. Thus our results for $d\sigma(\mathrm{std.})/dP_\LT$ and $d\sigma(\mathrm{full})/dP_\LT$ are strongly dependent on the choice of $\mu_\LF$ and $\mu_\LR$ that we made using knowledge of $d\sigma(\mathrm{NLO})/dP_\LT$.

\begin{figure}
\begin{center}
\ifusefigs 
\begin{tikzpicture}
\begin{semilogyaxis}[
   log ticks with fixed point,
   ytick = {0.4, 0.5, 0.6, 0.7, 0.8, 0.9, 1.0, 1.1, 1.2, 1.3, 1.4, 1.5},
   title = {$\mu_\Ls$ dependence},
   xlabel={$P_\LT\,\mathrm{[TeV]}$}, ylabel={$K(P_\LT)$},
   xmin=0, xmax=3800,
   ymin=0.38, ymax=1.6,
  legend cell align=left,
  every axis legend/.append style={
  at={(0.02,0.02)},
  anchor= south west,
  },
  width=14cm,	
  height=11cm,
  x coord trafo/.code={
    \pgflibraryfpuifactive{
    \pgfmathparse{(#1)/(1000)}
  }{
     \pgfkeys{/pgf/fpu=true}
     \pgfmathparse{(#1)/(1000)}
     \pgfkeys{/pgf/fpu=false}
   }
 },
 xminorgrids=false,
 yminorgrids=false,
 minor x tick num=4,
]

\addplot[black, very thick, densely dashed, domain=300:3500] 
{1.1019455579248745 - 0.0000632015510586682*x - 
5.76378578768999e-10*x^2};

\addlegendentry{\textsc{Deductor}(LO)}

\addplot[red, domain=300:3500] 
{0.9854211660494355 - 0.000018716423257807602*x + 
2.028986810514001e-8*x^2};

\addlegendentry{\textsc{Deductor}(full) $\mu_\Ls = (3/2)\,P_\LT^{\textrm{Born}}$}

\addplot[red, very thick, densely dashed, domain=300:3500] 
{0.9613726295306363 - 0.000014976398210343794*x + 
1.3611273398188686e-8*x^2};

\addlegendentry{\textsc{Deductor}(full) $\mu_\Ls = P_\LT^{\textrm{Born}}$}

\addplot[blue, domain=300:3500] 
{0.7780090775098257 - 0.000056369267929229*x - 3.068166458874673e-9*x^2};

\addlegendentry{\textsc{Deductor}(std.) $\mu_\Ls = (3/2)\,P_\LT^{\textrm{Born}}$}

\addplot[blue, very thick, densely dashed, domain=300:3500] 
{0.7602062106300053 - 0.00004864284006520231*x - 
4.007483823647723e-9*x^2};

\addlegendentry{\textsc{Deductor}(std.) $\mu_\Ls = P_\LT^{\textrm{Born}}$}
\end{semilogyaxis}
\end{tikzpicture}
\else 
\includegraphics[width = 14 cm]{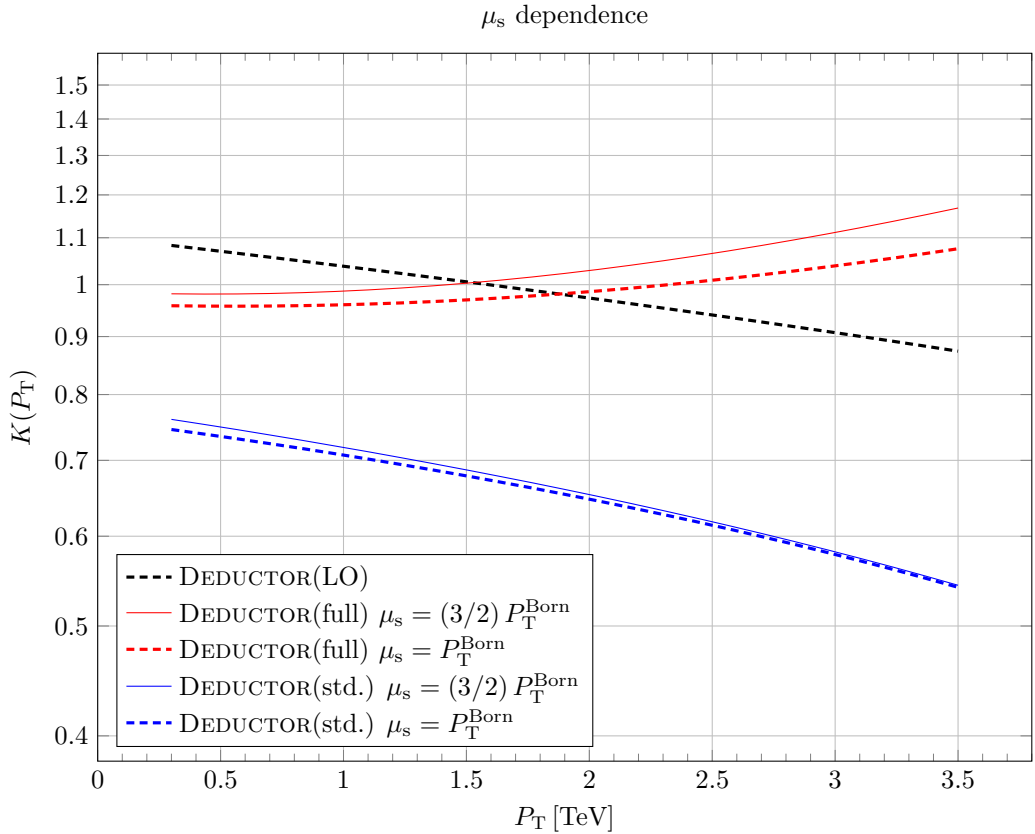}
\fi
\end{center}
\caption{
Ratios $K(P_\LT)$ of the one jet inclusive cross section $d\sigma/dP_\LT$ to the NLO cross section calculated using \textsc{Deductor} with the shower start scale set to $\mu_\Ls = P_\LT^{\textrm{Born}}$. These are compared to the results from Fig.~\ref{fig:JetKfactorsR04} in which we use our standard choice $\mu_\Ls = (3/2)\,P_\LT^{\textrm{Born}}$.
}
\label{fig:JetKfactorsR04altmus}
\end{figure}

We can also change the scale $\mu_\Ls$ at which $\Lambda$ of the shower starts. We argued at eq.~(\ref{eq:musforLambda}) that our choice $\mu_\Ls = (3/2)\,P_\LT^{\textrm{Born}}$ was sensible, but we can change this to $\mu_\Ls = P_\LT^{\textrm{Born}}$. The results from this choice are shown in Fig.~\ref{fig:JetKfactorsR04altmus}. We see that $d\sigma(\mathrm{std.})/dP_\LT$ is quite insensitive to $\mu_\Ls$ but that $d\sigma(\mathrm{full})/dP_\LT$ exhibits some sensitivity to $\mu_\Ls$.

Returning now to Figs.\ \ref{fig:JetKfactorsR04}, \ref{fig:JetKfactorsR0207}, and \ref{fig:JetKfactorsR04kT}, we believe that we can draw two robust conclusions from these results. First, at large $P_\LT$, the effect of applying the jet definition to the partons emerging from the hard scattering is substantial. Second, at large $P_\LT$, the threshold effects are substantial. These substantial effects act in opposite directions, so that they tend to cancel each other out. That is, the perturbative calculation of $d\sigma/d P_\LT$ beyond lowest order involves large effects that tend to cancel. 

These cancellations appear at all orders of perturbation theory.  In particular, they occur within the NLO calculation. Because of these cancellations at large $P_\LT$, we may regard the NLO calculation as more delicate than is at first apparent. 


\section{Gaps between jets}
\label{sec:gaps}

Consider an event with two high-$P_\LT$ jets that are separated by a large difference $\Delta y$ in rapidity. Let $\bar p_\LT$ be the average of the transverse momenta of these two jets. We can define the gap fraction $f(\bar p_\LT, \Delta y)$ to be the ratio of the cross section to produce the two jets while producing no more jets with transverse momenta above some cutoff $p_\LT^{\rm cut}$ in the rapidity interval between the two high-$P_\LT$ jets to the inclusive cross section to produce the two jets. 

It is of some importance to understand the gap fraction $f$ because it is often useful in experimental investigations to impose a requirement that there be some minimum number of high $P_\LT$ jets in an event but no jets beyond this that have  $P_\LT$ greater than some value $p_{\LT}^{\rm cut}$. In addition, the behavior of $f$ as a function of $\bar p_\LT$ and $\Delta y$ is a matter of substantial theoretical interest because it brings together several issues concerning the structure of QCD.

In this section, we address three questions. First, does \textsc{Deductor}, starting with LO matrix elements, come close to data for the gap fraction? Second, does it matter whether one uses a $\Lambda$ or $k_\LT$ ordered shower? Third, what happens if one uses NLO perturbation theory {\em without} a parton shower or any kind of summation of large logarithms? 

\subsection{The gap fraction and summation of large logarithms}
\label{sec:gaplogs}

The theory of the gap fraction $f$ involves two sorts of large logarithms. First, the logarithm $\log(\bar p_\LT/p_{\LT}^{\rm cut})$ can be large. Second, the rapidity separation $\Delta y$ can be large.\footnote{We can consider $\Delta y$ to be a logarithm. In fact, rapidities are logarithms of ratios of components of momenta.} At order $\as^N$, a perturbative calculation can give us a factor of $[\Delta y \times \log(\bar p_\LT/p_{\LT}^{\rm cut})]^N$, so a summation of large logarithms is called for. Many of the theoretical issues associated with are reviewed in Ref.~\cite{Manchester2009}. At a first level in an analytic summation of leading logarithms \cite{EarlyGap1, EarlyGap2}, one uses the exponential of a Sudakov exponent constructed from one loop graphs for the virtual exchange of a low transverse momentum gluon. There are further subtleties in the analytic treatment. The factors of $\Delta y$ in the exponent are especially important \cite{Manchester2005}, but $\Delta y$ is kinematically quite different from the factors $\log(P_\textrm{hard}/P_\textrm{soft})$ that are typically summed using the renormalization group.
Additionally, there can also be  ``non-global'' logarithms with a different structure than seen in the simplest analytic treatment \cite{NonGlobal1, NonGlobal2, NonGlobal3, NonGlobal4}. Furthermore, some of these logarithms are ``super-leading'' in the sense of having more powers of logarithms per power of $\alpha_\Ls$ than one gets in the simple analysis \cite{Manchester2009, SuperLeading1, SuperLeading2, Manchester2011}.

For the purpose of capturing the effects seen in analytical summations of logarithms, there may be an advantage to using a parton shower approach in place of analytical summations since, in principle, a parton shower can trace complex patterns involving the emission of real gluons and the subsequent virtual exchanges between these gluons. However, a better treatment of color within the shower is needed to realize this potential advantage. Furthermore, the analytic treatments suggest that exponentiated factors of $\Delta y$ are crucial. This suggests that a rapidity-ordered treatment might be desirable, as in Ref.~\cite{HEJgap}. Such a treatment could cover the case in which several gluons are emitted with very different rapidities $y_i$ but with similar transverse momenta $k_{\LT,i}$. We do not have a rapidity ordered shower, but we can choose $\Lambda$ ordering or $k_\LT$ ordering. 

The effects of shower ordering are not completely intuitive. For instance, suppose that the Born hard scattering produces partons 1 and 2 with $0 = \bm p_{\LT,1}^{\rm Born} + \bm p_{\LT, 2}^{\rm Born}$. Then an initial state emission can produce parton 3. But partons 1 and 2 must then recoil against parton 3 so that in the new state $0 = \bm p_{\LT,1} + \bm p_{\LT, 2} + \bm p_{\LT, 3}$. Then it is possible that partons 1 and 3 are the highest $P_\LT$ jets in the final state and a somewhat softer parton 2 lies in the gap between partons 1 and 3. It can also happen that parton 3 is created in a soft, wide angle emission and is the smallest $P_\LT$ jet and, furthermore, lies in the gap between partons 1 and 2. It is possible that $\Lambda$ ordering can give one shower history, say $(1,2) \to (1,2,3)$ while $k_\LT$ ordering can give a different shower history for the same parton momenta, say $(1,3) \to (1,2,3)$. Then the same state $(1,2,3)$ can be reached with different probabilities. We discuss some of the issues in Ref.~\cite{ShowerTime}. 

\subsection{Atlas data}
\label{sec:AtlasData}

We will compare to results from Atlas \cite{AtlasGap}. In the Atlas results that we use, $p_{\LT}^{\rm cut}$ is fixed and the gap fraction $f$ is plotted as a function of the transverse momentum of the hard jets. Specifically, Atlas uses a data sample at $\sqrt s = 7 \TeV$. Jets are defined using the anti-$k_\LT$ algorithm \cite{antikt} with $R = 0.6$. All jets in the rapidity window $-4.4 < y < 4.4$ are considered if they have $P_\LT > p_{\LT}^{\rm cut} = 20 \GeV$. Of these jets, the two jets with the highest $P_\LT$ are selected. Of the two leading jets, let jet 1 have the highest rapidity and let jet 2 have the lowest rapidity. The event is characterized by the rapidity difference $\Delta y = y_1 - y_2$ and the average transverse momentum $\bar p_{\LT} = (P_{\LT,1} + P_{\LT,2})/2$. The event has a gap if there is no jet (with $P_\LT > p_{\LT}^{\rm cut}$) in the rapidity range $y_1 < y < y_2$. Only a fraction $f$ of events with a given $\Delta y$ and $\bar p_\LT$ has a gap.

\subsection{Parton shower analyses}
\label{sec:ShowerAnalyses}

In the Atlas paper \cite{AtlasGap}, there are comparisons to predictions from \textsc{Pythia} and \textsc{Herwig}, which match the experimental results reasonably well. There are a number of other comparisons of the Atlas data to theory. Of these, we mention the approach in Ref.~\cite{HEJgap}, which looks for places in which the summation of $\Delta y$ factors are important. The paper \cite{HocheSchonherr} uses $\as^2 + \as^3$ perturbation theory matched to the \textsc{Sherpa} parton shower, obtaining a good match to the data.

We do not have code to match the \textsc{Deductor} parton shower to NLO perturbative matrix elements. For this reason, we cannot equal the accuracy of the Sherpa results \cite{HocheSchonherr}. However, our interest here is in the performance of the \textsc{Deductor} parton shower. If we were to use a matched calculation, we could not separate the contributions from the first shower emission and the NLO correction to the LO matrix element. With just the shower times a LO matrix element, we see clearly what the parton shower does. We do know, for instance, that the \textsc{Pythia} parton shower with a LO matrix element matches the data reasonably well \cite{AtlasGap}. However, as we have seen in Sec.~\ref{sec:onejetphysics}, \textsc{Pythia} and \textsc{Deductor} do not necessarily give the same results. Furthermore, with \textsc{Deductor}, we can use both $\Lambda$ ordering and $k_\LT$ ordering and see if either is much better than the other in capturing the large logarithmic factors of $\Delta y$.

\subsection{Perturbative analyses}
\label{sec:PerturbativeGap}

We also undertake a purely perturbative calculation. We write the gap fraction in the form
\begin{equation}
\label{eq:gapNLO}
f(\bar p_\LT,\Delta y) = 1
- \frac{d\sigma_3/[d\bar p_\LT\,d\Delta y]}{d\sigma_2/[d\bar p_\LT\,d\Delta y]}
\;.
\end{equation}
Here $d\sigma_2/[d\bar p_\LT\,d\Delta y]$ is the cross section to produce at least two jets in the rapidity window $-4.4 < y < 4.4$ such that the two jets in the rapidity window with the largest $P_\LT$ satisfy $\bar p_{\LT} = (P_{\LT,1} + P_{\LT,2})/2$ and $\Delta y = |y_1 - y_2|$. This is an infrared safe jet cross section for which the lowest order contribution has two partons in the final state. We calculate this cross section at NLO using \textsc{NLOJet++} \cite{NLOJet++}. In the numerator, $d\sigma_3/[d\bar p_\LT\,d\Delta y]$ is the cross section to produce at least three jets in the rapidity window $-4.4 < y < 4.4$ such that the two jets in the rapidity window with the largest $P_\LT$ satisfy $\bar p_{\LT} = (P_{\LT,1} + P_{\LT,2})/2$ and $\Delta y = |y_1 - y_2|$ and such that there is a third jet with 
$\min(y_1,y_2) < y_3 < \max(y_1,y_2)$ and $P_{\LT,3} > p_{\LT}^{\rm cut}$. This is an infrared safe jet cross section for which the lowest order contribution has three partons in the final state. Again, we calculate this cross section at NLO using \textsc{NLOJet++}. For these perturbative calculations, our primary choice for the factorization and renormalization scales is $\mu_\LF = \mu_\LR = 2 \bar p_\LT$, although we also investigate other choices.

A similar calculation was carried out in Ref.~\cite{Hatta:2013qj}. In this calculation, which used $\mu_\LR = \mu_\LF = \bar p_\LT$, $1 - f(\bar p_\LT,\Delta y)$ was expanded in powers of $\as$ and only the $\as^1$ and $\as^2$ terms were retained. The outcome was that for $\Delta y > 3$ the perturbative result became unstable and could not come close to the experimental result. Our approach is to keep the numerator and the denominator in Eq.~(\ref{eq:gapNLO}) as units that each represent sensible physical cross sections and should not be disassembled. Nevertheless, given that there are large logarithms in this problem, one may expect that the perturbative formula in Eq.~(\ref{eq:gapNLO}) will fail to match experiment.

\begin{figure}
\begin{center}
\ifusefigs 
\begin{tikzpicture}
\begin{axis}[
   title = {Dependence of gap fraction on non-perturbative input},
   xlabel={$\bar p_\LT\,\mathrm{[GeV]}$}, ylabel={$\Delta f(\bar p_\LT)$},
   xmin=50, xmax=500,
   ymin=-0.5, ymax=0.5,
  legend cell align=left,
  every axis legend/.append style={
  at={(0.02,0.02)},
  anchor= south west,
  },
  width=14cm,	
  height=11cm,
 xminorgrids=false,
 yminorgrids=false,
 minor x tick num=4,
]

\pgfplotstableread{
50.    -0.0157538
60.    -0.0430747
70.    -0.0323136
80.    -0.0533299
90.    -0.0317498
105.   -0.0312875
120.   -0.000725204
135.   -0.0317036
150.   -0.0288924
180.   -0.0156482
210.   -0.0221626
240.   -0.0151281
270.   -0.0107756
300.   -0.0125226
340.   -0.0098955
380.   -0.00936917
420.   -0.00845435
460.   -0.0104097
500.   -0.0104097
}\Deltaf

\addplot [const plot mark left, black, thick] table [x={0},y={1}]{\Deltaf};
\addlegendentry{(\textsc{Deductor} \& \textsc{Pythia}) $-$ \textsc{Deductor}}

\end{axis}
\end{tikzpicture}
\else 
\includegraphics[width = 14 cm]{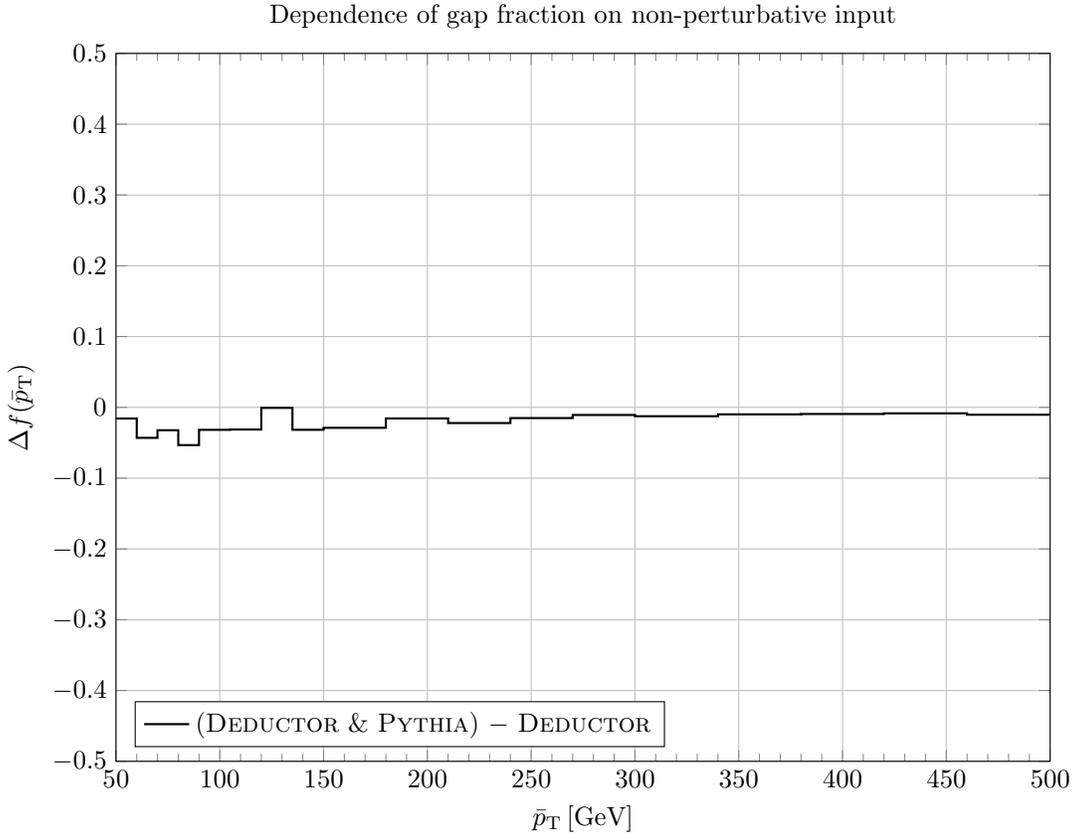}
\fi
\end{center}
\caption{
Change $\Delta f$ in the gap fraction for $2 < \Delta y < 3$ when nonperturbative effects are added.
}
\label{fig:gapfractionNonpert}
\end{figure}

\subsection{Non-perturbative effects}
\label{eq:GapNonPert}

Before proceeding with numerical results, we ask whether, for a measurement controlled by a parameter $p_{\LT}^{\rm cut} = 20 \GeV$, non-perturbative effects at a scale of 1 GeV might be important. We generate \textsc{Deductor} events with $\Lambda$ ordering using the default parameters of Eqs.~(\ref{eq:muFduct}) and (\ref{eq:musforLambda}). To these perturbative \textsc{Deductor} events we add an underlying event and then send the event to \textsc{Pythia} as described in Sec.~\ref{sec:nonperturbative}. We calculate the difference
\begin{equation}
\Delta f = f({\textsc{Deductor}\ \&\ \textsc{Pythia}})
- f(\textsc{Deductor})
\;.
\end{equation}
We display the results as a function of $\bar p_\LT$ in Fig.\ \ref{fig:gapfractionNonpert} for the case $2 < \Delta y < 3$. There are some evident statistical fluctuations but the result is clear: there is an effect, but it is no larger than 5\%. Since the non-perturbative effects make so little difference, in subsequent plots we work at just the partonic level.

\begin{figure}
\begin{center}
\ifusefigs 
\begin{tikzpicture}
 \begin{groupplot}[
      group style={
          group size=1 by 5,
          vertical sep=0pt,
          x descriptions at=edge bottom},
          xlabel={$\bar p_\LT\,\mathrm{[GeV]}$},
          width=14cm,
    ]

\nextgroupplot[
ylabel={$f(\bar p_\LT)$},
height=5cm,
xmin=50, xmax=500,
ymin=0.01, ymax=1.0,]

\pgfplotstableread{
50.    0.869041
60.    0.829231
70.    0.785057
80.    0.77467
90.    0.74635
105.   0.710944
120.   0.687418
135.   0.672407
150.   0.643561
180.   0.626222
210.   0.601456
240.   0.593203
270.   0.574926
300.   0.55345
340.   0.535905
380.   0.537309
420.   0.527018
460.   0.515959
500    0.515959
}\DuctLambdaA

\pgfplotstableread{
50.    0.835029
60.    0.783931
70.    0.76198
80.    0.739687
90.    0.716075
105.   0.649501
120.   0.64965
135.   0.640529
150.   0.606461
180.   0.582615
210.   0.555862
240.   0.53945
270.   0.527273
300.   0.521707
340.   0.493485
380.   0.497248
420.   0.493488
460.   0.481859
500    0.481859
}\DuctkTA

\pgfplotstableread{
x       y      ex-   ex+  ey-        ey+
55.  0.8488   5.  5.  0.0149933  0.0144838
65.  0.8154   5.  5.  0.0151119  0.0145066
75.  0.7874   5.  5.  0.0148933  0.0142864
85.  0.7638   5.  5.  0.0148  0.0144222
97.5  0.7457  7.5  7.5  0.0147221  0.0140431
112.5  0.7191 7.5  7.5  0.0151921  0.0148762
127.5  0.688  7.5  7.5  0.0146741  0.0139979
142.5  0.6713 7.5  7.5  0.014994  0.0144312
165.  0.658  15.  15.  0.014258  0.0137674
195.  0.6427 15.  15.  0.0148314  0.0141676
225.  0.627  15.  15.  0.0149452  0.014368
255.  0.6062 15.  15.  0.0153235  0.0148647
285.  0.582  15.  15.  0.0164125  0.0159
320.  0.563  20.  20.  0.0164012  0.0164012
360.  0.553  20.  20.  0.0205183  0.0198494
400.  0.56   20.  20.  0.0252389  0.0246982
440.  0.551  20.  20.  0.0317805  0.0317805
480.  0.487  20.  20.  0.0417612  0.04272
}\AtlasA

\pgfplotstableread{
 5.000000e+01     1.276384e+01   6.002985e-02     1.017511e+02   2.015555e-01
 6.000000e+01     6.809880e+00   2.964413e-02     4.372649e+01   9.723459e-02
 7.000000e+01     3.882134e+00   1.543605e-02     2.115231e+01   4.110488e-02
 8.000000e+01     2.272865e+00   9.258301e-03     1.096614e+01   2.140341e-02
 9.000000e+01     1.253814e+00   4.041815e-03     5.461076e+00   7.822730e-03
 1.050000e+02     6.453268e-01   2.121704e-03     2.510350e+00   3.681977e-03
 1.200000e+02     3.494667e-01   1.186815e-03     1.270695e+00   1.704026e-03
 1.350000e+02     2.043316e-01   7.014155e-04     6.863766e-01   1.104681e-03
 1.500000e+02     9.849068e-02   2.255296e-04     3.104434e-01   3.406551e-04
 1.800000e+02     4.037962e-02   9.866825e-05     1.171750e-01   1.156924e-04
 2.100000e+02     1.817985e-02   4.806298e-05     4.988608e-02   4.512034e-05
 2.400000e+02     8.926720e-03   2.119704e-05     2.333811e-02   1.916456e-05
 2.700000e+02     4.649947e-03   8.543375e-06     1.167630e-02   7.896564e-06
 3.000000e+02     2.328220e-03   2.586204e-06     5.673435e-03   1.941374e-06
 3.400000e+02     1.116276e-03   1.235061e-06     2.633317e-03   6.787543e-07
 3.800000e+02     5.656302e-04   6.410358e-07     1.302386e-03   3.377381e-07
 4.200000e+02     3.007991e-04   3.782459e-07     6.784407e-04   1.938798e-07
 4.600000e+02     1.658354e-04   2.230096e-07     3.687359e-04   1.061786e-07
 5.000000e+02     1.658354e-04   2.230096e-07     3.687359e-04   1.061786e-07
}\nloA

\addplot [const plot mark left, darkgreen, very thick, dash dot] table [x={0},y expr =1-\thisrow{1}/\thisrow{3}]{\nloA};

\addplot [const plot mark left, red, thick] table [x={0},y={1}]{\DuctLambdaA};

\addplot [const plot mark left, blue, thick] table [x={0},y={1}]{\DuctkTA};

\addplot [only marks, black,
        error bars/.cd,
            x dir=both, x explicit,
            y dir=both, y explicit,
    ] table [
        x error plus=ex+,
        x error minus=ex-,
        y error minus=ey-,
        y error plus=ey+,
]{\AtlasA};

\node at (420.0,0.8) {\fcolorbox{black}{white}{$1 < \Delta y < 2$}};


\nextgroupplot[
ylabel = $f(\bar p_\LT)$,
height = 5 cm,
xmin=50, xmax=500,
ymin=0.01, ymax=1.0,
]

\pgfplotstableread{
50.    0.794329
60.    0.750284
70.    0.698127
80.    0.661743
90.    0.618695
105.   0.57705
120.   0.559996
135.   0.515071
150.   0.504368
180.   0.461917
210.   0.433062
240.   0.420998
270.   0.405112
300.   0.400887
340.   0.376058
380.   0.383958
420.   0.371254
460.   0.367635
500    0.367635
}\DuctLambdaB

\pgfplotstableread{
50.    0.714705
60.    0.656902
70.    0.599321
80.    0.565371
90.    0.526022
105.   0.488308
120.   0.46278
135.   0.470926
150.   0.42264
180.   0.393418
210.   0.376604
240.   0.367215
270.   0.349589
300.   0.343097
340.   0.345882
380.   0.318498
420.   0.324449
460.   0.324163
500    0.324163
}\DuctkTB

\pgfplotstableread{
x       y      ex-   ex+  ey-        ey+
55.  0.7462   5.  5.  0.0193197  0.0186317
65.  0.7032   5.  5.  0.02022  0.0195494
75.  0.6624   5.  5.  0.0201792  0.0192094
85.  0.6315   5.  5.  0.0204362  0.0198736
97.5  0.6009   7.5  7.5  0.0200592  0.0194795
112.5  0.5913   7.5  7.5  0.0207793  0.0199462
127.5  0.5488   7.5  7.5  0.0198487  0.0187926
142.5  0.5302   7.5  7.5  0.0197396  0.0192769
165.  0.4921   15.  15.  0.0182439  0.0173669
195.  0.4838   15.  15.  0.0183469  0.0177924
225.  0.464   15.  15.  0.0173104  0.0167502
255.  0.4237   15.  15.  0.0185809  0.0179739
285.  0.401   15.  15.  0.0198494  0.0198494
320.  0.42   20.  20.  0.0219317  0.0212603
360.  0.411   20.  20.  0.0291548  0.0291548
400.  0.421   20.  20.  0.039  0.0395601
440.  0.327   20.  20.  0.0485077  0.0511957
480.  0.381   20.  20.  0.0696348  0.0733485
}\AtlasB

\pgfplotstableread{
 5.000000e+01     1.447666e+01   8.371886e-02     6.743162e+01   1.683374e-01
 6.000000e+01     7.529000e+00   4.041292e-02     2.802081e+01   7.654068e-02
 7.000000e+01     4.073357e+00   2.087238e-02     1.315372e+01   3.239661e-02
 8.000000e+01     2.283849e+00   1.142459e-02     6.711047e+00   1.578647e-02
 9.000000e+01     1.216392e+00   4.979215e-03     3.202420e+00   5.962303e-03
 1.050000e+02     5.999243e-01   2.419776e-03     1.426414e+00   2.717015e-03
 1.200000e+02     3.147199e-01   1.310285e-03     7.036697e-01   1.478661e-03
 1.350000e+02     1.727740e-01   7.290714e-04     3.629301e-01   7.588791e-04
 1.500000e+02     8.067320e-02   2.256136e-04     1.586991e-01   1.867345e-04
 1.800000e+02     3.086333e-02   7.254646e-05     5.679017e-02   5.538841e-05
 2.100000e+02     1.300961e-02   2.329581e-05     2.292092e-02   1.530285e-05
 2.400000e+02     6.035573e-03   8.748945e-06     1.015028e-02   4.075782e-06
 2.700000e+02     2.956964e-03   4.448595e-06     4.829843e-03   1.945175e-06
 3.000000e+02     1.397154e-03   1.961777e-06     2.213810e-03   8.076622e-07
 3.400000e+02     6.182698e-04   8.863389e-07     9.580256e-04   3.611616e-07
 3.800000e+02     2.903117e-04   4.431184e-07     4.413509e-04   1.720547e-07
 4.200000e+02     1.433706e-04   2.389864e-07     2.135586e-04   9.327106e-08
 4.600000e+02     7.266858e-05   1.341237e-07     1.071284e-04   5.314288e-08
 5.000000e+02     7.266858e-05   1.341237e-07     1.071284e-04   5.314288e-08
}\nloB

\addplot [const plot mark left, darkgreen, very thick, dash dot] table [x={0},y expr =1-\thisrow{1}/\thisrow{3}]{\nloB};

\addplot [const plot mark left, red, thick] table [x={0},y={1}]{\DuctLambdaB};

\addplot [const plot mark left, blue, thick] table [x={0},y={1}]{\DuctkTB};

\addplot [only marks, black,
        error bars/.cd,
            x dir=both, x explicit,
            y dir=both, y explicit,
    ] table [
        x error plus=ex+,
        x error minus=ex-,
        y error minus=ey-,
        y error plus=ey+,
]{\AtlasB};

\node at (420.0,0.8) {\fcolorbox{black}{white}{$2 < \Delta y < 3$}};


\nextgroupplot[
ylabel = $f(\bar p_\LT)$,
height = 5 cm,
xmin=50, xmax=500,
ymin=0.01, ymax=1.0,]

\pgfplotstableread{
50.    0.718585
60.    0.668641
70.    0.59565
80.    0.566885
90.    0.533686
105.   0.491389
120.   0.457264
135.   0.446529
150.   0.407818
180.   0.361995
210.   0.345479
240.   0.361706
270.   0.325814
300.   0.313919
340.   0.304646
380.   0.302332
420.   0.304916
460.   0.308353
500    0.308353
}\DuctLambdaC

\pgfplotstableread{
50.    0.593971
60.    0.494689
70.    0.478719
80.    0.439207
90.    0.373089
105.   0.343449
120.   0.320408
135.   0.311851
150.   0.303591
180.   0.260593
210.   0.260701
240.   0.250264
270.   0.245898
300.   0.244267
340.   0.247077
380.   0.233787
420.   0.253659
460.   0.265983
500    0.265983
}\DuctkTC

\pgfplotstableread{
x       y      ex-   ex+  ey-        ey+
55.     0.6499   5.    5.    0.0268896   0.0257932
65.     0.5818   5.    5.    0.0256673   0.0247348
75.     0.5529   5.    5.    0.0240027   0.0231534
85.     0.488    5.    5.    0.0237065   0.0228254
97.5    0.4794   7.5   7.5   0.0220239   0.021285
112.5   0.448    7.5   7.5   0.0233238   0.0224722
127.5   0.4226   7.5   7.5   0.0216585   0.0208432
142.5   0.405    7.5   7.5   0.0230217   0.0230217
165.    0.3775   15.   15.   0.0191269   0.0182923
195.    0.356    15.   15.   0.0216333   0.0208087
225.    0.331    15.   15.   0.0202485   0.0194165
255.    0.326    15.   15.   0.0247588   0.0254558
285.    0.248    15.   15.   0.0291548   0.0300167
320.    0.335    20.   20.   0.0398121   0.0412311
360.    0.287    20.   20.   0.0577235   0.0623699
400.    0.126    20.   20.   0.0563205   0.0762365
440.    0.23     20.   20.   0.110454    0.150333
480.    0.5      20.   20.   0.202237    0.202237
}\AtlasC

\pgfplotstableread{
 5.000000e+01     1.004637e+01   8.473179e-02     3.110063e+01   1.107757e-01
 6.000000e+01     4.885181e+00   3.407028e-02     1.235259e+01   4.651959e-02
 7.000000e+01     2.489220e+00   1.766415e-02     5.610313e+00   2.002472e-02
 8.000000e+01     1.341091e+00   9.238318e-03     2.738785e+00   1.040530e-02
 9.000000e+01     6.636424e-01   3.848243e-03     1.256141e+00   3.386371e-03
 1.050000e+02     3.049554e-01   1.488532e-03     5.258552e-01   1.370813e-03
 1.200000e+02     1.481090e-01   6.490609e-04     2.446479e-01   5.232353e-04
 1.350000e+02     7.828000e-02   2.804653e-04     1.220466e-01   2.048532e-04
 1.500000e+02     3.347284e-02   8.134888e-05     4.959597e-02   2.869600e-05
 1.800000e+02     1.142811e-02   2.431424e-05     1.601112e-02   1.051561e-05
 2.100000e+02     4.351312e-03   9.304968e-06     5.897164e-03   3.915825e-06
 2.400000e+02     1.800124e-03   4.008834e-06     2.359470e-03   1.817271e-06
 2.700000e+02     7.899684e-04   1.926692e-06     1.013409e-03   7.977835e-07
 3.000000e+02     3.260293e-04   7.625262e-07     4.111762e-04   2.723033e-07
 3.400000e+02     1.229109e-04   3.189302e-07     1.531783e-04   1.113403e-07
 3.800000e+02     4.884649e-05   1.381787e-07     5.968717e-05   5.096407e-08
 4.200000e+02     1.985104e-05   5.797023e-08     2.425604e-05   2.273134e-08
 4.600000e+02     8.398946e-06   3.033790e-08     1.010940e-05   1.072677e-08
 5.000000e+02     8.398946e-06   3.033790e-08     1.010940e-05   1.072677e-08
}\nloC

\addplot [const plot mark left, darkgreen, very thick, dash dot] table [x={0},y expr =1-\thisrow{1}/\thisrow{3}]{\nloC};

\addplot [const plot mark left, red, thick] table [x={0},y={1}]{\DuctLambdaC};

\addplot [const plot mark left, blue, thick] table [x={0},y={1}]{\DuctkTC};

\addplot [only marks, black,
        error bars/.cd,
            x dir=both, x explicit,
            y dir=both, y explicit,
    ] table [
        x error plus=ex+,
        x error minus=ex-,
        y error minus=ey-,
        y error plus=ey+,
]{\AtlasC};

\node at (420.0,0.8) {\fcolorbox{black}{white}{$3 < \Delta y < 4$}};


\nextgroupplot[
ylabel = $f(\bar p_\LT)$,
height = 5 cm,
xmin=50, xmax=500,
ymin=0.01, ymax=1.0,]

\pgfplotstableread{
50.    0.644343
60.    0.594773
70.    0.547274
80.    0.5242
90.    0.486543
105.   0.425727
120.   0.38814
135.   0.385221
150.   0.367883
180.   0.329235
210.   0.314834
240.   0.303757
270.   0.302437
300.   0.304755
340.   0.315029
380.   0.323796
420.   0.331982
460.   0.346462
500    0.346462
}\DuctLambdaD

\pgfplotstableread{
50.    0.483101
60.    0.390546
70.    0.348174
80.    0.305811
90.    0.287117
105.   0.252257
120.   0.202702
135.   0.230851
150.   0.22176
180.   0.19828
210.   0.214421
240.   0.215954
270.   0.214522
300.   0.225314
340.   0.249462
380.   0.271491
420.   0.283622
460.   0.271373
500    0.271373
}\DuctkTD

\pgfplotstableread{
x       y      ex-   ex+  ey-        ey+
55.     0.5515   5.    5.    0.0290998   0.0276921
65.     0.493    5.    5.    0.0313847   0.0305287
75.     0.467    5.    5.    0.0295296   0.0277849
85.     0.393    5.    5.    0.031241    0.0297321
97.5    0.389    7.5   7.5   0.0266271   0.0264008
112.5   0.344    7.5   7.5   0.031241    0.0320156
127.5   0.298    7.5   7.5   0.0269072   0.0261725
142.5   0.253    7.5   7.5   0.0286007   0.0294109
165.    0.296    15.   15.   0.0228035   0.0226716
195.    0.244    15.   15.   0.0308869   0.0322025
225.    0.259    15.   15.   0.0344819   0.0353553
255.    0.323    15.   15.   0.0550091   0.0576975
285.    0.203    15.   15.   0.065521    0.0781025
320.    0.3      20.   20.   0.111803    0.131529
360.    0.52     20.   20.   0.251794    0.251794
}\AtlasD

\pgfplotstableread{
 5.000000e+01     4.613036e+00   6.500245e-02     1.119498e+01   5.486533e-02
 6.000000e+01     2.163978e+00   2.474487e-02     4.199723e+00   2.110719e-02
 7.000000e+01     9.831736e-01   9.271330e-03     1.746631e+00   1.029496e-02
 8.000000e+01     5.104493e-01   3.978250e-03     8.130937e-01   3.034500e-03
 9.000000e+01     2.288680e-01   1.348444e-03     3.467440e-01   6.711638e-04
 1.050000e+02     9.385146e-02   5.062052e-04     1.346926e-01   2.047948e-04
 1.200000e+02     4.206217e-02   2.169628e-04     5.744764e-02   8.713355e-05
 1.350000e+02     1.991671e-02   1.034727e-04     2.628776e-02   4.512828e-05
 1.500000e+02     7.446864e-03   2.848235e-05     9.551073e-03   1.005218e-05
 1.800000e+02     2.124266e-03   8.736137e-06     2.574227e-03   3.200338e-06
 2.100000e+02     6.404649e-04   2.929481e-06     7.784499e-04   1.064908e-06
 2.400000e+02     2.120570e-04   1.074807e-06     2.523893e-04   3.709426e-07
 2.700000e+02     7.456880e-05   4.008784e-07     8.596386e-05   1.437068e-07
 3.000000e+02     2.300510e-05   1.240159e-07     2.648950e-05   4.634931e-08
 3.400000e+02     5.978590e-06   3.922455e-08     6.828515e-06   1.335500e-08
 3.800000e+02     1.561299e-06   1.270531e-08     1.781992e-06   4.605254e-09
 4.200000e+02     3.974319e-07   3.763422e-09     4.619386e-07   1.267787e-09
 4.600000e+02     1.031211e-07   1.100672e-09     1.164962e-07   4.415532e-10
 5.000000e+02     1.031211e-07   1.100672e-09     1.164962e-07   4.415532e-10
}\nloD

\addplot [const plot mark left, darkgreen, very thick, dash dot] table [x={0},y expr =1-\thisrow{1}/\thisrow{3}]{\nloD};

\addplot [const plot mark left, red, thick] table [x={0},y={1}]{\DuctLambdaD};

\addplot [const plot mark left, blue, thick] table [x={0},y={1}]{\DuctkTD};

\addplot [only marks, black,
        error bars/.cd,
            x dir=both, x explicit,
            y dir=both, y explicit,
    ] table [
        x error plus=ex+,
        x error minus=ex-,
        y error minus=ey-,
        y error plus=ey+,
]{\AtlasD};

\node at (420.0,0.8) {\fcolorbox{black}{white}{$4 < \Delta y < 5$}};

\nextgroupplot[
ylabel={$f(\bar p_\LT)$},
height=5cm,
xmin=50, xmax=500,
ymin=0.0, ymax=1.0,]

\pgfplotstableread{
50.    0.606419
60.    0.549602
70.    0.476753
80.    0.485873
90.    0.445351
105.   0.447636
120.   0.390705
135.   0.308101
150.   0.355184
180.   0.322212
210.   0.359146
240.   0.322348
270.   0.373466
300.   0.430126
340.   0.440527
380.   0.591407
420.   0.529998
460.   0.708377
500    0.708377
}\DuctLambdaE

\pgfplotstableread{
50.    0.438019
60.    0.369446
70.    0.331475
80.    0.241729
90.    0.253638
105.   0.189716
120.   0.229804
135.   0.158685
150.   0.207509
180.   0.234891
210.   0.202656
240.   0.303402
270.   0.31337
300.   0.444551
340.   0.395646
380.   0.442189
420.   0.658099
460.   0.748503
500    0.748503
}\DuctkTE

\pgfplotstableread{
x       y      ex-   ex+  ey-        ey+
55.     0.483   5.    5.    0.0358469   0.0341321
65.     0.4     5.    5.    0.0438634   0.0431393
75.     0.351   5.    5.    0.04245     0.0431393
85.     0.309   5.    5.    0.0463249   0.0470106
97.5    0.344   7.5   7.5   0.0420119   0.0430116
112.5   0.319   7.5   7.5   0.061327    0.064622
127.5   0.256   7.5   7.5   0.0492443   0.0514296
142.5   0.315   7.5   7.5   0.0720069   0.0757694
165.    0.246   15.   15.   0.0470106   0.0523259
195.    0.08    15.   15.   0.0608276   0.100499
225.    0.11    15.   15.   0.0806226   0.130384
255.    0.      15.   15.   0.          0.44
285.    0.      15.   15.   0.          0.44
}\AtlasE

\pgfplotstableread{
 5.000000e+01     1.610177e+00   2.474396e-02     3.143293e+00   1.771540e-02
 6.000000e+01     6.561692e-01   8.099210e-03     1.054869e+00   3.264473e-03
 7.000000e+01     2.733497e-01   3.158506e-03     4.118071e-01   1.125084e-03
 8.000000e+01     1.226383e-01   1.447625e-03     1.712666e-01   5.403791e-04
 9.000000e+01     4.852286e-02   4.924199e-04     6.628354e-02   1.774559e-04
 1.050000e+02     1.701294e-02   1.617354e-04     2.202625e-02   5.694162e-05
 1.200000e+02     6.504427e-03   6.272026e-05     8.013901e-03   2.633586e-05
 1.350000e+02     2.620166e-03   2.620145e-05     3.099602e-03   1.072948e-05
 1.500000e+02     7.313003e-04   6.001390e-06     8.871147e-04   2.208997e-06
 1.800000e+02     1.363876e-04   1.383167e-06     1.600909e-04   4.486784e-07
 2.100000e+02     2.663478e-05   3.132894e-07     3.054485e-05   1.238477e-07
 2.400000e+02     5.133012e-06   7.203211e-08     5.905573e-06   2.641496e-08
 2.700000e+02     9.102782e-07   1.736475e-08     1.092648e-06   6.048550e-09
 3.000000e+02     1.410968e-07   2.482408e-09     1.635084e-07   1.021568e-09
 3.400000e+02     1.163978e-08   2.630269e-10     1.392769e-08   1.177557e-10
 3.800000e+02     7.674511e-10   2.052675e-11     8.288852e-10   1.165416e-11
 4.200000e+02     2.813878e-11   1.198072e-12     3.481502e-11   5.993530e-13
 4.600000e+02     5.660128e-13   2.573277e-14     5.912549e-13   1.906549e-14
 5.000000e+02     5.660128e-13   2.573277e-14     5.912549e-13   1.906549e-14
}\nloE

\addplot [const plot mark left, darkgreen, very thick, dash dot] table [x={0},y expr =1-\thisrow{1}/\thisrow{3}]{\nloE};

\addplot [const plot mark left, red, thick] table [x={0},y={1}]{\DuctLambdaE};

\addplot [const plot mark left, blue, thick] table [x={0},y={1}]{\DuctkTE};

\addplot [only marks, black,
        error bars/.cd,
            x dir=both, x explicit,
            y dir=both, y explicit,
    ] table [
        x error plus=ex+,
        x error minus=ex-,
        y error minus=ey-,
        y error plus=ey+,
]{\AtlasE};

\node at (420.0,0.8) {\fcolorbox{black}{white}{$5 < \Delta y < 6$}};
\end{groupplot}
\end{tikzpicture}
\else 
\includegraphics[width = 14 cm]{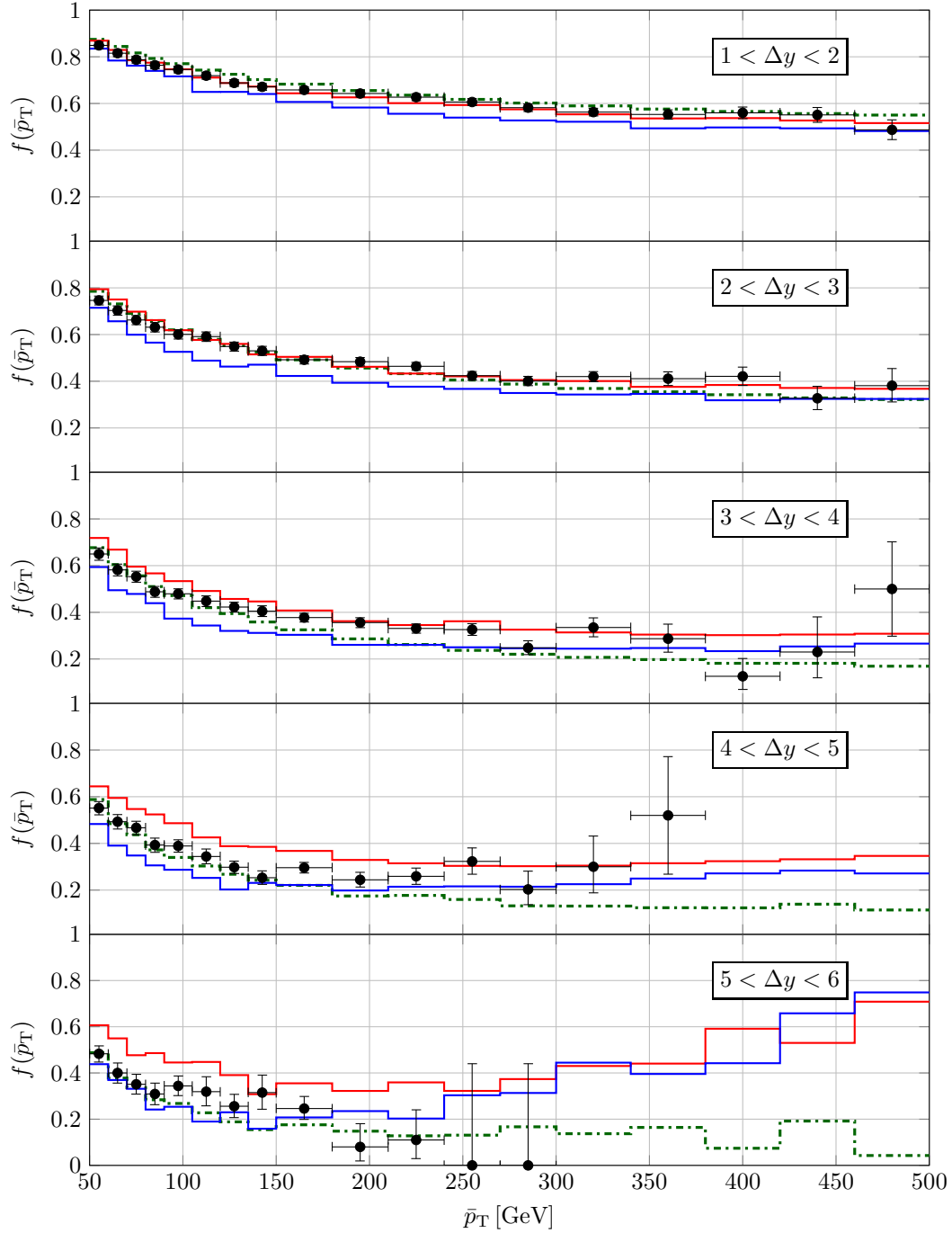}
\fi
\end{center}
\caption{
Gap fraction for various $\Delta y$ values. The black points with error bars are Atlas data \cite{AtlasGap}. The green dash-dotted histogram is NLO perturbation theory \cite{NLOJet++} as defined by Eq.~(\ref{eq:gapNLO}). The red (higher) solid histogram is \textsc{Deductor} with $\Lambda$ ordering, while the blue (lower) solid histogram is \textsc{Deductor} with $k_\LT$ ordering.
}
\label{fig:gapfractions}
\end{figure}

\subsection{\textsc{Deductor} and NLO results}
\label{sec:numerical}

We now generate perturbative \textsc{Deductor} results with both $\Lambda$ ordering and $k_\LT$ ordering. We use $\mu_\LR = \mu_\LF = P_\LT^{\rm Born}/\sqrt{2}$ as in Eq.~(\ref{eq:muFduct}), where  $P_\LT^{\rm Born}$ is the transverse momentum of either of the final state partons at the Born level. With a $\Lambda$-ordered shower, we choose the shower starting scale to be $\mu_\Ls = 3 P_\LT^{\rm Born}/2$ as in Eq.~(\ref{eq:musforLambda}). When we use a $k_\LT$-ordered shower, we choose the shower starting scale to be $\mu_\Ls = P_\LT^{\rm Born}$ as in Eq.~(\ref{eq:musforkT}). We also calculate $f$ at NLO according to Eq.~(\ref{eq:gapNLO}) with $\mu_\LF = \mu_\LR  =  2\,\bar p_\LT$. We show the results of these calculations in Fig.~\ref{fig:gapfractions} along with the data from Atlas \cite{AtlasGap}.

The purely perturbative result is plotted as a green dash-dotted line. For very large $\Delta y$ or $\log(\bar p_\LT/p_\LT^{\rm cut})$, the perturbative result must fail because it does not sum large logarithmic factors. Indeed, we see that the perturbative result lies under the data for $\as\,\Delta y\,\log(\bar p_\LT/p_\LT^{\rm cut})\gtrsim 1$, for instance, for $\bar p_\LT > 150 \GeV$ when $4 < \Delta y < 5$. However, it seems to us remarkable that the perturbative result is quite close to the data for smaller values of $\bar p_\LT$ and $\Delta y$ and that it is within 0.2 of the data for the whole range of $\bar p_\LT$ and $\Delta y$ for which data is available. 

The \textsc{Deductor} results with the default ordering variable $\Lambda$ are plotted in red while the results with $k_\LT$ ordering are plotted in blue. We see that the data are reasonably well described by \textsc{Deductor} with either $\Lambda$ or $k_\LT$ ordering. The match is closer with $\Lambda$ ordering, but given the inherent uncertainties of a calculation with a leading order parton shower based on a leading order hard scattering, we judge that there is not a clear preference of one ordering choice over the other. 

Ref.~\cite{HocheSchonherr} found that a \textsc{Sherpa} leading order shower matched to $\as^2 + \as^3$ perturbation theory for the hard scattering matched data very well.   Since since the \textsc{Deductor} shower with either ordering variable starting from just an $\as^2$ hard scattering is not to far from the higher order perturbative results, we expect that the \textsc{Deductor} showers matched to NLO perturbative results would also agree well with data.

\begin{figure}
\begin{center}
\ifusefigs 
\begin{tikzpicture}
 \begin{groupplot}[
      group style={
          group size=1 by 5,
          vertical sep=0pt,
          x descriptions at=edge bottom},
          xlabel={$\bar p_\LT\,\mathrm{[GeV]}$},
          width=14cm,
    ]

\nextgroupplot[
ylabel={$f(\bar p_\LT)$},
height=5cm,
xmin=50, xmax=500,
ymin=0.01, ymax=1.0,]

\pgfplotstableread{
x       y      ex-   ex+  ey-        ey+
55.  0.8488   5.  5.  0.0149933  0.0144838
65.  0.8154   5.  5.  0.0151119  0.0145066
75.  0.7874   5.  5.  0.0148933  0.0142864
85.  0.7638   5.  5.  0.0148  0.0144222
97.5  0.7457  7.5  7.5  0.0147221  0.0140431
112.5  0.7191 7.5  7.5  0.0151921  0.0148762
127.5  0.688  7.5  7.5  0.0146741  0.0139979
142.5  0.6713 7.5  7.5  0.014994  0.0144312
165.  0.658  15.  15.  0.014258  0.0137674
195.  0.6427 15.  15.  0.0148314  0.0141676
225.  0.627  15.  15.  0.0149452  0.014368
255.  0.6062 15.  15.  0.0153235  0.0148647
285.  0.582  15.  15.  0.0164125  0.0159
320.  0.563  20.  20.  0.0164012  0.0164012
360.  0.553  20.  20.  0.0205183  0.0198494
400.  0.56   20.  20.  0.0252389  0.0246982
440.  0.551  20.  20.  0.0317805  0.0317805
480.  0.487  20.  20.  0.0417612  0.04272
}\AtlasA

\pgfplotstableread{
 5.000000e+01     1.276384e+01   6.002985e-02     1.017511e+02   2.015555e-01     1.360009e+01   1.578634e-01     1.036975e+02   5.397427e-01     1.341033e+01   1.993239e-01     1.030925e+02   6.362840e-01
 6.000000e+01     6.809880e+00   2.964413e-02     4.372649e+01   9.723459e-02     7.173246e+00   7.826263e-02     4.540584e+01   2.442854e-01     7.034673e+00   9.877776e-02     4.525315e+01   2.889529e-01
 7.000000e+01     3.882134e+00   1.543605e-02     2.115231e+01   4.110488e-02     4.233917e+00   4.266417e-02     2.147629e+01   1.183129e-01     4.208052e+00   5.383136e-02     2.132413e+01   1.401992e-01
 8.000000e+01     2.272865e+00   9.258301e-03     1.096614e+01   2.140341e-02     2.459842e+00   2.465527e-02     1.146055e+01   6.012609e-02     2.445018e+00   3.112786e-02     1.141751e+01   7.146933e-02
 9.000000e+01     1.253814e+00   4.041815e-03     5.461076e+00   7.822730e-03     1.344381e+00   1.062820e-02     5.672500e+00   2.180281e-02     1.331052e+00   1.342983e-02     5.653043e+00   2.598852e-02
 1.050000e+02     6.453268e-01   2.121704e-03     2.510350e+00   3.681977e-03     6.969006e-01   5.597936e-03     2.602081e+00   9.562769e-03     6.925005e-01   7.070676e-03     2.587944e+00   1.142833e-02
 1.200000e+02     3.494667e-01   1.186815e-03     1.270695e+00   1.704026e-03     3.767858e-01   3.092888e-03     1.327124e+00   4.923526e-03     3.726049e-01   3.906112e-03     1.322915e+00   5.887962e-03
 1.350000e+02     2.043316e-01   7.014155e-04     6.863766e-01   1.104681e-03     2.196841e-01   1.673037e-03     7.074452e-01   2.901151e-03     2.181747e-01   2.112333e-03     7.031855e-01   3.473668e-03
 1.500000e+02     9.849068e-02   2.255296e-04     3.104434e-01   3.406551e-04     1.066897e-01   6.049669e-04     3.268930e-01   8.208498e-04     1.058062e-01   7.627547e-04     3.261545e-01   9.828753e-04
 1.800000e+02     4.037962e-02   9.866825e-05     1.171750e-01   1.156924e-04     4.397809e-02   2.671676e-04     1.231649e-01   3.082526e-04     4.376956e-02   3.368037e-04     1.228672e-01   3.699954e-04
 2.100000e+02     1.817985e-02   4.806298e-05     4.988608e-02   4.512034e-05     2.007680e-02   1.095570e-04     5.245718e-02   1.277786e-04     2.004288e-02   1.376693e-04     5.232410e-02   1.536694e-04
 2.400000e+02     8.926720e-03   2.119704e-05     2.333811e-02   1.916456e-05     9.658837e-03   5.167844e-05     2.453366e-02   5.636130e-05     9.595611e-03   6.502521e-05     2.448795e-02   6.779900e-05
 2.700000e+02     4.649947e-03   8.543375e-06     1.167630e-02   7.896564e-06     5.121710e-03   2.034383e-05     1.232925e-02   2.222417e-05     5.115206e-03   2.559152e-05     1.231503e-02   2.673834e-05
 3.000000e+02     2.328220e-03   2.586204e-06     5.673435e-03   1.941374e-06     2.553079e-03   6.417021e-06     6.006346e-03   5.536693e-06     2.546582e-03   8.086494e-06     6.004104e-03   6.672544e-06
 3.400000e+02     1.116276e-03   1.235061e-06     2.633317e-03   6.787543e-07     1.227405e-03   3.159987e-06     2.789803e-03   1.806411e-06     1.226259e-03   3.985703e-06     2.789839e-03   2.180801e-06
 3.800000e+02     5.656302e-04   6.410358e-07     1.302386e-03   3.377381e-07     6.194786e-04   1.729574e-06     1.384011e-03   9.604726e-07     6.185153e-04   2.181780e-06     1.385201e-03   1.162178e-06
 4.200000e+02     3.007991e-04   3.782459e-07     6.784407e-04   1.938798e-07     3.311679e-04   9.518942e-07     7.207243e-04   5.101888e-07     3.317035e-04   1.199167e-06     7.214642e-04   6.177043e-07
 4.600000e+02     1.658354e-04   2.230096e-07     3.687359e-04   1.061786e-07     1.838710e-04   5.436064e-07     3.934976e-04   2.846262e-07     1.845623e-04   6.849543e-07     3.943958e-04   3.444406e-07
 5.000000e+02     1.658354e-04   2.230096e-07     3.687359e-04   1.061786e-07     1.838710e-04   5.436064e-07     3.934976e-04   2.846262e-07     1.845623e-04   6.849543e-07     3.943958e-04   3.444406e-07
}\nloA

\addplot [const plot mark left, darkgreen, very thick, dashdotted] table [x={0},y expr =1-\thisrow{1}/\thisrow{3}]{\nloA};
\addplot [const plot mark left, brown, very thick] table [x={0},y expr =1-\thisrow{5}/\thisrow{7}]{\nloA};
\addplot [const plot mark left, blue, very thick, densely dashed] table [x={0},y expr =1-\thisrow{9}/\thisrow{11}]{\nloA};

\addplot [only marks, black,
        error bars/.cd,
            x dir=both, x explicit,
            y dir=both, y explicit,
    ] table [
        x error plus=ex+,
        x error minus=ex-,
        y error minus=ey-,
        y error plus=ey+,
]{\AtlasA};

\node at (420.0,0.8) {\fcolorbox{black}{white}{$1 < \Delta y < 2$}};


\nextgroupplot[
ylabel = $f(\bar p_\LT)$,
height = 5 cm,
xmin=50, xmax=500,
ymin=0.01, ymax=1,
]

\pgfplotstableread{
x       y      ex-   ex+  ey-        ey+
55.  0.7462   5.  5.  0.0193197  0.0186317
65.  0.7032   5.  5.  0.02022  0.0195494
75.  0.6624   5.  5.  0.0201792  0.0192094
85.  0.6315   5.  5.  0.0204362  0.0198736
97.5  0.6009   7.5  7.5  0.0200592  0.0194795
112.5  0.5913   7.5  7.5  0.0207793  0.0199462
127.5  0.5488   7.5  7.5  0.0198487  0.0187926
142.5  0.5302   7.5  7.5  0.0197396  0.0192769
165.  0.4921   15.  15.  0.0182439  0.0173669
195.  0.4838   15.  15.  0.0183469  0.0177924
225.  0.464   15.  15.  0.0173104  0.0167502
255.  0.4237   15.  15.  0.0185809  0.0179739
285.  0.401   15.  15.  0.0198494  0.0198494
320.  0.42   20.  20.  0.0219317  0.0212603
360.  0.411   20.  20.  0.0291548  0.0291548
400.  0.421   20.  20.  0.039  0.0395601
440.  0.327   20.  20.  0.0485077  0.0511957
480.  0.381   20.  20.  0.0696348  0.0733485
}\AtlasB

\pgfplotstableread{
 5.000000e+01     1.447666e+01   8.371886e-02     6.743162e+01   1.683374e-01     1.522359e+01   1.952498e-01     6.870082e+01   4.455708e-01     1.448822e+01   2.494465e-01     6.752813e+01   5.319874e-01
 6.000000e+01     7.529000e+00   4.041292e-02     2.802081e+01   7.654068e-02     7.749289e+00   1.014565e-01     2.851899e+01   1.912975e-01     7.333080e+00   1.296443e-01     2.797540e+01   2.293339e-01
 7.000000e+01     4.073357e+00   2.087238e-02     1.315372e+01   3.239661e-02     4.191898e+00   5.507095e-02     1.348050e+01   9.410941e-02     3.960403e+00   7.038721e-02     1.324897e+01   1.131120e-01
 8.000000e+01     2.283849e+00   1.142459e-02     6.711047e+00   1.578647e-02     2.428788e+00   3.053740e-02     6.719816e+00   4.830794e-02     2.309570e+00   3.899730e-02     6.572104e+00   5.813248e-02
 9.000000e+01     1.216392e+00   4.979215e-03     3.202420e+00   5.962303e-03     1.268696e+00   1.290392e-02     3.344309e+00   1.736560e-02     1.200152e+00   1.646598e-02     3.298584e+00   2.092778e-02
 1.050000e+02     5.999243e-01   2.419776e-03     1.426414e+00   2.717015e-03     6.068500e-01   6.281232e-03     1.486774e+00   7.715333e-03     5.671137e-01   8.008736e-03     1.463855e+00   9.337221e-03
 1.200000e+02     3.147199e-01   1.310285e-03     7.036697e-01   1.478661e-03     3.188988e-01   3.237818e-03     7.251083e-01   3.703021e-03     2.981645e-01   4.128459e-03     7.133332e-01   4.489285e-03
 1.350000e+02     1.727740e-01   7.290714e-04     3.629301e-01   7.588791e-04     1.753310e-01   1.889211e-03     3.768941e-01   2.034053e-03     1.627192e-01   2.408667e-03     3.703157e-01   2.466914e-03
 1.500000e+02     8.067320e-02   2.256136e-04     1.586991e-01   1.867345e-04     8.322478e-02   6.011154e-04     1.650566e-01   5.290440e-04     7.783256e-02   7.645493e-04     1.623064e-01   6.409456e-04
 1.800000e+02     3.086333e-02   7.254646e-05     5.679017e-02   5.538841e-05     3.202835e-02   1.754139e-04     5.879690e-02   1.427778e-04     3.001189e-02   2.225053e-04     5.778766e-02   1.730239e-04
 2.100000e+02     1.300961e-02   2.329581e-05     2.292092e-02   1.530285e-05     1.334584e-02   5.449195e-05     2.398291e-02   4.689672e-05     1.238222e-02   6.921998e-05     2.362578e-02   5.687329e-05
 2.400000e+02     6.035573e-03   8.748945e-06     1.015028e-02   4.075782e-06     6.237532e-03   2.194927e-05     1.059928e-02   1.209700e-05     5.808644e-03   2.794072e-05     1.043481e-02   1.473219e-05
 2.700000e+02     2.956964e-03   4.448595e-06     4.829843e-03   1.945175e-06     3.015606e-03   1.140875e-05     5.038683e-03   5.936679e-06     2.791190e-03   1.452091e-05     4.958681e-03   7.227173e-06
 3.000000e+02     1.397154e-03   1.961777e-06     2.213810e-03   8.076622e-07     1.429783e-03   4.638962e-06     2.320963e-03   2.243844e-06     1.325612e-03   5.904134e-06     2.286088e-03   2.733156e-06
 3.400000e+02     6.182698e-04   8.863389e-07     9.580256e-04   3.611616e-07     6.332501e-04   2.362866e-06     1.004946e-03   1.032305e-06     5.861630e-04   3.010111e-06     9.898622e-04   1.257906e-06
 3.800000e+02     2.903117e-04   4.431184e-07     4.413509e-04   1.720547e-07     2.977391e-04   1.180633e-06     4.624401e-04   5.427249e-07     2.759568e-04   1.501058e-06     4.553368e-04   6.618938e-07
 4.200000e+02     1.433706e-04   2.389864e-07     2.135586e-04   9.327106e-08     1.467533e-04   6.583571e-07     2.244843e-04   2.976060e-07     1.362463e-04   8.372719e-07     2.211475e-04   3.625242e-07
 4.600000e+02     7.266858e-05   1.341237e-07     1.071284e-04   5.314288e-08     7.612994e-05   3.345488e-07     1.125580e-04   1.449016e-07     7.124981e-05   4.253199e-07     1.107881e-04   1.768894e-07
 5.000000e+02     7.266858e-05   1.341237e-07     1.071284e-04   5.314288e-08     7.612994e-05   3.345488e-07     1.125580e-04   1.449016e-07     7.124981e-05   4.253199e-07     1.107881e-04   1.768894e-07
}\nloB

\addplot [const plot mark left, darkgreen, very thick, dashdotted] table [x={0},y expr =1-\thisrow{1}/\thisrow{3}]{\nloB};
\addplot [const plot mark left, brown, very thick] table [x={0},y expr =1-\thisrow{5}/\thisrow{7}]{\nloB};
\addplot [const plot mark left, blue, very thick, densely dashed] table [x={0},y expr =1-\thisrow{9}/\thisrow{11}]{\nloB};

\addplot [only marks, black,
        error bars/.cd,
            x dir=both, x explicit,
            y dir=both, y explicit,
    ] table [
        x error plus=ex+,
        x error minus=ex-,
        y error minus=ey-,
        y error plus=ey+,
]{\AtlasB};

\node at (420.0,0.8) {\fcolorbox{black}{white}{$2 < \Delta y < 3$}};


\nextgroupplot[
ylabel = $f(\bar p_\LT)$,
height = 5 cm,
xmin=50, xmax=500,
ymin=0.01, ymax=1.0,]

\pgfplotstableread{
x       y      ex-   ex+  ey-        ey+
55.     0.6499   5.    5.    0.0268896   0.0257932
65.     0.5818   5.    5.    0.0256673   0.0247348
75.     0.5529   5.    5.    0.0240027   0.0231534
85.     0.488    5.    5.    0.0237065   0.0228254
97.5    0.4794   7.5   7.5   0.0220239   0.021285
112.5   0.448    7.5   7.5   0.0233238   0.0224722
127.5   0.4226   7.5   7.5   0.0216585   0.0208432
142.5   0.405    7.5   7.5   0.0230217   0.0230217
165.    0.3775   15.   15.   0.0191269   0.0182923
195.    0.356    15.   15.   0.0216333   0.0208087
225.    0.331    15.   15.   0.0202485   0.0194165
255.    0.326    15.   15.   0.0247588   0.0254558
285.    0.248    15.   15.   0.0291548   0.0300167
320.    0.335    20.   20.   0.0398121   0.0412311
360.    0.287    20.   20.   0.0577235   0.0623699
400.    0.126    20.   20.   0.0563205   0.0762365
440.    0.23     20.   20.   0.110454    0.150333
480.    0.5      20.   20.   0.202237    0.202237
}\AtlasC

\pgfplotstableread{
 5.000000e+01     1.004637e+01   8.473179e-02     3.110063e+01   1.107757e-01     1.031324e+01   2.012462e-01     3.207889e+01   2.877976e-01     9.287026e+00   2.608413e-01     3.102790e+01   3.497867e-01
 6.000000e+01     4.885181e+00   3.407028e-02     1.235259e+01   4.651959e-02     4.794478e+00   9.392903e-02     1.247962e+01   1.157637e-01     4.217680e+00   1.216849e-01     1.197240e+01   1.410642e-01
 7.000000e+01     2.489220e+00   1.766415e-02     5.610313e+00   2.002472e-02     2.351320e+00   4.717845e-02     5.557195e+00   5.716476e-02     2.026668e+00   6.107936e-02     5.322258e+00   6.981695e-02
 8.000000e+01     1.341091e+00   9.238318e-03     2.738785e+00   1.040530e-02     1.327406e+00   2.580559e-02     2.785807e+00   2.938456e-02     1.159921e+00   3.341452e-02     2.685317e+00   3.595976e-02
 9.000000e+01     6.636424e-01   3.848243e-03     1.256141e+00   3.386371e-03     6.521547e-01   1.002757e-02     1.264210e+00   9.009218e-03     5.638899e-01   1.296228e-02     1.215093e+00   1.104008e-02
 1.050000e+02     3.049554e-01   1.488532e-03     5.258552e-01   1.370813e-03     2.916674e-01   4.361045e-03     5.320238e-01   4.042437e-03     2.464190e-01   5.633540e-03     5.102143e-01   4.949043e-03
 1.200000e+02     1.481090e-01   6.490609e-04     2.446479e-01   5.232353e-04     1.438772e-01   1.688672e-03     2.531645e-01   1.613113e-03     1.213375e-01   2.177197e-03     2.445050e-01   1.973123e-03
 1.350000e+02     7.828000e-02   2.804653e-04     1.220466e-01   2.048532e-04     7.654416e-02   7.611154e-04     1.247959e-01   6.064950e-04     6.477818e-02   9.813973e-04     1.202563e-01   7.440473e-04
 1.500000e+02     3.347284e-02   8.134888e-05     4.959597e-02   2.869600e-05     3.167614e-02   2.016766e-04     5.071507e-02   9.526606e-05     2.618555e-02   2.599062e-04     4.879254e-02   1.169858e-04
 1.800000e+02     1.142811e-02   2.431424e-05     1.601112e-02   1.051561e-05     1.059641e-02   6.860694e-05     1.655622e-02   2.814329e-05     8.571985e-03   8.858181e-05     1.595427e-02   3.462254e-05
 2.100000e+02     4.351312e-03   9.304968e-06     5.897164e-03   3.915825e-06     4.015596e-03   2.523115e-05     6.053112e-03   1.090551e-05     3.217144e-03   3.256173e-05     5.825402e-03   1.342631e-05
 2.400000e+02     1.800124e-03   4.008834e-06     2.359470e-03   1.817271e-06     1.652357e-03   1.167517e-05     2.431933e-03   4.812193e-06     1.312019e-03   1.507055e-05     2.339109e-03   5.931089e-06
 2.700000e+02     7.899684e-04   1.926692e-06     1.013409e-03   7.977835e-07     7.374858e-04   5.242406e-06     1.045521e-03   2.142891e-06     5.899720e-04   6.754860e-06     1.004557e-03   2.641051e-06
 3.000000e+02     3.260293e-04   7.625262e-07     4.111762e-04   2.723033e-07     3.000405e-04   2.107966e-06     4.239496e-04   8.404424e-07     2.363124e-04   2.712768e-06     4.065382e-04   1.035419e-06
 3.400000e+02     1.229109e-04   3.189302e-07     1.531783e-04   1.113403e-07     1.131339e-04   8.532924e-07     1.573190e-04   3.434750e-07     8.881394e-05   1.098970e-06     1.505335e-04   4.239816e-07
 3.800000e+02     4.884649e-05   1.381787e-07     5.968717e-05   5.096407e-08     4.449583e-05   3.695615e-07     6.165315e-05   1.462034e-07     3.474240e-05   4.759104e-07     5.888200e-05   1.807626e-07
 4.200000e+02     1.985104e-05   5.797023e-08     2.425604e-05   2.273134e-08     1.823154e-05   1.670794e-07     2.489014e-05   6.851736e-08     1.429160e-05   2.153243e-07     2.367082e-05   8.467439e-08
 4.600000e+02     8.398946e-06   3.033790e-08     1.010940e-05   1.072677e-08     7.652180e-06   8.292032e-08     1.037250e-05   3.029305e-08     5.980434e-06   1.068593e-07     9.836027e-06   3.747599e-08
 5.000000e+02     8.398946e-06   3.033790e-08     1.010940e-05   1.072677e-08     7.652180e-06   8.292032e-08     1.037250e-05   3.029305e-08     5.980434e-06   1.068593e-07     9.836027e-06   3.747599e-08
}\nloC

\addplot [const plot mark left, darkgreen, very thick, dashdotted] table [x={0},y expr =1-\thisrow{1}/\thisrow{3}]{\nloC};
\addplot [const plot mark left, brown, very thick] table [x={0},y expr =1-\thisrow{5}/\thisrow{7}]{\nloC};
\addplot [const plot mark left, blue, very thick, densely dashed] table [x={0},y expr =1-\thisrow{9}/\thisrow{11}]{\nloC};

\addplot [only marks, black,
        error bars/.cd,
            x dir=both, x explicit,
            y dir=both, y explicit,
    ] table [
        x error plus=ex+,
        x error minus=ex-,
        y error minus=ey-,
        y error plus=ey+,
]{\AtlasC};

\node at (420.0,0.8) {\fcolorbox{black}{white}{$3 < \Delta y < 4$}};


\nextgroupplot[
ylabel = $f(\bar p_\LT)$,
height = 5 cm,
xmin=50, xmax=500,
ymin=0.01, ymax=1.0,]

\pgfplotstableread{
x       y      ex-   ex+  ey-        ey+
55.     0.5515   5.    5.    0.0290998   0.0276921
65.     0.493    5.    5.    0.0313847   0.0305287
75.     0.467    5.    5.    0.0295296   0.0277849
85.     0.393    5.    5.    0.031241    0.0297321
97.5    0.389    7.5   7.5   0.0266271   0.0264008
112.5   0.344    7.5   7.5   0.031241    0.0320156
127.5   0.298    7.5   7.5   0.0269072   0.0261725
142.5   0.253    7.5   7.5   0.0286007   0.0294109
165.    0.296    15.   15.   0.0228035   0.0226716
195.    0.244    15.   15.   0.0308869   0.0322025
225.    0.259    15.   15.   0.0344819   0.0353553
255.    0.323    15.   15.   0.0550091   0.0576975
285.    0.203    15.   15.   0.065521    0.0781025
320.    0.3      20.   20.   0.111803    0.131529
360.    0.52     20.   20.   0.251794    0.251794
}\AtlasD

\pgfplotstableread{
 5.000000e+01     4.613036e+00   6.500245e-02     1.119498e+01   5.486533e-02     4.353042e+00   1.557865e-01     1.098229e+01   1.642102e-01     3.383367e+00   2.051223e-01     1.021714e+01   2.036592e-01
 6.000000e+01     2.163978e+00   2.474487e-02     4.199723e+00   2.110719e-02     1.977862e+00   5.953755e-02     4.231432e+00   5.521533e-02     1.541836e+00   7.829272e-02     3.969256e+00   6.856014e-02
 7.000000e+01     9.831736e-01   9.271330e-03     1.746631e+00   1.029496e-02     8.837711e-01   2.494972e-02     1.825909e+00   2.143798e-02     6.502134e-01   3.275166e-02     1.730366e+00   2.664673e-02
 8.000000e+01     5.104493e-01   3.978250e-03     8.130937e-01   3.034500e-03     4.700147e-01   1.036400e-02     8.284999e-01   8.353243e-03     3.571763e-01   1.360061e-02     7.799270e-01   1.037339e-02
 9.000000e+01     2.288680e-01   1.348444e-03     3.467440e-01   6.711638e-04     1.992868e-01   3.497770e-03     3.485914e-01   1.853051e-03     1.399213e-01   4.595262e-03     3.261555e-01   2.305459e-03
 1.050000e+02     9.385146e-02   5.062052e-04     1.346926e-01   2.047948e-04     8.178675e-02   1.343167e-03     1.355272e-01   5.764185e-04     5.549540e-02   1.762285e-03     1.267113e-01   7.184755e-04
 1.200000e+02     4.206217e-02   2.169628e-04     5.744764e-02   8.713355e-05     3.649244e-02   6.425679e-04     5.868241e-02   2.641190e-04     2.420580e-02   8.447774e-04     5.509576e-02   3.292763e-04
 1.350000e+02     1.991671e-02   1.034727e-04     2.628776e-02   4.512828e-05     1.623054e-02   2.851641e-04     2.665062e-02   1.379829e-04     9.933975e-03   3.742268e-04     2.493994e-02   1.722067e-04
 1.500000e+02     7.446864e-03   2.848235e-05     9.551073e-03   1.005218e-05     6.105392e-03   7.158464e-05     9.542007e-03   3.363068e-05     3.624877e-03   9.387809e-05     8.869160e-03   4.201088e-05
 1.800000e+02     2.124266e-03   8.736137e-06     2.574227e-03   3.200338e-06     1.672435e-03   2.197974e-05     2.600154e-03   9.063958e-06     9.391003e-04   2.881836e-05     2.417225e-03   1.132516e-05
 2.100000e+02     6.404649e-04   2.929481e-06     7.784499e-04   1.064908e-06     5.040185e-04   7.892886e-06     7.873345e-04   3.246879e-06     2.642882e-04   1.034303e-05     7.301295e-04   4.062579e-06
 2.400000e+02     2.120570e-04   1.074807e-06     2.523893e-04   3.709426e-07     1.631959e-04   2.783821e-06     2.506521e-04   1.173701e-06     8.109030e-05   3.645941e-06     2.295526e-04   1.469825e-06
 2.700000e+02     7.456880e-05   4.008784e-07     8.596386e-05   1.437068e-07     5.459304e-05   1.148289e-06     8.547685e-05   4.488108e-07     2.490331e-05   1.505472e-06     7.778833e-05   5.632207e-07
 3.000000e+02     2.300510e-05   1.240159e-07     2.648950e-05   4.634931e-08     1.702420e-05   3.580644e-07     2.617883e-05   1.311882e-07     7.591112e-06   4.694175e-07     2.360482e-05   1.647117e-07
 3.400000e+02     5.978590e-06   3.922455e-08     6.828515e-06   1.335500e-08     4.598772e-06   1.027408e-07     6.653215e-06   3.948604e-08     2.173088e-06   1.344965e-07     5.901257e-06   4.964251e-08
 3.800000e+02     1.561299e-06   1.270531e-08     1.781992e-06   4.605254e-09     1.142312e-06   3.061288e-08     1.729344e-06   1.256077e-08     4.752528e-07   4.018675e-08     1.512259e-06   1.584765e-08
 4.200000e+02     3.974319e-07   3.763422e-09     4.619386e-07   1.267787e-09     3.006759e-07   8.680678e-09     4.288069e-07   3.863473e-09     1.271184e-07   1.137903e-08     3.618495e-07   4.890970e-09
 4.600000e+02     1.031211e-07   1.100672e-09     1.164962e-07   4.415532e-10     7.285927e-08   2.609180e-09     1.124351e-07   1.094599e-09     2.752558e-08   3.432133e-09     9.503152e-08   1.389783e-09
 5.000000e+02     1.031211e-07   1.100672e-09     1.164962e-07   4.415532e-10     7.285927e-08   2.609180e-09     1.124351e-07   1.094599e-09     2.752558e-08   3.432133e-09     9.503152e-08   1.389783e-09
}\nloD

\addplot [const plot mark left, darkgreen, very thick, dashdotted] table [x={0},y expr =1-\thisrow{1}/\thisrow{3}]{\nloD};
\addplot [const plot mark left, brown, very thick] table [x={0},y expr =1-\thisrow{5}/\thisrow{7}]{\nloD};
\addplot [const plot mark left, blue, very thick, densely dashed] table [x={0},y expr =1-\thisrow{9}/\thisrow{11}]{\nloD};

\addplot [only marks, black,
        error bars/.cd,
            x dir=both, x explicit,
            y dir=both, y explicit,
    ] table [
        x error plus=ex+,
        x error minus=ex-,
        y error minus=ey-,
        y error plus=ey+,
]{\AtlasD};

\node at (420.0,0.8) {\fcolorbox{black}{white}{$4 < \Delta y < 5$}};

\nextgroupplot[
ylabel={$f(\bar p_\LT)$},
height=5cm,
xmin=50, xmax=500,
ymin=0.0, ymax=1.0,]

\pgfplotstableread{
x       y      ex-   ex+  ey-        ey+
55.     0.483   5.    5.    0.0358469   0.0341321
65.     0.4     5.    5.    0.0438634   0.0431393
75.     0.351   5.    5.    0.04245     0.0431393
85.     0.309   5.    5.    0.0463249   0.0470106
97.5    0.344   7.5   7.5   0.0420119   0.0430116
112.5   0.319   7.5   7.5   0.061327    0.064622
127.5   0.256   7.5   7.5   0.0492443   0.0514296
142.5   0.315   7.5   7.5   0.0720069   0.0757694
165.    0.246   15.   15.   0.0470106   0.0523259
195.    0.08    15.   15.   0.0608276   0.100499
225.    0.11    15.   15.   0.0806226   0.130384
255.    0.      15.   15.   0.          0.44
285.    0.      15.   15.   0.          0.44
}\AtlasE

\pgfplotstableread{
 5.000000e+01     1.610177e+00   2.474396e-02     3.143293e+00   1.771540e-02     1.224972e+00   4.946212e-02     3.178471e+00   3.540505e-02     6.466074e-01   6.647464e-02     2.897879e+00   4.470600e-02
 6.000000e+01     6.561692e-01   8.099210e-03     1.054869e+00   3.264473e-03     5.006665e-01   1.709092e-02     1.039520e+00   6.628865e-03     2.648418e-01   2.297284e-02     9.313936e-01   8.384624e-03
 7.000000e+01     2.733497e-01   3.158506e-03     4.118071e-01   1.125084e-03     2.107011e-01   6.904949e-03     4.047388e-01   2.534875e-03     1.040211e-01   9.252217e-03     3.647349e-01   3.208168e-03
 8.000000e+01     1.226383e-01   1.447625e-03     1.712666e-01   5.403791e-04     9.009677e-02   2.776739e-03     1.715891e-01   1.142343e-03     3.827232e-02   3.721185e-03     1.552155e-01   1.447463e-03
 9.000000e+01     4.852286e-02   4.924199e-04     6.628354e-02   1.774559e-04     3.494261e-02   8.975083e-04     6.479863e-02   3.646885e-04     1.265274e-02   1.198674e-03     5.815460e-02   4.625907e-04
 1.050000e+02     1.701294e-02   1.617354e-04     2.202625e-02   5.694162e-05     1.193119e-02   3.169814e-04     2.157821e-02   1.208205e-04     3.490369e-03   4.238569e-04     1.925593e-02   1.535757e-04
 1.200000e+02     6.504427e-03   6.272026e-05     8.013901e-03   2.633586e-05     4.074407e-03   1.313905e-04     7.949684e-03   4.989691e-05     5.993401e-04   1.757159e-04     7.100137e-03   6.353586e-05
 1.350000e+02     2.620166e-03   2.620145e-05     3.099602e-03   1.072948e-05     1.569611e-03   5.416480e-05     3.033966e-03   2.157845e-05     1.252836e-04   7.255683e-05     2.678467e-03   2.747844e-05
 1.500000e+02     7.313003e-04   6.001390e-06     8.871147e-04   2.208997e-06     4.126234e-04   1.175933e-05     8.597615e-04   4.867369e-06    -0.1            1.570018e-05     1.0            6.196330e-06
 1.800000e+02     1.363876e-04   1.383167e-06     1.600909e-04   4.486784e-07     6.871337e-05   2.840695e-06     1.530822e-04   1.096544e-06    -0.1            3.802921e-06     1.0            1.399741e-06
 2.100000e+02     2.663478e-05   3.132894e-07     3.054485e-05   1.238477e-07     1.235590e-05   5.937946e-07     2.910175e-05   2.419911e-07    -0.1            7.938326e-07     1.0            3.095899e-07
 2.400000e+02     5.133012e-06   7.203211e-08     5.905573e-06   2.641496e-08     2.293014e-06   1.324286e-07     5.457539e-06   5.320458e-08    -0.1            1.775161e-07     1.0            6.823661e-08
 2.700000e+02     9.102782e-07   1.736475e-08     1.092648e-06   6.048550e-09     4.401789e-07   3.053210e-08     9.555027e-07   1.304743e-08    -0.1            4.097775e-08     1.0            1.679813e-08
 3.000000e+02     1.410968e-07   2.482408e-09     1.635084e-07   1.021568e-09     6.519558e-08   5.009297e-09     1.399138e-07   2.273522e-09    -0.1            6.729342e-09     1.0            2.942302e-09
 3.400000e+02     1.163978e-08   2.630269e-10     1.392769e-08   1.177557e-10     7.032418e-09   5.208085e-10     1.023891e-08   2.697201e-10    -0.1            7.022696e-10     1.0            3.521095e-10
 3.800000e+02     7.674511e-10   2.052675e-11     8.288852e-10   1.165416e-11     2.171056e-10   4.352854e-11     5.383608e-10   2.721290e-11    -0.1            5.914200e-11     1.0            3.588336e-11
 4.200000e+02     2.813878e-11   1.198072e-12     3.481502e-11   5.993530e-13     9.215367e-12   2.064421e-12     1.473203e-11   1.539250e-12    -0.1            2.818314e-12     1.0            2.054176e-12
 4.600000e+02     5.660128e-13   2.573277e-14     5.912549e-13   1.906549e-14     2.555586e-13   4.034869e-14     2.354512e-13   4.684156e-14    -0.1            5.410401e-14     1.0            6.367390e-14
 5.000000e+02     5.660128e-13   2.573277e-14     5.912549e-13   1.906549e-14     2.555586e-13   4.034869e-14     2.354512e-13   4.684156e-14    -0.1            5.410401e-14     1.0            6.367390e-14 
}\nloE

\addplot [const plot mark left, darkgreen, very thick, dashdotted] table [x={0},y expr =1-\thisrow{1}/\thisrow{3}]{\nloE};
\addplot [const plot mark left, brown, very thick] table [x={0},y expr =1-\thisrow{5}/\thisrow{7}]{\nloE};
\addplot [const plot mark left, blue, very thick, densely dashed] table [x={0},y expr =1-\thisrow{9}/\thisrow{11}]{\nloE};

\addplot [only marks, black,
        error bars/.cd,
            x dir=both, x explicit,
            y dir=both, y explicit,
    ] table [
        x error plus=ex+,
        x error minus=ex-,
        y error minus=ey-,
        y error plus=ey+,
]{\AtlasE};

\node at (420.0,0.8) {\fcolorbox{black}{white}{$5 < \Delta y < 6$}};
\end{groupplot}
\end{tikzpicture}
\else 
\includegraphics[width = 14 cm]{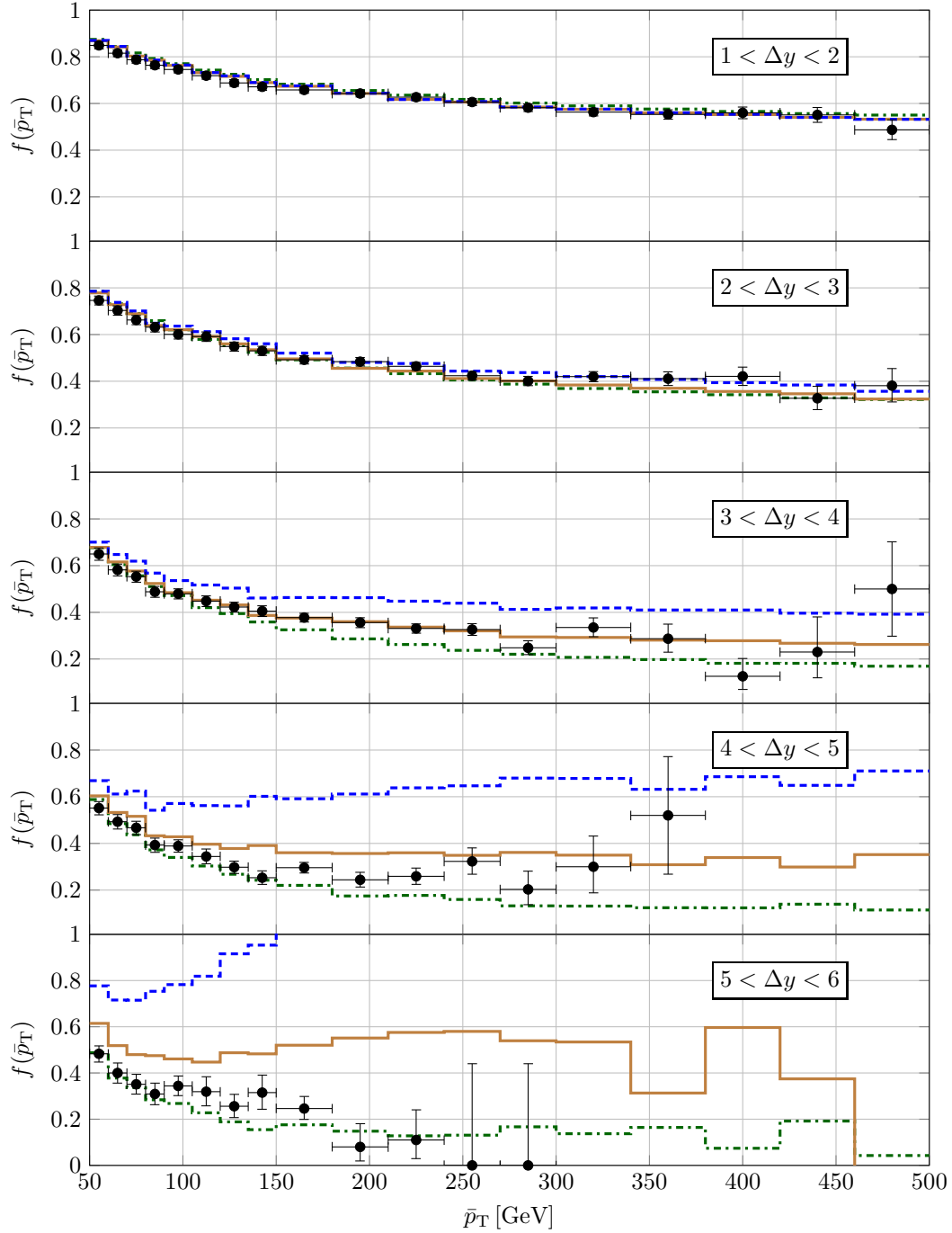}
\fi
\end{center}
\caption{
Gap fraction for various $\Delta y$ values. The black points with error bars are Atlas data \cite{AtlasGap}. The green dash-dotted, brown solid, and blue dashed histograms are NLO perturbation theory \cite{NLOJet++} as defined by Eq.~(\ref{eq:gapNLO}) at $\mu_\LR=\mu_\LF = 2 \bar p_\LT,\, \bar p_\LT,\, \bar p_\LT/2$ scales, respectively. (For $\mu_\LR=\mu_\LF = \bar p_\LT/2$, results in the $5 < \Delta y < 6$ plot for $\bar p_\LT > 150 \GeV$ are unphysical and are not shown.)
}
\label{fig:gapfractionsNLO}
\end{figure}

\subsection{NLO results with different scale choices}
\label{sec:NLOgapscales}

The perturbative result in Fig.~\ref{fig:gapfractions} seems somewhat surprising, since there are large logarithms $\Delta y$ and $\log(\bar p_\LT/p_\LT^{\rm cut})$ present. In Fig.~\ref{fig:gapfractions}, we chose $\mu_\LR = \mu_\LF = 2\,\bar p_\LT$. This is a compromise between choosing $\mu = \bar p_\LT$ and choosing $\mu = \hat s/2$, where $\hat s$ is the c.m.\ energy for the $2 \to 2$ process, $\hat s /2 = \cosh(\Delta y/2)\, \bar p_\LT$. (Compare to the scale choice in Ref.~\cite{EKS2}.) We have tried the choices $\mu_\LR = \mu_\LF = \bar p_\LT$ and $\mu_\LR = \mu_\LF = \bar p_\LT /2$ and present the results in Fig.~\ref{fig:gapfractionsNLO}. We see that for $\Delta y < 3$, the scale choice hardly matters. However, for the smaller values of the scales the NLO perturbative result at larger values of $\Delta y$ becomes unstable in the sense that $f(\bar p_\LT)$ increases with $\bar p_\LT$. In particular, for $\mu_\LR = \mu_\LF = \bar p_\LT/2$, the NLO result is unstable in this sense for $\Delta y > 4$. For $4 < \Delta y < 5$, the NLO result for $\mu_\LR = \mu_\LF = \bar p_\LT$ still appears stable and, in fact, is within about 0.1 of the data. However, for $\Delta y > 5$, the $\mu_\LR = \mu_\LF = \bar p_\LT$ NLO result has also become unstable.

We offer two conclusions about the NLO calculation based on Eq.~(\ref{eq:gapNLO}). First, as the scales $\bar p_\LT$ and $\hat s /2 = \cosh(\Delta y/2)\, \bar p_\LT$ become substantially different, the NLO result becomes more sensitive to the scale choice.  The increasing sensitivity to the scale choice suggests that the pure fixed order perturbative calculation without an all order summation of large logarithms is becoming less reliable. Second, it is remarkable how well the NLO result with $\mu_\LR = \mu_\LF = 2\bar p_\LT$ works. These two conclusions suggest that while a summation of factors of $\Delta y\,\log(\bar p_\LT/p_\LT^{\rm cut})$ is very interesting for QCD theory, it is not numerically dominant for $\Delta y < 5$, which covers most of the range of these gap survival data.

\subsection{\textsc{Deductor} results with different scale choices}
\label{sec:Deducttorgapscales}

\begin{figure}
\begin{center}
\ifusefigs 
\begin{tikzpicture}
 \begin{groupplot}[
      group style={
          group size=1 by 5,
          vertical sep=0pt,
          x descriptions at=edge bottom},
          xlabel={$\bar p_\LT\,\mathrm{[GeV]}$},
          width=14cm,
    ]

\nextgroupplot[
ylabel={$f(\bar p_\LT)$},
height=5cm,
xmin=50, xmax=500,
ymin=0.01, ymax=1.0,]

\pgfplotstableread{
50.    0.869041
60.    0.829231
70.    0.785057
80.    0.77467
90.    0.74635
105.   0.710944
120.   0.687418
135.   0.672407
150.   0.643561
180.   0.626222
210.   0.601456
240.   0.593203
270.   0.574926
300.   0.55345
340.   0.535905
380.   0.537309
420.   0.527018
460.   0.515959
500    0.515959
}\DuctLambdaA

\pgfplotstableread{
50.    0.91627
60.    0.895696
70.    0.862238
80.    0.83822
90.    0.811097
105.   0.783774
120.   0.761987
135.   0.731728
150.   0.701172
180.   0.67759
210.   0.646417
240.   0.630493
270.   0.614172
300.   0.593756
340.   0.57817
380.   0.569483
420.   0.553231
460.   0.555598
500    0.555598
}\DuctLambdaOneA

\pgfplotstableread{
50.    0.853672
60.    0.815545
70.    0.780091
80.    0.744085
90.    0.737704
105.   0.71494
120.   0.679563
135.   0.654686
150.   0.641269
180.   0.619933
210.   0.590949
240.   0.577
270.   0.558341
300.   0.547304
340.   0.538095
380.   0.526027
420.   0.521613
460.   0.513785
500    0.513785
}\DuctLambdaTwoA

\pgfplotstableread{
x       y      ex-   ex+  ey-        ey+
55.  0.8488   5.  5.  0.0149933  0.0144838
65.  0.8154   5.  5.  0.0151119  0.0145066
75.  0.7874   5.  5.  0.0148933  0.0142864
85.  0.7638   5.  5.  0.0148  0.0144222
97.5  0.7457  7.5  7.5  0.0147221  0.0140431
112.5  0.7191 7.5  7.5  0.0151921  0.0148762
127.5  0.688  7.5  7.5  0.0146741  0.0139979
142.5  0.6713 7.5  7.5  0.014994  0.0144312
165.  0.658  15.  15.  0.014258  0.0137674
195.  0.6427 15.  15.  0.0148314  0.0141676
225.  0.627  15.  15.  0.0149452  0.014368
255.  0.6062 15.  15.  0.0153235  0.0148647
285.  0.582  15.  15.  0.0164125  0.0159
320.  0.563  20.  20.  0.0164012  0.0164012
360.  0.553  20.  20.  0.0205183  0.0198494
400.  0.56   20.  20.  0.0252389  0.0246982
440.  0.551  20.  20.  0.0317805  0.0317805
480.  0.487  20.  20.  0.0417612  0.04272
}\AtlasA

\addplot [const plot mark left, red] table [x={0},y={1}]{\DuctLambdaA};

\addplot [const plot mark left, blue, very thick, dashdotted] table [x={0},y={1}]{\DuctLambdaOneA};

\addplot [const plot mark left, darkgreen, very thick, densely dashed] table [x={0},y={1}]{\DuctLambdaTwoA};

\addplot [only marks, black,
        error bars/.cd,
            x dir=both, x explicit,
            y dir=both, y explicit,
    ] table [
        x error plus=ex+,
        x error minus=ex-,
        y error minus=ey-,
        y error plus=ey+,
]{\AtlasA};

\node at (420.0,0.8) {\fcolorbox{black}{white}{$1 < \Delta y < 2$}};


\nextgroupplot[
ylabel = $f(\bar p_\LT)$,
height = 5 cm,
xmin=50, xmax=500,
ymin=0.01, ymax=1.0,
]

\pgfplotstableread{
50.    0.794329
60.    0.750284
70.    0.698127
80.    0.661743
90.    0.618695
105.   0.57705
120.   0.559996
135.   0.515071
150.   0.504368
180.   0.461917
210.   0.433062
240.   0.420998
270.   0.405112
300.   0.400887
340.   0.376058
380.   0.383958
420.   0.371254
460.   0.367635
500    0.367635
}\DuctLambdaB

\pgfplotstableread{
50.    0.861403
60.    0.830818
70.    0.796846
80.    0.78111
90.    0.732375
105.   0.693388
120.   0.665386
135.   0.639967
150.   0.596137
180.   0.551663
210.   0.526732
240.   0.498371
270.   0.477365
300.   0.455636
340.   0.446413
380.   0.432871
420.   0.423596
460.   0.412543
500    0.412543
}\DuctLambdaOneB

\pgfplotstableread{
50.    0.740564
60.    0.682157
70.    0.645955
80.    0.605554
90.    0.567956
105.   0.546534
120.   0.513372
135.   0.494063
150.   0.45127
180.   0.438188
210.   0.415428
240.   0.40188
270.   0.385322
300.   0.388338
340.   0.370099
380.   0.365491
420.   0.359793
460.   0.355608
500    0.355608
}\DuctLambdaTwoB

\pgfplotstableread{
x       y      ex-   ex+  ey-        ey+
55.  0.7462   5.  5.  0.0193197  0.0186317
65.  0.7032   5.  5.  0.02022  0.0195494
75.  0.6624   5.  5.  0.0201792  0.0192094
85.  0.6315   5.  5.  0.0204362  0.0198736
97.5  0.6009   7.5  7.5  0.0200592  0.0194795
112.5  0.5913   7.5  7.5  0.0207793  0.0199462
127.5  0.5488   7.5  7.5  0.0198487  0.0187926
142.5  0.5302   7.5  7.5  0.0197396  0.0192769
165.  0.4921   15.  15.  0.0182439  0.0173669
195.  0.4838   15.  15.  0.0183469  0.0177924
225.  0.464   15.  15.  0.0173104  0.0167502
255.  0.4237   15.  15.  0.0185809  0.0179739
285.  0.401   15.  15.  0.0198494  0.0198494
320.  0.42   20.  20.  0.0219317  0.0212603
360.  0.411   20.  20.  0.0291548  0.0291548
400.  0.421   20.  20.  0.039  0.0395601
440.  0.327   20.  20.  0.0485077  0.0511957
480.  0.381   20.  20.  0.0696348  0.0733485
}\AtlasB

\addplot [const plot mark left, red] table [x={0},y={1}]{\DuctLambdaB};

\addplot [const plot mark left, blue, very thick, dashdotted] table [x={0},y={1}]{\DuctLambdaOneB};

\addplot [const plot mark left, darkgreen, very thick, densely dashed] table [x={0},y={1}]{\DuctLambdaTwoB};

\addplot [only marks, black,
        error bars/.cd,
            x dir=both, x explicit,
            y dir=both, y explicit,
    ] table [
        x error plus=ex+,
        x error minus=ex-,
        y error minus=ey-,
        y error plus=ey+,
]{\AtlasB};

\node at (420.0,0.8) {\fcolorbox{black}{white}{$2 < \Delta y < 3$}};


\nextgroupplot[
ylabel = $f(\bar p_\LT)$,
height = 5 cm,
xmin=50, xmax=500,
ymin=0.01, ymax=1.0,]

\pgfplotstableread{
50.    0.718585
60.    0.668641
70.    0.59565
80.    0.566885
90.    0.533686
105.   0.491389
120.   0.457264
135.   0.446529
150.   0.407818
180.   0.361995
210.   0.345479
240.   0.361706
270.   0.325814
300.   0.313919
340.   0.304646
380.   0.302332
420.   0.304916
460.   0.308353
500    0.308353
}\DuctLambdaC

\pgfplotstableread{
50.    0.800552
60.    0.776352
70.    0.738736
80.    0.702789
90.    0.672344
105.   0.629706
120.   0.582539
135.   0.563663
150.   0.528921
180.   0.476224
210.   0.447759
240.   0.434141
270.   0.406018
300.   0.388037
340.   0.376469
380.   0.370824
420.   0.363288
460.   0.357696
500    0.357696
}\DuctLambdaOneC

\pgfplotstableread{
50.    0.638927
60.    0.580299
70.    0.532539
80.    0.504972
90.    0.47225
105.   0.41914
120.   0.389657
135.   0.357091
150.   0.358815
180.   0.327536
210.   0.31106
240.   0.288303
270.   0.288288
300.   0.286539
340.   0.274086
380.   0.287067
420.   0.285207
460.   0.284221
500    0.284221
}\DuctLambdaTwoC

\pgfplotstableread{
x       y      ex-   ex+  ey-        ey+
55.     0.6499   5.    5.    0.0268896   0.0257932
65.     0.5818   5.    5.    0.0256673   0.0247348
75.     0.5529   5.    5.    0.0240027   0.0231534
85.     0.488    5.    5.    0.0237065   0.0228254
97.5    0.4794   7.5   7.5   0.0220239   0.021285
112.5   0.448    7.5   7.5   0.0233238   0.0224722
127.5   0.4226   7.5   7.5   0.0216585   0.0208432
142.5   0.405    7.5   7.5   0.0230217   0.0230217
165.    0.3775   15.   15.   0.0191269   0.0182923
195.    0.356    15.   15.   0.0216333   0.0208087
225.    0.331    15.   15.   0.0202485   0.0194165
255.    0.326    15.   15.   0.0247588   0.0254558
285.    0.248    15.   15.   0.0291548   0.0300167
320.    0.335    20.   20.   0.0398121   0.0412311
360.    0.287    20.   20.   0.0577235   0.0623699
400.    0.126    20.   20.   0.0563205   0.0762365
440.    0.23     20.   20.   0.110454    0.150333
480.    0.5      20.   20.   0.202237    0.202237
}\AtlasC

\addplot [const plot mark left, red] table [x={0},y={1}]{\DuctLambdaC};

\addplot [const plot mark left, blue, very thick, dashdotted] table [x={0},y={1}]{\DuctLambdaOneC};

\addplot [const plot mark left, darkgreen, very thick, densely dashed] table [x={0},y={1}]{\DuctLambdaTwoC};

\addplot [only marks, black,
        error bars/.cd,
            x dir=both, x explicit,
            y dir=both, y explicit,
    ] table [
        x error plus=ex+,
        x error minus=ex-,
        y error minus=ey-,
        y error plus=ey+,
]{\AtlasC};

\node at (420.0,0.8) {\fcolorbox{black}{white}{$3 < \Delta y < 4$}};


\nextgroupplot[
ylabel = $f(\bar p_\LT)$,
height = 5 cm,
xmin=50, xmax=500,
ymin=0.01, ymax=1.0,]

\pgfplotstableread{
50.    0.644343
60.    0.594773
70.    0.547274
80.    0.5242
90.    0.486543
105.   0.425727
120.   0.38814
135.   0.385221
150.   0.367883
180.   0.329235
210.   0.314834
240.   0.303757
270.   0.302437
300.   0.304755
340.   0.315029
380.   0.323796
420.   0.331982
460.   0.346462
500    0.346462
}\DuctLambdaD

\pgfplotstableread{
50.    0.751773
60.    0.721383
70.    0.699117
80.    0.638314
90.    0.602855
105.   0.630195
120.   0.58373
135.   0.523842
150.   0.494395
180.   0.449492
210.   0.420123
240.   0.395742
270.   0.395501
300.   0.390995
340.   0.382894
380.   0.37942
420.   0.380471
460.   0.407084
500    0.407084
}\DuctLambdaOneD

\pgfplotstableread{
50.    0.572386
60.    0.514901
70.    0.447086
80.    0.429423
90.    0.392409
105.   0.346438
120.   0.323191
135.   0.320649
150.   0.271529
180.   0.278638
210.   0.264743
240.   0.261527
270.   0.243834
300.   0.256711
340.   0.273939
380.   0.301212
420.   0.308572
460.   0.336909
500    0.336909
}\DuctLambdaTwoD

\pgfplotstableread{
x       y      ex-   ex+  ey-        ey+
55.     0.5515   5.    5.    0.0290998   0.0276921
65.     0.493    5.    5.    0.0313847   0.0305287
75.     0.467    5.    5.    0.0295296   0.0277849
85.     0.393    5.    5.    0.031241    0.0297321
97.5    0.389    7.5   7.5   0.0266271   0.0264008
112.5   0.344    7.5   7.5   0.031241    0.0320156
127.5   0.298    7.5   7.5   0.0269072   0.0261725
142.5   0.253    7.5   7.5   0.0286007   0.0294109
165.    0.296    15.   15.   0.0228035   0.0226716
195.    0.244    15.   15.   0.0308869   0.0322025
225.    0.259    15.   15.   0.0344819   0.0353553
255.    0.323    15.   15.   0.0550091   0.0576975
285.    0.203    15.   15.   0.065521    0.0781025
320.    0.3      20.   20.   0.111803    0.131529
360.    0.52     20.   20.   0.251794    0.251794
}\AtlasD

\addplot [const plot mark left, red] table [x={0},y={1}]{\DuctLambdaD};

\addplot [const plot mark left, blue, very thick, dashdotted] table [x={0},y={1}]{\DuctLambdaOneD};

\addplot [const plot mark left, darkgreen, very thick, densely dashed] table [x={0},y={1}]{\DuctLambdaTwoD};

\addplot [only marks, black,
        error bars/.cd,
            x dir=both, x explicit,
            y dir=both, y explicit,
    ] table [
        x error plus=ex+,
        x error minus=ex-,
        y error minus=ey-,
        y error plus=ey+,
]{\AtlasD};

\node at (420.0,0.8) {\fcolorbox{black}{white}{$4 < \Delta y < 5$}};

\nextgroupplot[
ylabel={$f(\bar p_\LT)$},
height=5cm,
xmin=50, xmax=500,
ymin=0.0, ymax=1.0,]

\pgfplotstableread{
50.    0.606419
60.    0.549602
70.    0.476753
80.    0.485873
90.    0.445351
105.   0.447636
120.   0.390705
135.   0.308101
150.   0.355184
180.   0.322212
210.   0.359146
240.   0.322348
270.   0.373466
300.   0.430126
340.   0.440527
380.   0.591407
420.   0.529998
460.   0.708377
500    0.708377
}\DuctLambdaE

\pgfplotstableread{
50.    0.688317
60.    0.663754
70.    0.639224
80.    0.605875
90.    0.561263
105.   0.587383
120.   0.554214
135.   0.563262
150.   0.506416
180.   0.472846
210.   0.461514
240.   0.465452
270.   0.450842
300.   0.505074
340.   0.515852
380.   0.608499
420.   0.622213
460.   0.69885
500    0.69885
}\DuctLambdaOneE

\pgfplotstableread{
50.    0.522876
60.    0.466345
70.    0.424344
80.    0.361808
90.    0.343294
105.   0.313183
120.   0.314277
135.   0.317049
150.   0.285089
180.   0.283874
210.   0.286235
240.   0.296941
270.   0.337674
300.   0.356734
340.   0.395747
380.   0.425118
420.   0.59348
460.   0.736458
500    0.736458
}\DuctLambdaTwoE

\pgfplotstableread{
x       y      ex-   ex+  ey-        ey+
55.     0.483   5.    5.    0.0358469   0.0341321
65.     0.4     5.    5.    0.0438634   0.0431393
75.     0.351   5.    5.    0.04245     0.0431393
85.     0.309   5.    5.    0.0463249   0.0470106
97.5    0.344   7.5   7.5   0.0420119   0.0430116
112.5   0.319   7.5   7.5   0.061327    0.064622
127.5   0.256   7.5   7.5   0.0492443   0.0514296
142.5   0.315   7.5   7.5   0.0720069   0.0757694
165.    0.246   15.   15.   0.0470106   0.0523259
195.    0.08    15.   15.   0.0608276   0.100499
225.    0.11    15.   15.   0.0806226   0.130384
255.    0.      15.   15.   0.          0.44
285.    0.      15.   15.   0.          0.44
}\AtlasE

\addplot [const plot mark left, red] table [x={0},y={1}]{\DuctLambdaE};

\addplot [const plot mark left, blue, very thick, dashdotted] table [x={0},y={1}]{\DuctLambdaOneE};

\addplot [const plot mark left, darkgreen, very thick, densely dashed] table [x={0},y={1}]{\DuctLambdaTwoE};

\addplot [only marks, black,
        error bars/.cd,
            x dir=both, x explicit,
            y dir=both, y explicit,
    ] table [
        x error plus=ex+,
        x error minus=ex-,
        y error minus=ey-,
        y error plus=ey+,
]{\AtlasE};

\node at (420.0,0.8) {\fcolorbox{black}{white}{$5 < \Delta y < 6$}};
\end{groupplot}
\end{tikzpicture}
\else 
\includegraphics[width = 14 cm]{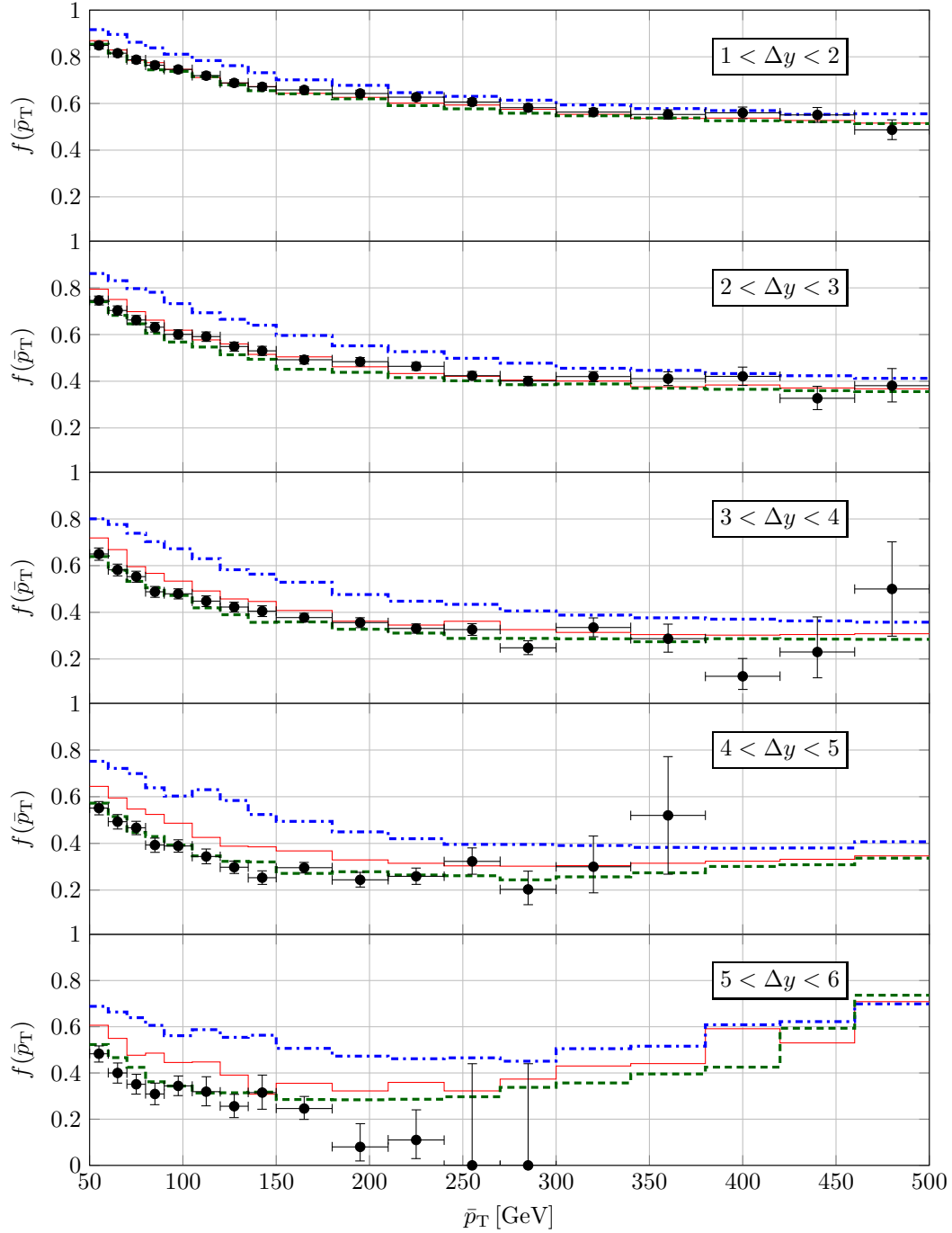}
\fi
\end{center}
\caption{
Gap fraction for various $\Delta y$ values using \textsc{Deductor} with $\Lambda$ ordering, showing the dependence on the shower start scale $\mu_\Ls$. The black points with error bars are Atlas data \cite{AtlasGap}. The red solid (middle) histogram uses $\mu_\Ls = (3/2)\, P_\LT^\textrm{Born}$, as in Fig.~\ref{fig:gapfractions}. The dash-dotted (higher) histogram uses $\mu_\Ls = P_\LT^\textrm{Born}$. The green dashed (lower) histogram uses $\mu_\Ls = 2\,P_\LT^\textrm{Born}$.
}
\label{fig:gapfractionsmus}
\end{figure}

For the results from \textsc{Deductor} with $\Lambda$ ordering in Fig.~\ref{fig:gapfractions}, we set the shower start scale to $\mu_\Ls = (3/2) P_\LT^{\rm Born}$, as in Eq.~(\ref{eq:musforLambda}). This choice was based on the idea that the $P_\LT$ of the first shower splitting should range up to $P_\LT^{\rm Born}$ and $\Lambda$ for in a shower splitting is larger than $P_\LT$ in the splitting. We could, however, choose a different value for $\mu_\Ls$. We found in Fig.~\ref{fig:JetKfactorsR04altmus} that the one jet cross section is rather insensitive to $\mu_\Ls$. However, the calculated gap fraction is quite sensitive to 
$\mu_\Ls$, as we see in Fig.~\ref{fig:gapfractionsmus}, where we compare the gap fraction $f$ calculated with  $\mu_\Ls = P_\LT^{\rm Born}$ and $\mu_\Ls = 2\,P_\LT^{\rm Born}$ to $f$ calculated with $\mu_\Ls = (3/2) P_\LT^{\rm Born}$.

There is not much change between $\mu_\Ls = (3/2) P_\LT^{\rm Born}$ and $\mu_\Ls = 2\, P_\LT^{\rm Born}$ because, in the $\Lambda$ ordered shower, the splitting kernel restricts the splitting $P_\LT$ to be no greater than $P_\LT^{\rm Born}$ in order to ensure that the scattering that initiates the shower is the hardest scattering, in the sense of $P_\LT$, in the event. 

The result obtained with $\mu_\Ls = P_\LT^{\rm Born}$ deviates substantially from the data, suggesting that the physical motivation for the choice $\mu_\Ls = (3/2) P_\LT^{\rm Born}$ was sensible. 

\section{Conclusions}
\label{sec:conclusions}

We have modified \textsc{Deductor} to include 
\begin{itemize}

\item the ability to change between the default virtuality based ordering parameter $\Lambda$ and a transverse momentum ordering parameter $k_\LT$;

\item the ability to include an estimate of the effect of non-perturbative physics on an infrared safe cross section by adding a simple underlying event and sending events to \textsc{Pythia} for hadronization;

\item a summation of threshold logarithms according to the lowest order version of the formulation in Ref.~\cite{NSallorder}.

\end{itemize}
A version of the threshold factors was given in Ref.~\cite{DuctThreshold}, but in this version, there was an unphysical dependence on low momentum physics. With this earlier formulation one had to use an {\it ad hoc} cutoff parameter to get physically reasonable results. The all-order formulation of Ref.~\cite{NSallorder} shows how the threshold factor should appear.

The formulation of Ref.~\cite{NSallorder} also includes the appropriate subtractions to match the shower to the perturbative matrix elements for a hard scattering that initiates the shower. For a lowest order shower, with splitting functions at order $\as$, this means that the shower is matched to the perturbative calculation at next-to-leading order. Thus complete cross sections with a leading order shower naturally contain a probability preserving parton shower, a factor that sums threshold logarithms at this order, and NLO matching to the hard scattering matrix elements. 

The treatment in this paper includes the probability preserving shower and the threshold factor. However, we lack the computer code for the NLO matching. Matching would improve the precision of the results and would remove much of the sensitivity to the starting conditions of the shower and to the choice of $\Lambda$ or $k_\LT$ ordering. Without matching, we can look for large effects, but we miss the finer details. Nevertheless, we find that there are interesting large effects. 

We first investigated the one jet inclusive cross section at large values of the jet transverse momentum $P_\LT$. We find that non-perturbative physics is not important for this cross section. We find that the difference between a jet and a parton is numerically highly significant: the jet cross section is substantially affected by parton showering. The cross section after showering is smaller than it was before showering, particularly for smaller values of the jet radius $R$. This effect depends on what is built into the parton shower. We see a more pronounced effect with \textsc{Deductor} than with \textsc{Dire} or \textsc{Pythia}. It appears that \textsc{Deductor} has more wide angle emissions that are away from the strict soft emission limit. We also find that the threshold factor is important at large $P_\LT$. In fact, with \textsc{Deductor}, the threshold factor is large enough to cancel the loss of jet cross section from showering, leaving a cross section that is close to the NLO jet cross section. Other parton shower programs lack the threshold summation. We also find that with $k_\LT$ ordering, the threshold factor is smaller than with $\Lambda$ ordering, but is still substantial.

We next investigated the gap fraction: in events with two two high-$P_\LT$ jets that are separated by a large difference $\Delta y$ in rapidity, this is the fraction of events in which there are no jets above a minimum $p_\LT$ in the rapidity interval between the jets. There are a number of subtle theoretical issues associated with the gap fraction, so it is not obvious how well a parton shower should describe the physics. We find that non-perturbative physics is not important for the gap fraction. We then find that \textsc{Deductor} with $\Lambda$ ordering reproduces reasonably well Atlas data for the gap fraction. With $k_\LT$ ordering, the  agreement is also reasonably good. Finally, an NLO perturbative calculation using $\mu_\LR = \mu_\LF = 2 \bar p_\LT$ without showering matches the experimental results well when $\bar p_\LT/p_\LT^{\rm min}$ and $\Delta y$ are not too large and is not far off in the entire range of $\bar p_\LT/p_\LT^{\rm min}$ and $\Delta y$ for which there are data.

\begin{acknowledgments}
This work was supported in part by the United States Department of Energy under grant DE-SC0011640. Some of the project was carried out while the authors were at the Munich Institute for Astro- and Particle Physics program ``Automated, resummed and effective: precision computations for the LHC and beyond.'' We thank the MIAPP for providing a stimulating research environment. We thank Mrinal Dasgupta and Gavin Salam for helpful conversations about the relation between parton evolution and jets. We thank Werner Vogelsang for providing us with the de Florian, Hinderer, Mukherjee, Ringer, Vogalsang code for threshold corrections to the jet cross section. This work benefited from access to the University of Oregon high performance computer, Talapas.
\end{acknowledgments}

\appendix

\section{The threshold factor}
\label{sec:thresholdmods}

The threshold factor, as given in Eq.~(\ref{eq:UVexponential00}) is
\begin{equation}
\label{eq:UVexponential00bis}
\cU_\cV(\mu_\Lf^2,\mu_\scH^{2})
=\mathbb{T} \exp\!\left(
\int_{\mu_\Lf^2}^{\mu_\scH^{2}}\!\frac{d\mu^2}{\mu^2}\,\cS_\cV(\mu^2)
\right)
\;.
\end{equation}
This is in the notation of Ref.~\cite{NSallorder}. It is instructive to see how Eq.~(\ref{eq:UVexponential00bis}) is translated to the notation of Ref.~\cite{DuctThreshold}. The first step is to approximate Eq.~(\ref{eq:UVexponential00bis}) by using only the first order version, $\cS_\cV^{(1)}(\mu^2)$, of $\cS_\cV(\mu^2)$. According to Eq.~(D.26) of Ref.~\cite{NSallorder}, we have
\begin{equation}
\label{eq:SVdefencore1}
\cS_\cV^{(1)}(\mur) = \left[\mus\,\frac{\partial}{\partial\mus}\cV_{\!\rm ao}^{(1)}(\mur,\mus)\right]_{\mus=\mur}
  - \cF^{-1}(\mur)\ \left[\mur\, \frac{d\cF(\mur)}{d\mur}\right]
  \;.
\end{equation}
Here $\cV_{\!\rm ao}^{(1)}(\mur,\mus)$ is the first order version of the operator called $\cV$ in Ref.~\cite{NSallorder}. Since there is a different operator defined in Ref.~\cite{DuctThreshold} with the name $\cV$, we add a subscript ``ao'' to indicate that it is the operator called $\cV$ in the {\em all order} notation. This operator has two scale arguments: $\mur$ is the factorization and renormalization scale while $\mus$ is the shower scale, to be discussed below. After differentiating with respect to $\mus$, we set $\mus = \mur$. The operator $\cF(\mur)$ multiplies by the right product of parton distribution functions and a parton luminosity factor to make a cross section \cite{DuctThreshold, NSallorder}. The parton distribution functions depend on $\mur$. Differentiating with respect to $\mur$ gives the evolution kernel for the parton distribution functions convolved with the parton distribution functions. Here, working to lowest order, we would use the order $\as$ evolution kernel.

The operator $\cV_{\!\rm ao}^{(1)}(\mur,\mus)$ has three parts,
\begin{equation}
\cV_{\!\rm ao}^{(1)}(\mur,\mus)
= \cV_{\!\rm ao}^{(1,0)}(\mur,\mus) + \cV_{\!\rm ao}^{(0,1)}(\mur,\mus)
+ \cV_{\!\rm ao}^{{\rm pdf}}(\mur)
\;.
\end{equation}
Here $\cV_{\!\rm ao}^{{\rm pdf}}(\mur)$ is related to the definition of the parton distribution functions. It does not depend on $\mus$, so it does not contribute once we differentiate with respect to $\mus$. The operator $\cV_{\!\rm ao}^{(1,0)}$ comes from real emission graphs and $\cV_{\!\rm ao}^{(0,1)}$ comes from virtual graphs. These operators leave the number of partons, their momenta, and their flavors unchanged. They are defined by specifying what $\sbra{1} \cV_{\!\rm ao}^{(1)}$ is, where multiplying by $\sbra{1}$ indicates making a completely inclusive measurement, in which we sum over flavors, integrate over momenta, and take the trace over spins and colors. The definition (at first order) is
\begin{equation}
\begin{split}
\sbra{1}\cV_{\!\rm ao}^{(1,0)}(\mur,\mus)
={}& \sbra{1}
\cF(\mur)  \cD^{(1,0)}(\mur,\mus)  \cF^{-1}(\mur)
\;,
\\
\sbra{1}\cV_{\!\rm ao}^{(0,1)}(\mur,\mus)
={}& \sbra{1}
 \cD^{(0,1)}(\mur,\mus) 
\;.
\end{split}
\end{equation}
The operator $\cD^{(1,0)}(\mur,\mus)$ represents real emission Feynman graphs in which we integrate over the scale of the emission up to an upper limit $\mu_\Ls^2$. For a $\Lambda$ ordered shower, the emission scale is $\Lambda^2$ as defined in Eq.~(\ref{eq:Lambdadef}). Then the integrals in $\cD^{(1,0)}(\mur,\mus)$ contain a factor $\Theta(\Lambda^2 - \mu_\Ls^2)$. Differentiating with respect to $\mu_\Ls^2$ as in Eq.~(\ref{eq:SVdefencore1}) then gives a factor $\delta(\Lambda^2 - \mu_\Ls^2)$. Similarly $\cD^{(0,1)}(\mur,\mus)$ represents virtual Feynman graphs in which we integrate over a scale variable $\Lambda^2$ \cite{DuctThreshold} with a factor $\Theta(\Lambda^2 < \mu_\Ls^2)$. Again, differentiation gives a factor $\delta(\Lambda^2 - \mu_\Ls^2)$.

In the notation of Ref.~\cite{DuctThreshold}, the names are different: 
\begin{equation}
\begin{split}
\label{eq:V01}
\left[\mus\,\frac{\partial}{\partial\mus}\cV_{\!\rm ao}^{(1,0)}(\mur,\mus)\right]_{\mus=\mur} 
\to{}& \cV(\mu^2)
\;,
\\
\left[\mus\,\frac{\partial}{\partial\mus}\cV_{\!\rm ao}^{(0,1)}(\mur,\mus)\right]_{\mus=\mur}
\to{}& 
- \{\cS(\mu^2) - \cS_{\mi \pi}(\mu^2)\}
+ \cF^{-1}(\mur)\ \left[\mur\, \frac{d\cF(\mur)}{d\mur}\right]
\;.
\end{split}
\end{equation}
There are two things to note. First, as defined in Ref.~\cite{DuctThreshold}, $\cS(\mu^2)$ includes both the contribution from virtual graphs, called $\cS^{\rm pert}(\mu^2)$, and the contribution from the evolution of the parton distribution functions that we remove in Eq.~(\ref{eq:V01}). See Eq.~(\ref{eq:VfromVpertandFXX}) below. Second, some of the virtual graphs have an imaginary part. Then $\cS(\mu^2)$ includes a contribution $\cS_{\mi \pi}(\mu^2)$ from the imaginary parts. However, $\cS_{\mi \pi}(\mu^2)$ does not contribute to $\cV_{\!\rm ao}^{(0,1)}$ because $\sbra{1}\cS_{\mi \pi}(\mu^2) = 0$. We include only the real part of the one loop graphs in the exponent of the threshold factor.

We are thus able to represent the threshold factor $\cU_\cV$ from Eq.~(\ref{eq:UVexponential00}) in the notation of ref.~\cite{DuctThreshold} and our earlier papers. We use $\Lambda^2$ defined in Eq.~(\ref{eq:Lambdadef}) as the hardness scale $\mu^2$ and use the shower time
\begin{equation}
t = \log(Q_0^2/\Lambda^2)
\end{equation}
as the integration variable instead of $\mu^2$. This gives the representation
\begin{equation}
\label{eq:UVexponential02}
\cU_\cV(t_\Lf^2,t_\scH^{2})
=\mathbb{T} \exp\!\left(
\int_{t_\Lf^2}^{t_\scH^{2}}\!dt\,
\left[
\cV(t) - \left\{\cS(t) - \cS_{\mi\pi}(t)\right\}
\right]
\right)
\;.
\end{equation}
Here the operator $\cS(\mu^2)$ has two contributions: 
\begin{equation}
\begin{split}
\label{eq:VfromVpertandFXX}
\cS(t)
 ={}& 
\cS^{\rm pert}(t)
- {\cal F}(t)^{-1}\left[\frac{d}{dt}\,{\cal F}(t)\right]
\;.
\end{split}
\end{equation}
The operator $\cS^{\rm pert}(t)$ is calculated from one loop virtual Feynman graphs. We remove the contribution $\cS_{\mi\pi}(\mu^2)$ from the imaginary part of the one loop graphs.

We could simply use $\cS(t)$ and $\cV(t)$ as given in Ref.~\cite{DuctThreshold} to construct the threshold factor (\ref{eq:UVexponential01}). However, we have found that some of the integrations that go into these operators can be performed so that they are accurate in a wider range of the kinematic variables compared to Ref.~\cite{DuctThreshold}. Thus we use the improved versions of $\cS(t)$ and $\cV(t)$ in \textsc{Deductor} v.~2.1.1. We explain the changes relative to Ref.~\cite{DuctThreshold} in the subsections that follow.


\subsection{Initial state virtual contribution}

In this subsection, we examine the contribution to $\cS$ from a virtual graph in which a gluon is emitted from the initial state line and absorbed by a final state line. We modify the calculation in Appendix C.3 of Ref.~\cite{DuctThreshold} to make it accurate in a wider range, as explained below. 

\subsubsection{The momenta}

The exchanged gluon carries momentum $q$ from line ``a,'' which carries momentum $p_\La$ into the graph, to line $k$, which carries momentum $p_k$ out of the graph, so that, inside the loop, line ``a'' carries momentum $p_\La - q$ and line $k$ carries momentum $p_k - q$.

We denote the components of $q$ and $p_k$ in the rest frame of $Q$ 
\begin{equation}
\begin{split}
\label{eq:qpknullplane}
q ={}& (1-z')\, p_\La + \xi\, p_\Lb + q_\perp
\;,
\\
p_k ={}& (1 - z_k)\, p_\La + \xi_k\, p_\Lb + p_{k,\perp}
\;,
\end{split}
\end{equation}
where $0 < z_k < 1$ and
\begin{equation}
\label{eq:xik1}
\xi_k = \frac{\bm p_{k,\perp}^{\,2}}{(1 - z_k)Q^2}
\;.
\end{equation}
Here we need $(1-z_k) > 0$ so that $p_k$ has positive + momentum (momentum along $p_\La$). Also, we need $(1-z_k) < 1$ because no final state particle can have more + momentum than is contained in $p_\La$. Then also
\begin{equation}
\begin{split}
\label{eq:internallinesnullplane}
p_\La - q ={}& z'\, p_\La - \xi\, p_\Lb - q_\perp
\;,
\\
p_k - q ={}& (z' - z_k)\, p_\La + (\xi_k - \xi)\, p_\Lb + p_{k,\perp} - q_\perp
\;.
\end{split}
\end{equation}

\subsubsection{The integral}

We start with integral representing the exchange in Coulomb gauge,
\begin{equation}
\begin{split}
\label{eq:Gak0}
\int\!dt\ [S^{\LL}_{\La k}&(\{p,f\}_{m};t) 
+ S^{\LL}_{k \La}(\{p,f\}_{m};t)]
\\
={}&
\mi\, \frac{\as}{(2\pi)^{3}}\,
\int\! d^4 q\
\frac{2 J_\La(p_\La,q)\cdot D(q)\cdot J_k(p_k,q)}
{(-(q - p_\La)^2 - \mi\epsilon)(-(q - p_k)^2 - \mi \epsilon)(q^2 + \mi\epsilon)}\,
\;.
\end{split}
\end{equation}
Here, following the notation in Ref.~\cite{DuctThreshold}, the superscript L refers to a virtual graph to the left of the final state cut. In Ref.~\cite{DuctThreshold} we used the eikonal approximation, in which $J_\La(p_\La,q) \to 2 p_\La$, $J_k(p_k,q) \to 2 p_k$, $-(q - p_\La)^2 \to 2 q \cdot p_\La$, and $-(q - p_k)^2 \to 2 q \cdot p_k$. This is a good approximation if $q$ is small, but $q$ is perhaps not always small, and if we make this approximation we may even allow $q$ to become much larger than it becomes in the exact integral. For this reason, we do not make the eikonal approximation to start with here.

For $S^{\LL}_{\La k}$ use the dimensionless integration variable $y = -(q - p_\La)^2/Q^2$, which is used to define the shower time for the virtual splitting through $t = - \log[y\, Q^2/(2 p_\La \cdot Q_0)]$, where $Q_0$ is the total momentum of the final state at the start of the shower. As in Ref.~\cite{DuctThreshold}, we use the approximation $y \ll 1$. The calculation in Ref.~\cite{DuctThreshold} also used the approximation $y \ll 1 - \cos \theta_{\La k}$. However, it is certainly possible to have a final state parton $k$ that is very nearly collinear with the momentum $p_\La$ of the incoming beam parton. For this reason, in this appendix we seek to modify the calculation in Ref.~\cite{DuctThreshold} so that it is valid also when $1 - \cos \theta_{\La k} \lesssim y$. We thus suppose that $p_k$ is nearly collinear with $p_\La$ and concentrate on the integration region in which $q$ is nearly collinear with $p_\La$. In Ref.~\cite{DuctThreshold}, we first performed the integration over $q^0$ by contour integration, then separated $S^{\LL}_{\La k}$ and $S^{\LL}_{k \La}$. In $S^{\LL}_{\La k}$, we inserted a factor $\delta(y + (q - p_\La)^2/Q^2)$  to eliminate one dimension of the integration over $\vec q$, then performed the rest of the integration over $\vec q$ analytically in the small $y$ limit. We will see that a very simple change is needed in the integral that represents $S^{\LL}_{\La k}(\{p,f\}_{m};t)$ in \cite{DuctThreshold}. To motivate this change, the most straightforward path would be to expand the denominators in Eq.~(\ref{eq:Gak0}) in powers of the angles of $\vec q$ and $\vec p_k$ with respect to $\vec p_a$, then perform the $q^0$ integration and proceed along the lines of Ref.~\cite{DuctThreshold}. However, we find it more instructive to introduce null-plane coordinates for the momenta, as we have done in Eqs. (\ref{eq:qpknullplane}) and (\ref{eq:internallinesnullplane}). Then in the collinear limits, the component of $q$ along $p_\La$ is large while the component along $p_\Lb$ is small. We then start by performing the integral over the small component of $q$ by contour integration.

\subsubsection{Performing the $\xi$ integration}

We are particularly interested in the denominators in Eq.~(\ref{eq:Gak0}):
\begin{equation}
\begin{split}
\frac{1}{q^2 + \mi \epsilon} ={}&
\frac{1}{(1-z') \xi Q^2 - \bm q_\perp^2 + \mi \epsilon}
\;,
\\
\frac{1}{-(q - p_\La)^2 - \mi\epsilon} ={}&
\frac{1}{z' \xi Q^2 + \bm q_\perp^2 - \mi \epsilon} 
\;,
\\
\frac{1}{-(q - p_k)^2 - \mi\epsilon} ={}&
\frac{1}{(z' - z_k)(\xi - \xi_k)Q^2 + (\bm p_{k,\perp} - \bm q_\perp)^2 - \mi \epsilon} 
\;.
\end{split}
\end{equation}
We will integrate over $z'$. We examine the integration region in which the components along $p_\La$ of $q$, $p_\La - q$, and $p_k - q$ are all positive. That is, we integrate over the region $z_k < z' < 1$. Other regions for $z'$ give qualitatively different results. We first integrate over $\xi$, noting that in the region $z_k < z' < 1$, the first denominator factor has a pole in the lower half $\xi$ plane, while the other two poles are in the upper half $\xi$ plane. We close the $\xi$ contour in the lower half plane so that we pick up the pole at $\xi = \xi_q$ where
\begin{equation}
\xi_q = \frac{\bm q_\perp^2}{(1-z') Q^2}
\;.
\end{equation}
This gives
\begin{equation}
\begin{split}
\label{eq:Gak1}
\int\!dt\ [S^{\LL}_{\La k}&(\{p,f\}_{m};t) 
+ S^{\LL}_{k \La}(\{p,f\}_{m};t)]
\\
={}&
\frac{\as}{(2\pi)^{2}}\,
\int_{z_k}^1\!\frac{dz'}{1-z'}\,
\int\! d\bm q_\perp\
\frac{J_\La(p_\La,q)\cdot D(q)\cdot J_k(p_k,q)}
{D_\La D_k}\,
\;.
\end{split}
\end{equation}
Here
\begin{equation}
\begin{split}
\label{eq:DaDk1}
D_\La ={}& z' \xi_q Q^2 + \bm q_\perp^2 
\;,
\\ 
D_k ={}&
(z' - z_k)(\xi_q - \xi_k)Q^2 + (\bm p_{k,\perp} - \bm q_\perp)^2
\;.
\end{split}
\end{equation}

\subsubsection{Structure of the result}

The two denominators are
\begin{equation}
\begin{split}
\label{eq:DaDk2}
D_\La ={}& \frac{1}{1-z'}\,\bm q_\perp^2 
\;,
\\ 
D_k ={}&
\frac{z' - z_k}{1-z'}\,\bm q_\perp^2
- \frac{z' - z_k}{1-z_k}\,\bm p_{k,\perp}^2 
+ (\bm p_{k,\perp} - \bm q_\perp)^2
\;.
\end{split}
\end{equation}

We define the virtuality variable $y$ by
\begin{equation}
y Q^2 = D_\La
\;.
\end{equation}
Then there is a relation between $\bm q_\perp^2$, $y$, and $z$ 
\begin{equation}
\bm q_\perp^2 = (1-z') y Q^2
\;.
\end{equation}

We could have tried this using the eikonal approximation. Then
\begin{equation}
\begin{split}
\label{eq:eikonaldenoms}
-(q-p_\La)^2 ={}& 2 q \cdot p_\La - q^2 \to 2 q \cdot p_\La
\;,
\\
-(q-p_k)^2 ={}& 2 q \cdot p_k - q^2 \to 2 q \cdot p_k
\;.
\end{split}
\end{equation}
We then evaluate this by setting $\xi$ to $\xi_q$. But with $\xi \to \xi_q$, $q^2 \to 0$. Thus we get exactly the same result for $D_\La$ and $D_k$. However, if we make the replacements Eq.~(\ref{eq:eikonaldenoms}) before performing the $\xi$ integration, the locations of poles can shift between the upper and lower half $\xi$ planes, so that the results change. 

The result for $D_k$ emerges in the form
\begin{equation}
D_k = \frac{1-z'}{1-z_k}\,\bm p_{k,\perp}^2
+ \frac{1-z_k}{1-z'}\,\bm q^2
- 2 \bm q_\perp \cdot \bm p_{k,\perp}
\;.
\end{equation}
This is the same as the result that we had, simply expanded differently. In this form, it is evident that $D_k$ is linear in $q$ and $p_k$. That is, $D_k$ is proportional to $\lambda$ under the scaling $(1-z_k) \to \lambda (1-z_k)$, $\bm p_{k,\perp} \to \lambda \bm p_{k,\perp}$. It is also proportional to $\lambda$ under the scaling $(1-z) \to \lambda (1-z)$, $\bm q_{\perp} \to \lambda \bm q_{\perp}$.

It is perhaps also worthwhile to note that
\begin{equation}
D_k = (1-z_k)(1-z')\left(
\frac{\bm p_{k,\perp}}{1-z_k}
-\frac{\bm q_\perp}{1-z'}\right)^2
\;.
\end{equation}
With this form, we see that $D_k$ is invariant under a null-plane boost: $\bm p_{k,\perp} \to \bm p_{k,\perp} + (1-z_k)\bm v$,  $\bm q_{\perp} \to \bm q_{\perp} + (1-z')\bm v$. It is also invariant under a $z$ boost: $(1-z_k) \to \lambda (1-z_k)$, $(1-z') \to \lambda (1-z')$.

\subsubsection{Results for the integral}

We can now make use of our results in Ref.~\cite{DuctThreshold}. Our integral is in Appendix C.3. There, we used the eikonal approximation, where we should have used the full energy denominators. However, we have seen that using the full energy denominators gives the same result as using the eikonal approximation except that using the full energy denominators tells us where to put bounds on the integration over the component of $q$ along $p_\La$.

From Eq.~(C.63) of Ref.~\cite{DuctThreshold}, we have
\begin{equation}
\begin{split}
\label{eq:Sab}
S^{\LL}_{\La\Lb}(\{p,f\}_{m};t)
\approx{}&
\frac{\as}{2\pi}\,\left[
- 1
+ \mi \pi
\right]
\;.
\end{split}
\end{equation}
Also, (noting that $2|\vec p_\La| = E_Q$ in the result from Eq.~(C.91) of Ref.~\cite{DuctThreshold}), we have
\begin{equation}
\label{eq:Saa}
S^{\LL}_{\La \La}(\{p,f\}_{m};t) = 
-\frac{\as}{2\pi}\,
\left(
\frac{\gamma_{f_\La}}{2 C_{f_\La}}
+\log\left(y\right)
+ 1
\right)
\;.
\end{equation}

In Appendix C.3 of Ref.~\cite{DuctThreshold}, there are two parts of the result for $S^{\LL}_{\La k}$,
\begin{equation}
\begin{split}
\label{eq:CoulombSplitak}
S^{\LL}_{\La k}(\{p,f\}_{m};t) 
={}& 
S^{\LL}_{\La k}(\{p,f\}_{m};t;{\rm dipole})  +
S^{\LL}_{\La\La}(\{p,f\}_{m};t;{\rm eikonal})
\;.
\end{split}
\end{equation}

The first part is
\begin{equation}
\begin{split}
\label{eq:GlkRe}
S^{\LL}_{\La k}(\{p,f\}_{m};t;{\rm dipole})
\approx{}&
-\frac{\as}{2\pi}\,\int_{1 - M/|\vec p_\La|}^{1 - y}\!\frac{dz}{\sqrt{(1-z)^2 +  y^2/\psi_{\La k}^2}}
\\
={}& 
-\frac{\as}{2\pi}
\log\!\left(
\frac{M/|\vec p_\La| + \sqrt{M^2/|\vec p_\La|^2 + y^2/\psi_{\La k}^2}}
{y  \left(1 + \sqrt{1 + 1/\psi_{\La k}^2}\right)}
\right)
\;,
\end{split}
\end{equation}
where, according to Eq.~(A.13) of Ref.~\cite{DuctThreshold},
\begin{equation}
\label{eq:psiak}
\psi_{ak} = \frac{1 -  \cos\theta_{ak}}{\sqrt{8(1 +  \cos\theta_{ak})}}
\;.
\end{equation}
This definition gives
\begin{equation}
1 + \sqrt{1 + 1/\psi_{\La k}^2} = \frac{4}{1-\cos\theta_{\La k}}
\;.
\end{equation}
The variable $z$ in Eq.~(\ref{eq:GlkRe}) is almost the same as the variable $z'$ of this appendix: $z = z' - y$. The $z$ integration has a lower bound $1 - M/|\vec p_\La|$. In Ref.~\cite{DuctThreshold}, we took $M$ to be a large positive number. However, we now recognize that, at least when $1-\cos\theta_{\La k} \ll 1$, the lower bound on the $z$ integration should be $z_k + y$, which we can approximate by just $z_k$. Thus we should set $M/|\vec p_\La| = 1 - z_k$. 
This gives
\begin{equation}
\begin{split}
\label{eq:Saknew}
S^{\LL}_{\La k}(\{p,f\}_{m};t;{\rm dipole})
={}& 
-\frac{\as}{2\pi}
\log\!\left(
\frac{1-z_k + \sqrt{(1-z_k)^2 + y^2/\psi_{\La k}^2}}
{4y /(1-\cos\theta_{\La k})}
\right)
\;.
\end{split}
\end{equation}

For $S^{\LL}_{ll}(\{p,f\}_{m};t;{\rm eikonal})$, we have from Eq.~(C.58) of Ref.~\cite{DuctThreshold}
\begin{equation}
\begin{split}
\label{eq:Glleik6}
S^{\LL}_{\La\La}(\{p,f\}_{m};t;{\rm eikonal}) 
={}&
\frac{\as}{2\pi}\,
\int_{1- M/|\vec p_La|}^{1 -y}\! dz\ 
\frac{(1-z) - y}
{(1-z)^2}
\\={}&
\frac{\as}{2\pi}\,
\left[
\log\left(
\frac{M/|\vec p_\La|}{y}
\right)
-1 
+ \frac{y}{M/|\vec p_\La|}
\right]
\;.
\end{split}
\end{equation}
We set $M/|\vec p_\La| = 1 - z_k$ to obtain
\begin{equation}
\begin{split}
\label{eq:Glleik7}
S^{\LL}_{\La\La}(\{p,f\}_{m};t;{\rm eikonal}) 
={}&
\frac{\as}{2\pi}\,
\left[
\log\left(
\frac{1 - z_k}{y}
\right)
-1 
+ \frac{y}{1 - z_k}
\right]
\;.
\end{split}
\end{equation}

When we add $S^{\LL}_{\La k}(\{p,f\}_{m};t;{\rm dipole})$ and $S^{\LL}_{\La\La}(\{p,f\}_{m};t;{\rm eikonal})$ according to Eq.~(\ref{eq:CoulombSplitak}), we obtain
\begin{equation}
\begin{split}
\label{eq:Saktot1}
S^{\LL}_{\La k}(\{p,f\}_{m};t)
={}& 
-\frac{\as}{2\pi}
\left\{
\log\!\left(
\frac{1-z_k + \sqrt{(1-z_k)^2 + y^2/\psi_{\La k}^2}}
{4(1-z_k)/(1-\cos\theta_{\La k})}
\right)
+ 1 -  \frac{y}{1 - z_k}
\right\}
\;.
\end{split}
\end{equation}
We treat $z_k$ as some finite number, not close to 1. We suppose that $y\ll 1$ but we do not assume that $y$ is small compared to $\psi_{\La k}$. Then we can neglect $y/(1-z_k)$. Also, we can simplify the argument of the logarithm. Then we get
\begin{equation}
\begin{split}
\label{eq:Saktot}
S^{\LL}_{\La k}(\{p,f\}_{m};t)
={}& 
-\frac{\as}{2\pi}
\left\{
\log\!\left(
\frac{1 + \sqrt{1 + y^2/[\psi_{\La k}^2(1-z_k)^2]}}
{4/(1-\cos\theta_{\La k})}
\right)
+ 1
\right\}
\;.
\end{split}
\end{equation}
When $y \ll 1-\cos\theta_{\La k}$ we recover the result of Ref.~\cite{DuctThreshold}, but now we have a result that works also for $1-\cos\theta_{\La k} \gtrsim y$.

For later use, we write the log term as an integral :
\begin{equation}
\begin{split}
\label{eq:Sakintegral}
\log\!\left(
\frac{1 + \sqrt{1 + y^2/[\psi_{\La k}^2(1-z_k)^2]}}
{4/(1-\cos\theta_{\La k})}
\right)
\approx{}& 
\int_{z_k}^{1/(1+y)}\!dz
\left[
\frac{1}{\sqrt{(1-z)^2 +  y^2/\psi_{\La k}^2}} - \frac{1}{1-z}
\right]
\;.
\end{split}
\end{equation}
(Here we have replaced $1-y$ by $1/(1+y)$ for  $y \ll 1$ in the upper limit of the integral.)

\subsubsection{Assembling the result}

We now write the total contribution to $\cS$ from virtual emissions from the initial state parton ``a'' as (See Eqs.~(C.3) and (C.4) of Ref.~\cite{DuctThreshold})
\begin{equation}
\begin{split}
\label{eq:Spert1}
\cS_{\La}^{\rm pert}(t)&\sket{\{p,f,c',c\}_{m}} 
\\={}& 
\bigg\{
\sum_{k \ne \La,\Lb}
 S^\LL_{ak}
\big(
[(\bm{T}_\La\cdot \bm{T}_k)\otimes 1]
+ [1 \otimes (\bm{T}_\La\cdot \bm{T}_k)]
\big)
\\&\ \; 
+
{\mathrm {Re}}\, S^\LL_{ab}\,
\big(
[(\bm{T}_\La\cdot \bm{T}_\Lb)\otimes 1]
+ [1 \otimes (\bm{T}_\La\cdot \bm{T}_\Lb)]
\big)
\\&\ \; 
+
{\mathrm {Im}}\, S^\LL_{ab}\,
\big(
[(\bm{T}_\La\cdot \bm{T}_\Lb)\otimes 1]
- [1 \otimes (\bm{T}_\La\cdot \bm{T}_\Lb)]
\big)
\\&\ \; 
+
S^\LL_{aa}
\big(
[(\bm{T}_\La\cdot \bm{T}_\La)\otimes 1]
+ [1 \otimes (\bm{T}_\La\cdot \bm{T}_\Lb)]
\big)
\bigg\}
\\&\times
\sket{\{p,f,c',c\}_{m}}
\;.
\end{split}
\end{equation}
We use Eq.~(\ref{eq:Saktot}) for $S^\LL_{ak}$ and Eq.~(\ref{eq:Saa}) for $S^\LL_{aa}$ and Eq.~(\ref{eq:Sab}) for $S^\LL_{ab}$. Then
\begin{equation}
\begin{split}
\label{eq:Spert2}
\cS_{\La}^{\rm pert}(t)&\sket{\{p,f,c',c\}_{m}} 
\\={}& 
\frac{\as}{2\pi} \bigg\{
\sum_{k \ne \La,\Lb}
\left[
-\log\!\left(
\frac{1 + \sqrt{1 + y^2/[\psi_{\La k}^2(1-z_k)^2]}}
{4/(1-\cos\theta_{\La k})}
\right)
- 1
\right]
\\&\qquad\quad\times
\big(
[(\bm{T}_\La\cdot \bm{T}_k)\otimes 1]
+ [1 \otimes (\bm{T}_\La\cdot \bm{T}_k)]
\big)
\\&\qquad 
-
\big(
[(\bm{T}_\La\cdot \bm{T}_\Lb)\otimes 1]
+ [1 \otimes (\bm{T}_\La\cdot \bm{T}_\Lb)]
\big)
\\&\qquad 
+
\mi \pi\,
\big(
[(\bm{T}_\La\cdot \bm{T}_\Lb)\otimes 1]
- [1 \otimes (\bm{T}_\La\cdot \bm{T}_\Lb)]
\big)
\\&\qquad 
-
\left[
\frac{\gamma_a}{2 C_a}
+\log\left(y\right)
+ 1
\right]
\big(
[(\bm{T}_\La\cdot \bm{T}_\La)\otimes 1]
+ [1 \otimes (\bm{T}_\La\cdot \bm{T}_\Lb)]
\big)
\bigg\}
\\&\times
\sket{\{p,f,c',c\}_{m}}
\;.
\end{split}
\end{equation}
The terms $-1$ times color operators cancel because $\sum_k (\bm{T}_\La\cdot \bm{T}_k)$ = 0. Also $(\bm{T}_\La\cdot \bm{T}_\La) = C_a 1$. Also, for $k = \Lb$ we have $1/\psi_{\La\Lb}  = 0$, so
\begin{equation}
\log\!\left(
\frac{1 + \sqrt{1 + y^2/[\psi_{\La \Lb}^2(1-z_k)^2]}}
{4/(1-\cos\theta_{\La \Lb})}
\right) 
= 0
\;.
\end{equation}
This means that the sum over $k$ in the first term can include $k = \Lb$. Thus
\begin{equation}
\begin{split}
\label{eq:Spert3}
\cS^{\rm pert}_{\La}(t)&\sket{\{p,f,c',c\}_{m}} 
\\={}& 
\frac{\as}{2\pi} \Bigg\{
-\sum_{k \ne \La}
\log\!\left(
\frac{1 + \sqrt{1 + y^2/[\psi_{\La k}^2(1-z_k)^2]}}
{4/(1-\cos\theta_{\La k})}
\right)
\\&\qquad\quad\times
\big(
[(\bm{T}_\La\cdot \bm{T}_k)\otimes 1]
+ [1 \otimes (\bm{T}_\La\cdot \bm{T}_k)]
\big)
\\&\qquad  
+
\mi \pi\,
\big(
[(\bm{T}_\La\cdot \bm{T}_\Lb)\otimes 1]
- [1 \otimes (\bm{T}_\La\cdot \bm{T}_\Lb)]
\big)
\\&\qquad  
- 2 C_a
\left[
\frac{\gamma_a}{2 C_a}
+\log\left(y\right)
\right]
[1\otimes 1]
\Bigg\}
\\&\times
\sket{\{p,f,c',c\}_{m}}
\;.
\end{split}
\end{equation}
To this we have to add the contribution from parton evolution, using Eqs.~(6.6), (6.7), and (6.9) of Ref.~\cite{DuctThreshold}. This gives us
\begin{equation}
\begin{split}
\label{eq:S1}
\cS_{\La}(t)&\sket{\{p,f,c',c\}_{m}} 
\\={}& 
\frac{\as}{2\pi} \Bigg\{
-\sum_{k \ne \La}
\log\!\left(
\frac{1 + \sqrt{1 + y^2/[\psi_{\La k}^2(1-z_k)^2]}}
{4/(1-\cos\theta_{\La k})}
\right)
\\&\qquad\quad\times
\big(
[(\bm{T}_\La\cdot \bm{T}_k)\otimes 1]
+ [1 \otimes (\bm{T}_\La\cdot \bm{T}_k)]
\big)
\\&\qquad  
+
\mi \pi\,
\big(
[(\bm{T}_\La\cdot \bm{T}_\Lb)\otimes 1]
- [1 \otimes (\bm{T}_\La\cdot \bm{T}_\Lb)]
\big)
\\&\qquad  
- 2 C_a
\left[
\frac{\gamma_a}{2 C_a}
+\log\left(y\right)
\right][1\otimes 1]
\\&\qquad  
+ \sum_{\hat a} 
\int_0^{1}\!dz\
\bigg(
\frac{1}{z}
P_{a\hat a}\!\left(z\right)
\,
\frac{f_{\hat a/A}(\eta_{\La}/z, yQ^2)}
{f_{a/A}(\eta_{\La}, yQ^2)}
-
\delta_{a\hat a}\,
\left[
\frac{2 C_a}{1-z}-\gamma_a
\right]
\bigg)
[1\otimes 1]
\Bigg\}
\\&\times
\sket{\{p,f,c',c\}_{m}}
\;.
\end{split}
\end{equation}
The terms proportional to $\gamma_a$ cancel. Also, we can use
\begin{equation}
2 C_a [1\otimes 1] = 
\big(
[(\bm{T}_\La\cdot \bm{T}_\La)\otimes 1]
+ [1 \otimes (\bm{T}_\La\cdot \bm{T}_\La)]
\big) 
=
-\sum_{k \ne a}
\big(
[(\bm{T}_\La\cdot \bm{T}_k)\otimes 1]
+ [1 \otimes (\bm{T}_\La\cdot \bm{T}_k)]
\big)
\end{equation}
to associate the $\log(y)$ term with the first line. This gives us
\begin{equation}
\begin{split}
\label{eq:S2}
\cS_{\La}(t)&\sket{\{p,f,c',c\}_{m}} 
\\={}& 
\frac{\as}{2\pi} \Bigg\{
-\sum_{k \ne \La,\Lb}
\log\!\left(
\frac{1 + \sqrt{1 + y^2/[\psi_{\La k}^2(1-z_k)^2]}}
{4y/(1-\cos\theta_{\La k})}
\right)
\\&\qquad\quad\times
\big(
[(\bm{T}_\La\cdot \bm{T}_k)\otimes 1]
+ [1 \otimes (\bm{T}_\La\cdot \bm{T}_k)]
\big)
\\&\qquad  
+
\mi \pi\,
\big(
[(\bm{T}_\La\cdot \bm{T}_\Lb)\otimes 1]
- [1 \otimes (\bm{T}_\La\cdot \bm{T}_\Lb)]
\big)
\\&\qquad  
+ \sum_{\hat a} 
\int_0^{1}\!\frac{dz}{z}\
\bigg(
P_{a\hat a}\!\left(z\right)
\,
\frac{f_{\hat a/A}(\eta_{\La}/z, yQ^2)}
{f_{a/A}(\eta_{\La}, yQ^2)}
-
\delta_{a\hat a}\,
\frac{2 z C_a}{1-z}
\bigg)
[1\otimes 1]
\Bigg\}
\\&\times
\sket{\{p,f,c',c\}_{m}}
\;.
\end{split}
\end{equation}

\subsection{Initial state real contribution}

We need $\cV_\La(t)$. From Eq.~(B.63) of Ref.~\cite{DuctThreshold}, we have the probability for emitting a parton from initial state gluon ``a'' at shower time $t$,  assuming $y \ll 1$, 
\begin{equation}
\begin{split}
\label{eq:V1}
\cV_{\La}(t)&\sket{\{p,f,c',c\}_{m}}
\\
={}&
\frac{\as}{2\pi}\,
\int_0^{1/(1+y)}\!\frac{dz}{z}\
\sum_{\hat a}
\frac{
f_{\hat a/A}(\eta_{\La}/z, y Q^2)}
{f_{a/A}(\eta_{\La}, y Q^2)}
\\&\times \Bigg\{
\frac{1}{2C_a}
\left(P_{\hat a a}(z)
- \delta_{\hat a a}\frac{2zC_a}{1-z}
\right)
\big([(\bm T_\La\cdot \bm T_\La)\otimes 1] 
+ [1 \otimes (\bm T_\La\cdot \bm T_\La)]\big)
\\&\qquad -
\delta_{\hat a a}
\sum_{k\ne \La}\,
z\,v(y,z,\theta_{\La k})\,
\big([(\bm T_\La\cdot \bm T_k)\otimes 1] + [1 \otimes (\bm T_\La\cdot \bm T_k)]\big)
\Bigg\}
\\&\times
\sket{\{p,f,c',c\}_{m}}
\;.
\end{split}
\end{equation}
Here
\begin{equation}
\label{eq:vdef}
v(y,z,\theta_{\La k}) = \int_0^{2\pi}\frac{d\phi}{2\pi}\,
\frac{\hat p_k\cdot \hat p_\La }
{\hat p_{m+1}\cdot \hat p_k 
+ yz\,\hat p_k\cdot \hat Q} 
\;.
\end{equation}
In Eq.~(B.63) of Ref.~\cite{DuctThreshold}, we used an approximate form for $v(y,z,\xi_{\La k})$, but here, we calculate it exactly:
\begin{equation}
\label{eq:vresult}
v(y,z,\theta_{\La k}) = 
\frac{z}{1-z}\,\frac{1+y}{1+zy}\ 
\frac{1-\delta}{\sqrt{(1-\delta)^2 + 4 x^2 \delta}}
+ \frac{1}{1 + z y} 
\;.
\end{equation}
Here
\begin{equation}
\label{eq:xdef}
x = \frac{zy}{1-z}
\end{equation}
and
\begin{equation}
\label{eq:deltadef}
\delta = (1+zy)\left(1+y\right)(1+\cos\theta_{\La k})/2
\;.
\end{equation}
We note that $x$ runs from 0 to 1 when $z$ ranges from 0 to its upper limit, $1/(1+y)$, and that $\delta > 0$. However, $\delta$ can be larger than 1 when $\theta_{\La k}$ is small.

We can simplify Eq.~(\ref{eq:V1}) a little by using $ \bm T_\La\cdot \bm T_\La =  C_a$, giving us
\begin{equation}
\begin{split}
\label{eq:V2}
\cV_{\La}(t)&\sket{\{p,f,c',c\}_{m}}
\\
={}&
\frac{\as}{2\pi}\,
\int_0^{1/(1+y)}\!\frac{dz}{z}\
\sum_{\hat a}
\frac{
f_{\hat a/A}(\eta_{\La}/z, y Q^2)}
{f_{a/A}(\eta_{\La}, y Q^2)}
\\&\times \Bigg\{
\left(P_{\hat a a}(z)
- \delta_{\hat a a}\frac{2zC_a}{1-z}
\right)
[1 \otimes 1]
\\&\qquad -
\delta_{\hat a a}
\sum_{k\ne \La}\,
z\,v(y,z,\theta_{\La k})\,
\big([(\bm T_\La\cdot \bm T_k)\otimes 1] + [1 \otimes (\bm T_\La\cdot \bm T_k)]\big)
\Bigg\}
\\&\times
\sket{\{p,f,c',c\}_{m}}
\;.
\end{split}
\end{equation}

\subsection{The initial state cross section changing exponent}

Now we need $\cV_\La(t) - \cS_\La(t)$. Using Eqs.~(\ref{eq:S2}) and (\ref{eq:V2}), we have
\begin{equation}
\begin{split}
\label{eq:VS1}
[\cV_{\La}(t) - &\cS_{\La}(t)]\sket{\{p,f,c',c\}_{m}} 
\\={}& 
\frac{\as}{2\pi} \Bigg\{
\sum_{k \ne \La,\Lb}
\log\!\left(
\frac{1 + \sqrt{1 + y^2/[\psi_{\La k}^2(1-z_k)^2]}}
{4y/(1-\cos\theta_{\La k})}
\right)
\\&\qquad\quad\times
\big(
[(\bm{T}_\La\cdot \bm{T}_k)\otimes 1]
+ [1 \otimes (\bm{T}_\La\cdot \bm{T}_k)]
\big)
\\&\qquad  
-
\mi \pi\,
\big(
[(\bm{T}_\La\cdot \bm{T}_\Lb)\otimes 1]
- [1 \otimes (\bm{T}_\La\cdot \bm{T}_\Lb)]
\big)
\\&\qquad  
- \sum_{\hat a} 
\int_0^{1}\!\frac{dz}{z}\
\bigg(
P_{a\hat a}\!\left(z\right)
\,
\frac{f_{\hat a/A}(\eta_{\La}/z, yQ^2)}
{f_{a/A}(\eta_{\La}, yQ^2)}
-
\delta_{a\hat a}\,
\frac{2 zC_a}{1-z}
\bigg)
[1\otimes 1]
\\&\qquad+
\sum_{\hat a}\int_0^{1/(1+y)}\!\frac{dz}{z}\
\frac{
f_{\hat a/A}(\eta_{\La}/z, y Q^2)}
{f_{a/A}(\eta_{\La}, y Q^2)}\,
\left(P_{\hat a a}(z)
- \delta_{\hat a a}\frac{2zC_a}{1-z}
\right)[1\otimes 1]
\\&\qquad
-
\int_0^{1/(1+y)}\!\frac{dz}{z}\
\frac{
f_{a/A}(\eta_{\La}/z, y Q^2)}
{f_{a/A}(\eta_{\La}, y Q^2)}\,
\sum_{k\ne \La}\,
z\,v(y,z,\theta_{\La k})
\\&\qquad\quad\times
\big([(\bm T_\La\cdot \bm T_k)\otimes 1] + [1 \otimes (\bm T_\La\cdot \bm T_k)]\big)
\Bigg\}
\\&\times
\sket{\{p,f,c',c\}_{m}}
\;.
\end{split}
\end{equation}

We can simplify this. We replace
\begin{equation}
P_{\hat a a}(z) = P_{\hat a a}^{\rm reg}(z) 
+ \delta_{\hat a a}\frac{2 z C_a}{1-z}
\;.
\end{equation}
This gives
\begin{equation}
\begin{split}
\label{eq:VS2}
[\cV_{\La}(t) - &\cS_{\La}(t)]\sket{\{p,f,c',c\}_{m}} 
\\={}& 
\frac{\as}{2\pi} \Bigg\{
\sum_{k \ne \La,\Lb}
\log\!\left(
\frac{1 + \sqrt{1 + y^2/[\psi_{\La k}^2(1-z_k)^2]}}
{4y/(1-\cos\theta_{\La k})}
\right)
\\&\qquad\quad\times
\big(
[(\bm{T}_\La\cdot \bm{T}_k)\otimes 1]
+ [1 \otimes (\bm{T}_\La\cdot \bm{T}_k)]
\big)
\\&\qquad  
-
\mi \pi\,
\big(
[(\bm{T}_\La\cdot \bm{T}_\Lb)\otimes 1]
- [1 \otimes (\bm{T}_\La\cdot \bm{T}_\Lb)]
\big)
\\&\qquad  
- \sum_{\hat a} 
\int_0^{1}\!\frac{dz}{z}\
P_{a\hat a}^{\rm reg}\!\left(z\right)
\,
\frac{f_{\hat a/A}(\eta_{\La}/z, yQ^2)}
{f_{a/A}(\eta_{\La},yQ^2)}\,
[1\otimes 1]
\\&\qquad  
+
\int_0^{1}\!\frac{dz}{z}\
\bigg(
1
-\frac{f_{a/A}(\eta_{\La}/z, \mu_\La^2(t))}
{f_{a/A}(\eta_{\La}, \mu_\La^2(t))}
\bigg)
\frac{2 zC_a}{1-z}\,
[1\otimes 1]
\\&\qquad+
\sum_{\hat a}\int_0^{1/(1+y)}\!\frac{dz}{z}\
\frac{
f_{\hat a/A}(\eta_{\La}/z, y Q^2)}
{f_{a/A}(\eta_{\La}, y Q^2)}\,
P_{\hat a a}^{\rm reg}(z)\,
[1\otimes 1]
\\&\qquad
-
\int_0^{1/(1+y)}\!\frac{dz}{z}\
\frac{
f_{a/A}(\eta_{\La}/z, y Q^2)}
{f_{a/A}(\eta_{\La}, y Q^2)}\,
\sum_{k\ne \La}\,
z\,v(y,z,\theta_{\La k})
\\&\qquad\quad\times
\big([(\bm T_\La\cdot \bm T_k)\otimes 1] + [1 \otimes (\bm T_\La\cdot \bm T_k)]\big)
\Bigg\}
\\&\times
\sket{\{p,f,c',c\}_{m}}
\;.
\end{split}
\end{equation}
The two terms involving $P_{\hat a a}^{\rm reg}$ cancel except for not having the same limits of integration. We divide the last term into three terms by using
\begin{equation}
\begin{split}
\frac{
f_{a/A}(\eta_{\La}/z, y Q^2)}
{f_{a/A}(\eta_{\La}, y Q^2)}\,
v(y,z,\xi_k) ={}& 
-\left(1 - \frac{f_{a/A}(\eta_{\La}/z, y Q^2)}
{f_{a/A}(\eta_{\La}, y Q^2)}
\right)\frac{1}{1-z}
\\ & +
\left(
1 - \frac{f_{a/A}(\eta_{\La}/z, y Q^2)}
{f_{a/A}(\eta_{\La}, y Q^2)}
\right)
\left(\frac{1}{1-z} - v(y,z,\theta_{\La k})\right)
\\ & + v(y,z,\theta_{\La k})
\;.
\end{split}
\end{equation}
This gives
\begin{equation}
\begin{split}
\label{eq:VS3}
[\cV_{\La}(t) - &\cS_{\La}(t)]\sket{\{p,f,c',c\}_{m}} 
\\={}& 
\frac{\as}{2\pi} \Bigg\{
\sum_{k \ne \La,\Lb}
\log\!\left(
\frac{1 + \sqrt{1 + y^2/[\psi_{\La k}^2(1-z_k)^2]}}
{4y/(1-\cos\theta_{\La k})}
\right)
\\&\qquad\quad\times
\big(
[(\bm{T}_\La\cdot \bm{T}_k)\otimes 1]
+ [1 \otimes (\bm{T}_\La\cdot \bm{T}_k)]
\big)
\\&\qquad  
-
\mi \pi\,
\big(
[(\bm{T}_\La\cdot \bm{T}_\Lb)\otimes 1]
- [1 \otimes (\bm{T}_\La\cdot \bm{T}_\Lb)]
\big)
\\&\qquad  
- \sum_{\hat a} 
\int_{1/(1+y)}^1\!\frac{dz}{z}\
P_{a\hat a}^{\rm reg}\!\left(z\right)
\,
\frac{f_{\hat a/A}(\eta_{\La}/z, yQ^2))}
{f_{a/A}(\eta_{\La}, yQ^2)}\,
[1\otimes 1]
\\&\qquad  
+
\int_{0}^{1}\!\frac{dz}{z}\
\bigg(
1
-\frac{f_{a/A}(\eta_{\La}/z, yQ^2)}
{f_{a/A}(\eta_{\La}, yQ^2)}
\bigg)
\frac{2 zC_a}{1-z}\,
[1\otimes 1]
\\&\qquad  
+
\int_0^{1/(1+y)}\!\frac{dz}{z}\
\left(
1
-\frac{
f_{a/A}(\eta_{\La}/z, y Q^2)}
{f_{a/A}(\eta_{\La}, y Q^2)}
\right)
\frac{z}{1-z}
\\&\qquad\quad\times
\sum_{k\ne \La}\,
\big([(\bm T_\La\cdot \bm T_k)\otimes 1] + [1 \otimes (\bm T_\La\cdot \bm T_k)]\big)
\\&\qquad  
-
\int_0^{1/(1+y)}\!\frac{dz}{z}\
\left(
1
-\frac{
f_{a/A}(\eta_{\La}/z, y Q^2)}
{f_{a/A}(\eta_{\La}, y Q^2)}
\right)
\sum_{k\ne \La}\,
\left(
\frac{z}{1-z}
-z\,v(y,z,\theta_{\La k})
\right)
\\&\qquad\quad\times
\big([(\bm T_\La\cdot \bm T_k)\otimes 1] + [1 \otimes (\bm T_\La\cdot \bm T_k)]\big)
\\&\qquad  
-
\int_0^{1/(1+y)}\!\frac{dz}{z}\
\sum_{k\ne \La}\,
z\,v(y,z,\theta_{\La k})
\\&\qquad\quad\times
\big([(\bm T_\La\cdot \bm T_k)\otimes 1] + [1 \otimes (\bm T_\La\cdot \bm T_k)]\big)
\Bigg\}
\\&\times
\sket{\{p,f,c',c\}_{m}}
\;.
\end{split}
\end{equation}
In the first of the new terms, we can use
\begin{equation}
\sum_{k\ne \La}\,
\big([(\bm T_\La\cdot \bm T_k)\otimes 1] + [1 \otimes (\bm T_\La\cdot \bm T_k)]\big)
= -2 C_a [1 \otimes 1]
\end{equation}
Then this term almost cancels the term that precedes it, leaving
\begin{equation}
\begin{split}
\label{eq:VS4}
[\cV_{\La}(t) - &\cS_{\La}(t)]\sket{\{p,f,c',c\}_{m}} 
\\={}& 
\frac{\as}{2\pi} \Bigg\{
\sum_{k \ne \La,\Lb}
\log\!\left(
\frac{1 + \sqrt{1 + y^2/[\psi_{\La k}^2(1-z_k)^2]}}
{4y/(1-\cos\theta_{\La k})}
\right)
\\&\qquad\quad\times
\big(
[(\bm{T}_\La\cdot \bm{T}_k)\otimes 1]
+ [1 \otimes (\bm{T}_\La\cdot \bm{T}_k)]
\big)
\\&\qquad  
-
\mi \pi\,
\big(
[(\bm{T}_\La\cdot \bm{T}_\Lb)\otimes 1]
- [1 \otimes (\bm{T}_\La\cdot \bm{T}_\Lb)]
\big)
\\&\qquad  
- \sum_{\hat a} 
\int_{1/(1+y)}^1\!\frac{dz}{z}\
P_{a\hat a}^{\rm reg}\!\left(z\right)
\,
\frac{f_{\hat a/A}(\eta_{\La}/z, yQ^2)}
{f_{a/A}(\eta_{\La}, yQ^2)}\,
[1\otimes 1]
\\&\qquad  
+
\int_{1/(1+y)}^{1}\!\frac{dz}{z}\
\bigg(
1
-\frac{f_{a/A}(\eta_{\La}/z, yQ^2)}
{f_{a/A}(\eta_{\La}, yQ^2))}
\bigg)
\frac{2 zC_a}{1-z}\,
[1\otimes 1]
\\&\qquad  
-
\int_0^{1/(1+y)}\!\frac{dz}{z}\
\left(
1
-\frac{
f_{a/A}(\eta_{\La}/z, y Q^2)}
{f_{a/A}(\eta_{\La}, y Q^2)}
\right)
\sum_{k\ne \La}\,
\left(
\frac{z}{1-z}
-z\,v(y,z,\theta_{\La k})
\right)
\\&\qquad\quad\times
\big([(\bm T_\La\cdot \bm T_k)\otimes 1] + [1 \otimes (\bm T_\La\cdot \bm T_k)]\big)
\\&\qquad  
-
\int_0^{1/(1+y)}\!\frac{dz}{z}\
\sum_{k\ne \La}\,
z\,v(y,z,\theta_{\La k})
\big([(\bm T_\La\cdot \bm T_k)\otimes 1] + [1 \otimes (\bm T_\La\cdot \bm T_k)]\big)
\Bigg\}
\\&\times
\sket{\{p,f,c',c\}_{m}}
\;.
\end{split}
\end{equation}

Finally, we combine the first and last terms, using the representation (\ref{eq:Sakintegral}) of the logarithm in the first term as an integral over $z$. We have not yet specified the scale argument of $\as$. We note that the virtuality of an initial state splitting is $yQ^2$ and its transverse momentum (as defined in \textsc{Deductor}) is $(1-z) y Q^2$. We set the $\as$ scale to either $\lambda_\LR y Q^2$ or $(1-z)\lambda_\LR y Q^2$, where $\lambda_\LR = \exp\left(- [ C_\LA(67 - 3\pi^2)- 10\, n_\Lf]/[3\, (33 - 2\,n_\Lf)]\right) \approx 0.4$ \cite{lambdaR}. (In $\lambda_\LR$,  the number $n_\Lf$ of active flavors depends on the scale.) We also insert an infrared cutoff $(1-z) y Q^2 > m_\perp^2(a)$ where $m_\perp(a)$ is the quark mass when $a$ is a bottom or charm flavor and is otherwise of order 1 GeV. The result is not sensitive to the infrared cutoff. This gives
\begin{equation}
\begin{split}
\label{eq:VS5alt}
[\cV_{\La}(t) - &\cS_{\La}(t)]\sket{\{p,f,c',c\}_{m}} 
\\={}& 
\Bigg\{
\int_{1/(1+y)}^{1}\!\frac{dz}{z}\ 
\frac{\as((1-z)\lambda_\LR y Q^2)}{2\pi}\,
\theta((1-z) y Q^2 > m_\perp^2(a))
\\&\quad\times
\bigg(
1
-\frac{f_{a/A}(\eta_{\La}/z, y Q^2)}
{f_{a/A}(\eta_{\La}, y Q^2)}
\bigg)
\frac{2 zC_a}{1-z}\,
[1\otimes 1]
\\&  
- \sum_{\hat a} 
\int_{1/(1+y)}^1\!\frac{dz}{z}\ 
\frac{\as(\lambda_\LR y Q^2)}{2\pi}\,
\theta((1-z) y Q^2 > m_\perp^2(a))
\\&\quad\times
P_{a\hat a}^{\rm reg}\!\left(z\right)
\,
\frac{f_{\hat a/A}(\eta_{\La}/z, y Q^2)}
{f_{a/A}(\eta_{\La}, y Q^2)}\,
[1\otimes 1]
\\&  
-\int_0^{1/(1+y)}\!\frac{dz}{z}\ 
\frac{\as((1-z)\lambda_\LR y Q^2)}{2\pi}\,
\theta((1-z) y Q^2 > m_\perp^2(a))
\\&\quad\times
\left(
1
-\frac{
f_{a/A}(\eta_{\La}/z, y Q^2)}
{f_{a/A}(\eta_{\La}, y Q^2)}
\right)
\sum_{k\ne \La}\,
\left(
\frac{z}{1-z}
-z\,v(y,z,\theta_{\La k})
\right)
\\&\quad\times
\big([(\bm T_\La\cdot \bm T_k)\otimes 1] + [1 \otimes (\bm T_\La\cdot \bm T_k)]\big)
\\&  
+ 
\sum_{k \ne \La,\Lb}
I_k(y,\xi_k,z_k)
\big(
[(\bm{T}_\La\cdot \bm{T}_k)\otimes 1]
+ [1 \otimes (\bm{T}_\La\cdot \bm{T}_k)]
\big)
\\&  
-\mi \pi\,
\frac{\as(\lambda_\LR y Q^2))}{2\pi}
\big(
[(\bm{T}_\La\cdot \bm{T}_\Lb)\otimes 1]
- [1 \otimes (\bm{T}_\La\cdot \bm{T}_\Lb)]
\big)
\Bigg\}
\\&\times
\sket{\{p,f,c',c\}_{m}}
\;,
\end{split}
\end{equation}
where we have defined
\begin{equation}
\begin{split}
I_k(y,\xi_k,z_k)
={}&
\int_0^{1/(1+y)}\!dz\
\frac{\as((1-z)\lambda_\LR y Q^2)}{2\pi}\,
\theta((1-z) y Q^2 > m_\perp^2(a))
\\&\times
\left[
\frac{\theta(z > z_k)}{\sqrt{(1-z)^2 +  y^2/\psi_{\La k}^2}} 
- \frac{\theta(z > z_k)}{1-z}
- v(y,z,\theta_{\La k})
\right]
\;.
\end{split}
\end{equation}

Eq.~(\ref{eq:VS5alt}) replaces Eq.~(7.21) of Ref.~\cite{DuctThreshold}.
The first term is the main threshold term. The second is a correction from $P_{a\hat a}^{\rm reg}$. The third term is corrected from what we had because we use the function $v(y,z,\xi_k)$ instead of our previous approximation to it. The fourth term was approximated by zero in Ref.~\cite{DuctThreshold}. The fifth term is from the $\mi\pi$ part of the virtual corrections. This is probability preserving and, according to Eq.~(\ref{eq:UVexponential02}), is not included in the threshold correction.


\end{document}